\documentclass[useAMS,usenatbib]{mn2e}

\def\aj{{AJ}}                  
            
\def\apj{{ApJ}}                
\def\apjl{{ApJ}}               
\def\apjs{{ApJS}}

\def\aap{{A\&A}}               
         
\def\aaps{{A\&AS}}

\def\mnras{{MNRAS}}

\def\pasa{{PASA}}

\def\nat{{Nature}}

\usepackage{txfonts,amssymb}
\usepackage[pdftex]{graphicx}
\usepackage{lscape}
\usepackage{longtable}

\newcommand{\HI}[0]{H\,{\sevensize I} }

\title[A blind HI survey in the Ursa Major region]{A blind HI survey in the Ursa Major region \thanks{The data cubes were obtained with the Lovell telescope at the Jodrell Bank observatory, UK.}}
\author[K. Wolfinger et al.]{K.~Wolfinger$^{1,2}$\thanks{E-mail:
kwolfinger@astro.swin.edu.au}, V.~A.~Kilborn$^1$, B.~S.~Koribalski$^2$, R.~F.~Minchin$^3$, P.~J.~Boyce$^4$, \newauthor M.~J.~Disney$^5$, R.~H.~Lang$^5$, C.~A.~Jordan$^6$\\
$^1$ Centre for Astrophysics \& Supercomputing, Swinburne University of Technology, Mail H39, PO Box 218, Hawthorn, VIC 3122, Australia\\
$^2$ CSIRO Astronomy and Space Science, PO Box 76, Epping, NSW 1710, Australia\\
$^3$ Arecibo Observatory, HC03 Box 53995, Arecibo, Puerto Rico 00612\\
$^4$ Registry, Cardiff University, Cardiff, CF10 3UA, United Kingdom\\
$^5$ School of Physics and Astronomy, Cardiff University, Cardiff, CF24 3YB, United Kingdom\\ 
$^6$ Jodrell Bank Observatory, Macclesfield, Cheshire, SK11 9DL, United Kingdom}
\begin{document}

\date{Accepted 2012 October 4.  Received 2012 September 30; in original form 2012 January 31}
\pagerange{\pageref{firstpage}--\pageref{lastpage}} \pubyear{2012}
\maketitle
\label{firstpage}


\begin{abstract}

We have conducted the first blind \HI survey covering 480~deg$^2$ and a heliocentric velocity range from 300--1900~km~s$^{-1}$ to investigate the \HI content of the nearby spiral-rich Ursa Major region and to look for previously uncatalogued gas-rich objects. Here we present the catalog of \HI sources. The \HI data were obtained with the 4-beam receiver mounted on the 76.2-m Lovell telescope (FWHM 12~arcmin) at the Jodrell Bank Observatory (UK) as part of the \HI Jodrell All Sky Survey (HIJASS). We use the automated source finder {\small DUCHAMP} and identify 166~\HI sources in the data cubes with \HI masses in the range of 10$^{7}$--10$^{10.5}$~M$_{\odot}$. Our Ursa Major \HI catalogue includes 10 first time detections in the 21-cm emission line. 

We identify optical counterparts for 165~\HI sources (99~per~cent). For 54~\HI sources ($\sim 33$~per~cent) we find numerous optical counterparts in the HIJASS beam, indicating a high density of galaxies and likely tidal interactions. Four of these \HI systems are discussed in detail.

We find only one \HI source (1~per~cent) without a visible optical counterpart out of the 166~\HI detections. Green Bank Telescope (FWHM 9~arcmin) follow-up observations confirmed this \HI source and its \HI properties. The nature of this detection is discussed and compared to similar sources in other \HI surveys.

\end{abstract}


\section{Introduction}

The distribution of galaxies in the Universe is not uniform -- galaxies are found in filamentary structures, groups and clusters (e.g. \citealt{colless2001}). The observed `cosmic web' structure is in good agreement with numerical simulations according to Cold Dark Matter cosmology (e.g. \citealt{springel2005}; \citealt{simon2007}), whereby gravity drives the formation of structure from small, primordial, Gaussian fluctuations. Dark matter haloes grow through merging and accretion of smaller ones. Within the dark matter haloes, galaxies form by cooling of baryons (e.g. \citealt{white1978}). Hence, small dark matter haloes are the primordial building blocks of larger structures in the hierarchical model of structure formation. 

Dwarf irregular galaxies with low stellar luminosity and low surface brightness galaxies are typically discovered in neutral hydrogen (H\,{\sevensize I}) surveys (\citealt{schombert1992}, \citealt{minchin2003}, \citealt{giovanelli2005}), as they are rich in neutral gas but difficult to detect in optical or infrared surveys. Whilst optical surveys are most sensitive to high surface brightness early-type galaxies \citep{disney1976}, \HI surveys are unbiased by stellar populations and are most likely to detect gas-rich late-type spirals and irregular galaxies. Hence optical and \HI surveys complement each other as there is a rough correlation between increasing \HI content from early to late type galaxies and decreasing optical surface brightness with morphological type \citep{toribio2011}. Discoveries in \HI surveys also include intriguing objects, e.g. the detection of an H\,{\sevensize I}-massive cloud without stellar counterpart, within the NGC2442 group \citep{ryder2001}, extended \HI arms or tidal tails \citep{koribalski_lopez2009} and \HI between galaxies in groups (\citealt{koribalski_manthey2005}, \citealt{kilborn2006}, \citealt{english2010}). Large-area blind \HI surveys have been conducted such as the Arecibo HI Strip Survey (AHISS, \citealt{zwaan1997}), the Arecibo Dual Beam Survey (ADBS, \citealt{rosenberg2000}), the \HI Jodrell All-Sky Survey (HIJASS, \citealt{lang2003}), the \HI Parkes All-Sky Survey (HIPASS, \citealt{staveley-smith1996}, \citealt{barnes2001}, \citealt{koribalski2004}, \citealt{meyer2004}), the Arecibo Legacy Fast ALFA survey (ALFALFA, \citealt{giovanelli2005}) and the Arecibo Galaxy Environment Survey (AGES, \citealt{auld2006}). Upcoming \HI surveys such as the ASKAP \HI All-Sky Survey (known as WALLABY, \citealt{koribalski_staveley2009}\footnote{http://www.atnf.csiro.au/research/WALLABY/proposal.html}) will further contribute to our knowledge of gas-rich galaxies in the local Universe. 

Isolated galaxies tend to have noticeably different properties from galaxies residing in dense regions. Observations have shown that early-type galaxies are common in clusters whereas late-type galaxies dominate the less dense environments (morphology-density relation; e.g. \citealt{dressler1980}). The importance of environment and the idea of transformation of galaxies in dense environments arose when \citet{butcher1978} discovered that galaxy clusters at higher redshift contain a population of blue galaxies which nearby galaxy cluster counterparts are missing. However, it is still a matter of debate, which properties correlate with the decrease of blue galaxies from high- to low-redshift galaxy clusters, e.g. X-ray luminosity (\citealt{andreon1999}; \citealt{fairley2002}), richness (\citealt{margoniner2001}; \citealt{propris2004}), cluster concentration (\citealt{butcher1984}; \citealt{propris2004}), presence of substructure (\citealt{metevier2000}), cluster velocity dispersion (\citealt{propris2004}). \citet{lewis2002} and \citet{gomez2003} discuss the correlation between star formation rate and local projected density. The authors find low star formation rates at 2-3 virial cluster radii -- which is equivalent to poor galaxy group densities. Therefore, galaxies may be preprocessed in groups, as physical processes in the studied clusters can not solely account for the observed variations in galaxy properties. Galaxy transformation as a smooth function of environment is also suggested by \citet{cappellari2011} in terms of a \textit{kinematic} morphology-density relation, i.e. separation into fast and slow rotators.

There have been a number of studies on the effect of environment on the \HI content of galaxies in clusters (e.g. \citealt{giovanelli1985}, \citealt{solanes2001}, \citealt{gavazzi2005}, \citealt{chung2009}), and loose groups (e.g. \citealt{kilborn2005}; \citealt{omar2005}). Through observations of isolated galaxies, we can define the expected \HI mass in normal spiral galaxies of a particular linear optical diameter and morphological type. In dense regions, such as clusters, there is often an \HI deficiency observed, indicating that \HI gas is removed from galaxies in these environments (e.g. \citealt{solanes2001}). Recent observations have found a similar \HI deficiency in the less dense group environments (\citealt{omar2005}, \citealt{kilborn2009}) indicating that the spirals may begin to lose their \HI reservoir well before reaching the dense centre of a cluster. 

An ideal target to study the effect of environment on the evolution of gas-rich galaxies, is the nearby Ursa Major cluster and its surroundings as most of the known member galaxies are late-type galaxies. The Ursa Major cluster is quite different in its nature compared to the well-known and similarly distant Virgo and Fornax clusters. Generally clusters are composed of hundreds of gas-poor early-type galaxies clustered around a central larger elliptical (e.g. \citealt{dressler1980}). The Ursa Major cluster is smaller -- \citet{tully1996} identified 79 high-probability cluster members. The cluster is essentially composed of late-type galaxies, in particular spiral galaxies with normal gas content, that are not centrally concentrated \citep{tully1996}. This lack of a concentration towards any core and the overlap with many groups nearby the cluster might explain why the Ursa Major cluster is less studied compared to the Virgo and Fornax clusters. The several groups nearby the Ursa Major cluster make studying its dynamics a complex problem and the cluster difficult to define (\citealt{appleton1982}, \citealt{tully1996}). However, \citet{tully1996} define the cluster members to lie within a projected 7.5$^{\circ}$ circle centered at $ \alpha = 11^h56.9^m, \delta = +49^{\circ}22'$ and with heliocentric velocities between 700 and 1210~km~s$^{-1}$.

The cluster lies at a distance of 17.1~Mpc \citep{tully2008}. The few previous studies of the Ursa Major cluster have found that it has a low velocity dispersion of only 148~km s$^{-1}$, a virial radius of 880~kpc \citep{tully1996} and no X-ray emitting intra-cluster gas has yet been detected \citep{verheijen2001}. The Ursa Major cluster has a total mass of $4.4 \times 10^{13}$~M$_{\odot}$ \citep{tully1987} and a B-band absolute luminosity of $5.5 \times 10^{11}$~L$_{\odot}$ \citep{tully2008}. Compared to the Virgo cluster this is about 30~per~cent of its light but only 5~per~cent of its mass (\citealt{tully1987}, \citealt{tully2008}). The number of H\,{\sevensize I}-rich galaxies in Ursa Major is comparable to those in Virgo \citep{tully1996} and the Ursa Major cluster almost meets the Abell richness 0 standard. To qualify as an Abell richness class 0 the criteria are (\textit{i}) 30 galaxies in B or (\textit{ii}) 26 in R within 2~mag of the third brightest galaxy. The Ursa Major cluster is slightly short - 29 galaxies in B and 25 in R \citep{tully1996}. Besides Ursa Major, only the Virgo cluster has such a high abundance of H\,{\sevensize I}-rich galaxies on a 2~Mpc-scale in the nearby Universe ($cz <$~3000~km~s$^{-1}$) \citep{tully1996}.

\citet{zwaan1999} note that the mostly late-type galaxies in the Ursa Major cluster do not seem to be seriously affected by tidal interactions indicating that the system is in an \textit{early evolutionary state}. However, \citet{verheijen2001} studied 43 spiral galaxies in detail and found 10 interacting systems that show distorted morphologies and tidal debris. There is a deficiency of known dwarf galaxies in the Ursa Major cluster -- there are only two dozen dwarfs, whereas up to $10^3$ could be expected to be found if the same steep luminosity function claimed for Virgo was valid for the Ursa Major cluster \citep{trentham2001}. To date, (\textit{i}) the small number of dwarf galaxies, (\textit{ii}) the low velocity dispersion, (\textit{iii}) the non-detection of X-ray emitting intra-cluster gas, (\textit{iv}) the low abundance of elliptical galaxies in the cluster and (\textit{v}) the total mass $< 10^{14} M_{\odot}$ challenge the question "why not call it a group?". Clusters usually have a broader velocity distribution, a hot intra-cluster gas which can strip the galaxies off their \HI gas and an enhanced population of both the early-type (S0 and E) galaxies and the \HI depleted spiral galaxies. As the majority of the Ursa Major members are H\,{\sevensize I}-rich spiral galaxies such as those one finds in the field, the distinction between the `Ursa Major cluster' as opposed to an `Ursa Major group' isn't clear. One hypothesis is that Ursa Major is a \textit{newly forming cluster} given that the estimated crossing time is approximately half a Hubble time, in contrast to Virgo and Fornax where the crossing times are 0.08 and 0.07 $H_{0}^{-1}$, respectively \citep{tully1996}. 

In this paper we present the catalogue of \HI sources as detected in the first blind \HI survey of the Ursa Major region (approximately 480~deg$^2$). The analyzed region overlaps with the Canes Venatici region for which \citet{kovac2009} have conducted a deep blind \HI survey (\HI mass sensitivity $\sim10^6$~M$_{\odot}$) using the Westerbork Synthesis Radio Telescope (WSRT) in an area of 86~deg$^2$ ($v\mathrm{_{sys}} \leq 1330$~km~s$^{-1}$).  Our \HI data were obtained as part of the \HI Jodrell All Sky Survey (HIJASS). The data acquisition and determination of the \HI sources are described in Section~\ref{sec:obs} including a discussion of the catalogue completeness and reliability. The \HI properties of the sources are presented in a catalogue in Section~\ref{sec:cat}. The \HI mass limit of the presented catalogue is 2.5$\times 10^8$~M$_{\odot}$ assuming a velocity width of 100~km~s$^{-1}$ and a distance of 17.1~Mpc. First results and basic property distributions of the HIJASS sources are presented in Section~\ref{sec:res}. Furthermore, we identify galaxy pairs/systems that are high-probability candidates for galaxy-galaxy interactions and discuss four systems in detail, two of which show strikingly extended \HI content. 

In a forthcoming paper we will investigate the nature of the Ursa Major cluster/group and probe the hypothesis that it is a \textit{newly forming cluster}. We will study substructures and their dynamics, and present dynamical masses, total mass-to-light ratios etc. for both individual galaxies and the subregions. Furthermore, to investigate the \HI properties of galaxies residing in different density regions, we will analyze the \HI sources in the Ursa Major cluster and compare them to galaxies residing in foreground groups and the less dense filament connecting the Ursa Major with the Virgo cluster, which lies south of Canes Venatici. 

\section{Observations and Data Analysis}
\label{sec:obs}

\subsection{Data acquisition and processing}
\label{sec:acquisition}

The \HI data cubes analyzed in this paper were obtained in 2001--2002 as part of the \HI Jodrell All Sky Survey (HIJASS) \citep{lang2003}. The blind \HI survey was conducted using the 4-beam receiver mounted on the 76.2-m Lovell telescope at Jodrell Bank observatory, UK. HIJASS covered more than 1000~deg$^{2}$ in the northern sky ($\delta\geq+22^{\circ}$) complementary to (\textit{i}) the \HI Parkes All Sky Survey (HIPASS) which surveyed the entire southern sky ($\delta\leq+2^{\circ}$) from 1997 to 2000 \citep{staveley-smith1996,barnes2001,koribalski2004} and (\textit{ii}) the northern extension to HIPASS ($+2^{\circ}\leq\delta\leq+25^{\circ} 30'$) carried out in 2000 to 2002 \citep{wong2006}. HIJASS provides the first blind \HI survey of the nearby Ursa Major cluster and its surroundings. To investigate the Ursa Major region we only consider velocities in the range 300 to 1900~km~s$\mathrm{^{-1}}$ (all velocities are given in the optical convention $cz$). Sources with systemic velocities below 300~km~s$^{-1}$ are excluded to minimize confusion with high-velocity clouds and Galactic emission -- an exception are 15 obviously real galaxies with systemic velocities between 150 and 300~km~s$^{-1}$ that are included in the catalogue. We note that HIJASS data suffers from a broad band of radio-frequency interference between 4500 and 7500~km~s$\mathrm{^{-1}}$ \citep{lang2003}, which lies outside our studied velocity range.

The Ursa Major region is covered by seven HIJASS data cubes measuring $8^{\circ}\times8^{\circ}$ and one HIJASS cube measuring $5^{\circ}\times8^{\circ}$ in size (the sky coverage is approximately 480 deg$^2$). The cubes span a range in right ascension (RA) between 10$^{h}$45$^{m}$ and 13$^{h}$04$^{m}$ and a range in declination (Dec.) between 22$^{\circ}$ and 54$^{\circ}$. We note that the analyzed region overlaps with the Canes Venatici region for which \citet{kovac2009} have conducted a deep blind \HI survey in an area of 86~deg$^2$ ($v\mathrm{_{sys}} \leq 1330$~km~s$^{-1}$). The HIJASS cubes have a spatial pixel size of $4 \times 4$~arcmin$^2$. In the third dimension, the channel spacing is 62.5~kHz which corresponds to an increment of 13.2~km~s$^{-1}$ at the rest frequency of H\,{\sevensize I}. The velocity resolution is 18~km~s$^{-1}$ after smoothing to reduce `ringing' caused by Galactic emission (applying 25 per cent Tukey filter). The root mean square (rms) noise level in the analyzed data is 12--14~mJy~beam$\mathrm{^{-1}}$. To minimize edge effects the overlapping HIJASS data cubes are mosaiced to one mega-cube. For further information on data reduction, the Lovell telescope and HIJASS see \citet{lang2003}.

The angular resolution of the Lovell telescope at 1.4~GHz is 12.0~arcmin \citep{lang2003}. The gridding process (using similar parameters as for HIPASS) slightly degrades the angular resolution (see \citealt{barnes2001}). According to \citet{barnes2001} the resultant beam size is not well-defined and depends on smoothing radius, pixel size, source position, shape and strength (signal-to-noise ratio) as the gridding is a non-linear process. One can only assess the gridded beam size by analysing simulations and synthetic source injections. To determine the gridded beam size, we inject synthetic point sources into the raw data and measure their properties after the gridding process. We determine the full width at half-maximum beamwidth (FWHM) of a two-dimensional Gaussian fit to the synthetic sources, analogous to \citet{barnes2001}. The measured beamwidths vary as a function of source peak brightness (in Jy~beam$^{-1}$, which corresponds to Jy for point sources). Given a wide range of peak flux densities in the presented catalogue -- the majority of sources have $33 \leq S\mathrm{_{p}} \leq 300$~mJy and a few strong sources range to 3.5~Jy in peak flux density -- we determine two appropriate beamwidths: (\textit{i}) 13.1$\pm$0.9~arcmin for the majority of sources with $S\mathrm{_{p}} \leq$~300~mJy and (\textit{ii})12.4$\pm$0.6~arcmin for sources with $S\mathrm{_{p}} >$~300~mJy.

The peak and integrated fluxes of point sources are preserved during the gridding process but for extended sources, the peak and integrated fluxes require a correction: We calculated the correction factors by similarly injecting synthetic extended sources into the raw data and measure their properties after the gridding process (analogous to \citealt{barnes2001}). The correction factors for extended sources as a function of FWHM are given in Table~\ref{tab:corr_factors}. We obtain this function by fitting a third order polynomial to the median data points (with $R^2$ of 99.9\%).  In our catalogue (Section~\ref{sec:catalogue}), we quote both the measured and corrected fluxes for extended sources.

\begin{table}
\begin{center}
\caption{The peak and integrated correction factors ($f\mathrm{_p}$ and $f\mathrm{_{int}}$) for extended sources as a function of FWHM ($x$).}
\begin{tabular}{| c | c | c |}
\multicolumn{3}{c}{$f(x)=a_3x^3+a_2x^2+a_1x+a_0$}\\
\hline\hline\\[-2ex]
Coefficient & $f\mathrm{_p}(x)$ & $f\mathrm{_{int}}(x)$ \\
\hline\hline\\[-2ex]
$a_3$ & $-4.78\times10^{-6}$ & $7.12\times10^{-6}$ \\
$a_2$ & $5.54\times10^{-4}$ & $-2.27\times10^{-4}$ \\
$a_1$ & $-2.26\times10^{-2}$ & $1.05\times10^{-2}$ \\
$a_0$ & 1.112 & 1.003 \\
\hline\\[-2ex]
\end{tabular}
\label{tab:corr_factors}
\end{center}
\end{table}

\subsection{Data analysis using {\small DUCHAMP}}
\label{sec:Duchamp}

We use the automated source finder {\small DUCHAMP-1.1.8} \citep{whiting2012}\footnote{http://www.atnf.csiro.au/people/Matthew.Whiting/Duchamp} on the \HI data cube to compile our initial candidate galaxy list. {\small DUCHAMP} is based on a threshold method and searches for groups of adjacent pixels in position and velocity that are above the given threshold, which is estimated from the cube statistics (in Jy~beam$^{-1}$) or alternatively specified from the user as a peak flux density value. Furthermore the user can (\textit{i}) reduce the noise prior to the searching via wavelet reconstruction and/or smoothing (spatial or spectral) and (\textit{ii}) control various parameters such as the minimum extent of the detections in pixels and/or channels. {\small DUCHAMP} provides a list of the \HI detections and their properties, the \HI spectra and integrated \HI intensity maps (moment zero). The {\small DUCHAMP} outputs need to be inspected carefully as \HI sources are occasionally merged. We can identify such sources by eye and separate them when they do not overlap both spatially and spectrally.

We made a comprehensive test of various input parameters and methods to facilitate {\small DUCHAMP} detections of low peak flux and extended \HI sources without disproportionally increasing noise detections (see Section~\ref{sec:reliability} and \ref{sec:comp}). We first apply a spectral Hanning smoothing to the HIJASS data and then reconstruct them with the `\`a trous' wavelet method. Hanning smoothing correlates neighbouring pixels in the spectral domain and determines the weighted average flux over a certain number of channels -- the channel spacing remains 13.2~km~s$^{-1}$. We determine that the optimal Hanning smoothing width for the data is seven channels (see Section~\ref{sec:reliability}). The wavelet method reconstructs each spectrum separately and we chose the following criteria when compiling the {\small DUCHAMP} detection list: (\textit{i}) the detection must be contiguous in two channels and (\textit{ii}) the peak flux density of the candidate galaxies must be above 16~mJy. This is an approximate 2.5$\sigma$ peak flux density threshold as the mean noise level in the smoothed and reconstructed cube is $\sim$6.6~mJy~beam$^{-1}$.

In this paper, we only consider 5$\sigma$ detections in peak flux in the edge-masked HIJASS megacube to present a reliable and complete catalogue of \HI sources in the Ursa Major region within 300 to 1900~km~s$^{-1}$. This corresponds to a minimum in peak flux density of 33~mJy and an \HI mass limit of $2.5\times10^8$ M$_{\odot}$ (assuming a velocity width of 100~km~s$^{-1}$ and a distance of 17.1~Mpc). These selection criteria lead to a {\small DUCHAMP} detection list with 140 candidate galaxies including merged detections (doubles to quads) that are labelled in Section~\ref{sec:catalogue}.

\subsection{Catalogue reliability -- Negative troughs}
\label{sec:reliability}

To test the reliability of the $5\sigma$ peak flux {\small DUCHAMP} detection list, we investigate the negative troughs in the edge-masked HIJASS cube. When using the selection criteria as discussed in Section~\ref{sec:Duchamp}, we identify 25 negative troughs with $S\mathrm{_p} \leq -33$~mJy and $w_{50} \geq$~26.4~km~s$^{-1}$. All 25 negative troughs can be associated with negative bandpass sidelobes, i.e. an artifact of bandpass removal which is apparent north and south of strong HI sources (see e.g. \citealt{barnes2001}). This lack of negative detections (noise peaks) indicates that the positive {\small DUCHAMP} candidate galaxy list with $S\mathrm{_p} \geq $~33~mJy and $w_{50} \geq$~26.4~km~s$^{-1}$contains high-probability real \HI sources.

Hanning smoothing is known to enhance the brightness of sources -- on the scale of the smoothing width -- with respect to the noise in each channel map, but reduces the peak strength (signal-to-noise) of sources on other scales. \citet{whiting2012} shows that an appropriately chosen smoothing width can still increase the recovery rate of sources on different scales than the smoothing width as the noise level is suppressed. However, to prevent loss of information on different scales, we also use the wavelet reconstruction to further suppress the noise and to enhance structures on various scales (narrow and broad profiles). Using wavelet reconstruction also suppresses noise peaks, baseline ripples or radio-frequency interferences as discussed in \citet{westmeier2011}. For further information on {\small DUCHAMP} and the noise suppression methods see \citet{whiting2012}.

Some restrictions on spatial and spectral extent of {\small DUCHAMP} detections greatly improve the reliability, such as the applied minimum number of channels ($\geq 2$) in which the detection must be contiguous \citep{whiting2012}. The selection criteria as listed in Section~\ref{sec:Duchamp} are chosen to limit the number of false detections and to present a reliable catalogue of \HI sources in the Ursa Major region in this paper.

\subsection{Catalogue reliability -- Follow-up observations}
\label{sec:reliability2}

The reliability of the catalogue is increased by follow-up observations and rejection of unconfirmed sources from the catalogue. We re-observed 17 apparent first time detections in HIJASS with the Green Bank Telescope (GBT, FWHM 9~arcmin) -- nine of which have an associate optical counterpart that match in position and velocity to the \HI detection and eight candidate galaxies/\HI clouds without a previously catalogued optical counterpart. The rms noise level is between 4 and 7~mJy for \HI sources with an associate optical counterpart and approximately 3~mJy for the candidate galaxies/\HI clouds. Given that the lowest peak flux density of the candidate galaxies/\HI clouds is 33~mJy, this will lead to signal-to-noise detections $> 10$. The velocity resolution of the GBT data is 3.3~km~s$^{-1}$ after boxcar smoothing.

The GBT follow-up observations confirmed all nine \HI sources with an associate optical counterpart. However, only one out of the eight candidate galaxies/\HI clouds was confirmed, namely HIJASS~J1219+46. Therefore, 7 initial {\small DUCHAMP} detections are excluded from the presented catalogue in this paper. Inspecting the original as well as the smoothed and resonstructed data cubes by eye reveal that the unconfirmed detections are noise peaks on top of large-scale baseline ripples. {\small DUCHAMP} provides an option `flag baseline' which removes the baseline from each spectrum prior to compiling the detection list. However, we did not utilize this option as not only does it remove all candidate galaxies/\HI clouds from the detection list, some of the faint \HI detections with an associate optical counterpart were not included as well.

\subsection{Catalogue completeness -- Synthetic source injection}
\label{sec:comp}

\begin{figure*}
\begin{center}
\begin{tabular}{p{8.5cm} p{8.5cm}}
\includegraphics[width=10 cm]{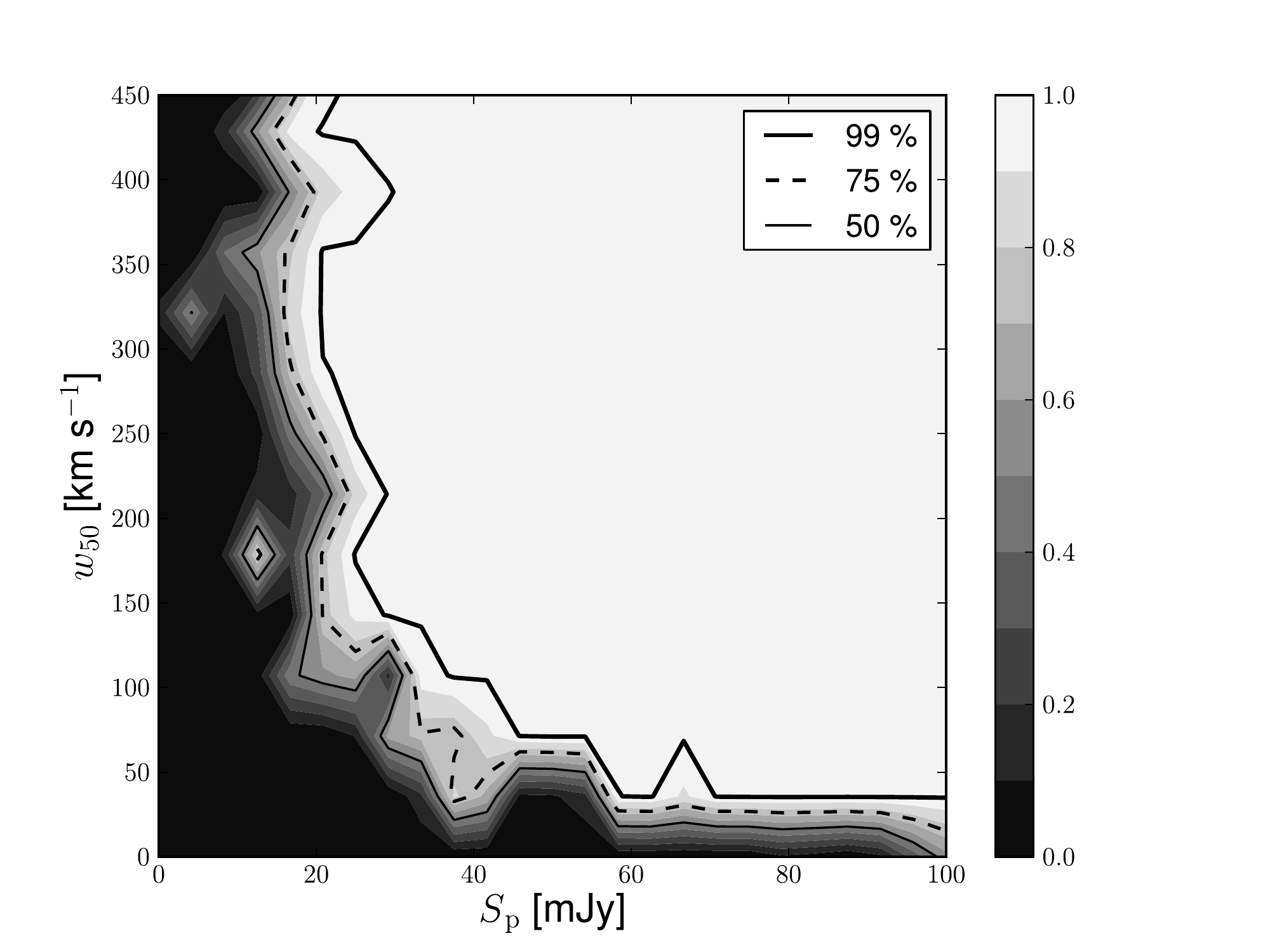} & 
\includegraphics[width=10 cm]{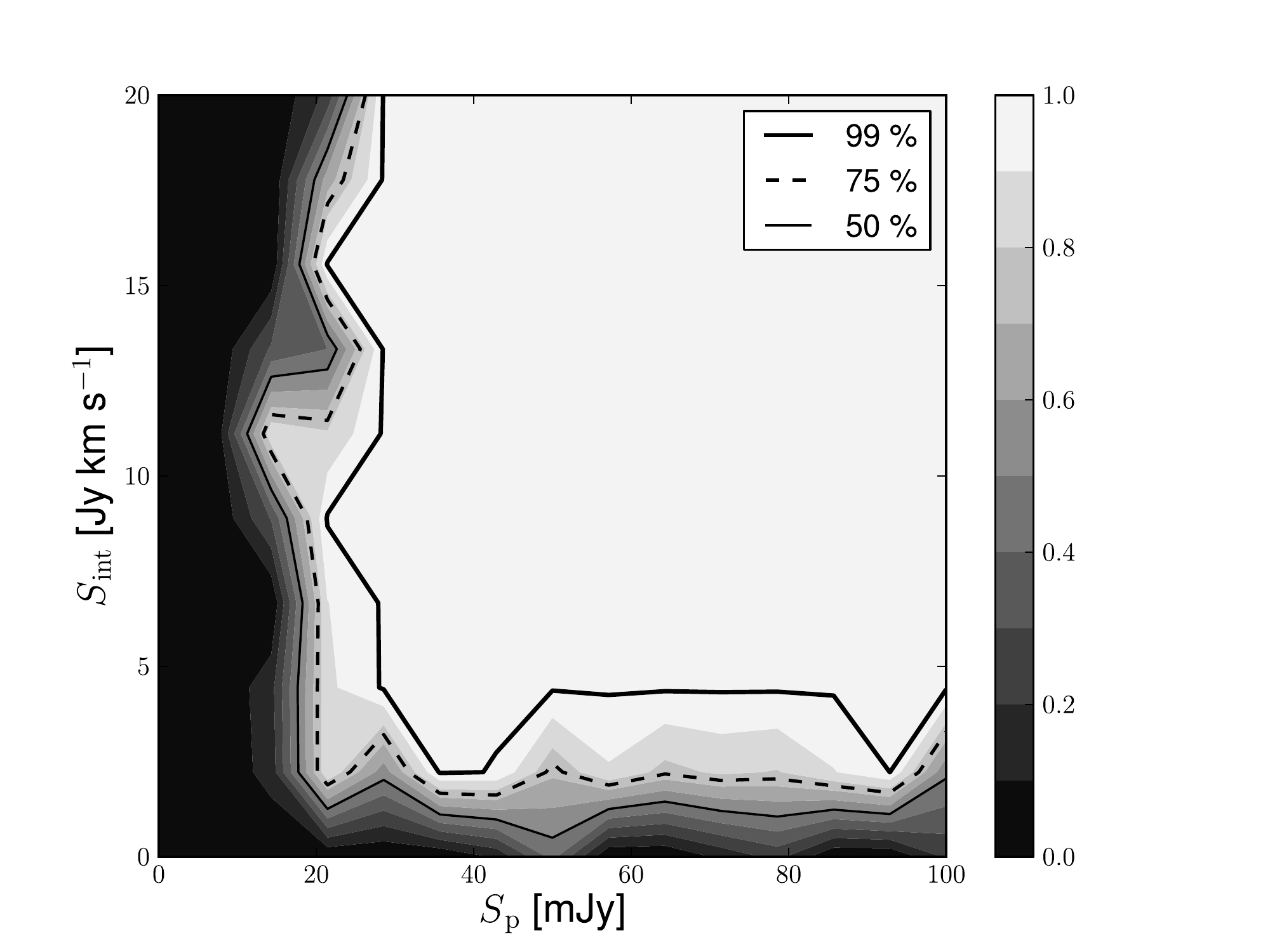}\\
\end{tabular}
\end{center}
\caption{The \small{DUCHAMP} completeness in a representative HIJASS cube as a function of $S\mathrm{_{p}}$ and $w_{50}$ (left) and as a function of $S\mathrm{_{p}}$ and $S\mathrm{_{int}}$ (right). The brighter colours correspond to high completeness levels and the contour lines show the 99, 75 and 50~per~cent completeness levels.}
\label{fig:comp_one}
\end{figure*}

\begin{figure*}
\includegraphics[width=18cm]{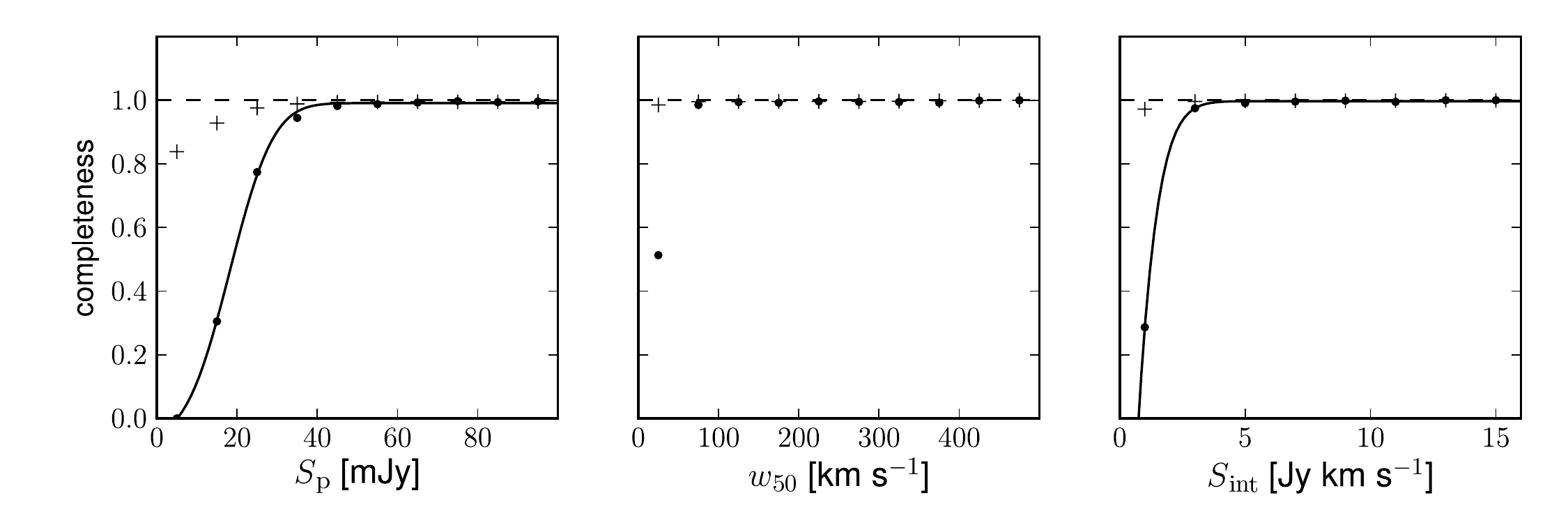}
\caption{The \small{DUCHAMP} completeness as a function of $S\mathrm{_p}$, $w_{50}$ and $S\mathrm{_{int}}$. The differential completeness is marked with circles whereas the cumulative completeness is marked with crosses. The solid lines are the error function fits to the data.}
\label{fig:comp_two}
\end{figure*}

To measure the catalogue completeness we insert numerous synthetic point sources into the eight HIJASS cubes, prior to running {\small DUCHAMP}. The rate at which the synthetic sources are recovered reflects the completeness, as discussed explicitly in \citet{zwaan2004} (also see \citealt{wong2006}). The criteria for a recovered synthetic source are a {\small DUCHAMP} detection within 10~arcmin and 200~km~s$^{-1}$ of the original position and velocity of the synthetic source (the average position and velocity accuracies for \HI sources with peak flux densities below 100~mJy are $\sim$1.2~arcmin and $\sim$14~km~s$^{-1}$, respectively). 

The synthetic point sources are given a Gaussian spectral shape and have random positions, 50~per~cent velocity widths and peak flux densities \citep{kilborn2009}. The 50~per~cent velocity widths are $w_{50}\leq$~500~km~s$^{-1}$ and the peak flux densities are $S\mathrm{_{p}} \leq$~100~mJy. The synthetic sources are placed away from the noisy edges of the data cube and are not placed on top of other sources either synthetic or real (the positions of the real sources are according to an initial {\small DUCHAMP} output list). 

A total of 2000 synthetic sources are inserted into each HIJASS data cube to obtain the {\small DUCHAMP} completeness. For a representative HIJASS cube the completeness is shown in Figure~\ref{fig:comp_one}. We use natural neighbor interpolation to achieve the grey-scale representation of the completeness depending on $S\mathrm{_{p}}$ and $w_{50}$ or $S\mathrm{_{p}}$ and integrated flux $S\mathrm{_{int}}$ (which is a combination of both $S\mathrm{_{p}}$ and $w_{50}$). 

To measure the catalogue completeness as a function of one parameter only - $S\mathrm{_{p}}$, $w_{50}$ or $S\mathrm{_{int}}$ - we integrate along the other variable of the completeness as discussed in \citet{zwaan2004}. We integrate above the {\small DUCHAMP} selection parameters which are $w_{50}=$~26.4~km~s$^{-1}$ and $S\mathrm{_{p}}=$~16~mJy. These values mark the {\small DUCHAMP} detection limit in the smoothed and reconstructed cube -- below these values the completeness drops to zero. Figure~\ref{fig:comp_two} shows this differential completeness (circles) as well as the cumulative (crosses) completeness as defined by \citet{zwaan2004}. Figure~\ref{fig:comp_two} also shows the error function fit to the data to determine the 95~per~cent confidence levels which lie at $S\mathrm{_{p}}=$~33~mJy and $S\mathrm{_{int}}=$~2.63~Jy~km~s$^{-1}$. The 99~per~cent completeness levels are achieved at $S\mathrm{_{p}} =$~45~mJy and $S\mathrm{_{int}} =$~3.55~Jy~km~s$^{-1}$. Therefore, the peak flux density limited catalogue (at 33~mJy) presented in Section~\ref{sec:cat} is 95~per~cent complete.

\subsection{Catalogue completeness -- Literature comparison}
\label{sec:comp2}

As an additional measure of completeness we compare our results (Section~\ref{sec:catalogue}) to an overlapping deep blind \HI survey in the Canes Venatici region using WSRT (area of 86~deg$^2$, $v\mathrm{_{sys}} \leq 1330$~km~s$^{-1}$ with an \HI mass sensitivity of $\sim10^6$~M$_{\odot}$; \citealt{kovac2009}). We find that our presented catalogue contains all but one \HI sources in the velocity range 300 to 1330~km~s$\mathrm{^{-1}}$, with $S\mathrm{_{p}} \geq$~33~mJy and $S\mathrm{_{int}} \geq$~2.63~Jy~km~s$^{-1}$ (95~per~cent completeness level) as listed in \citet{kovac2009} -- the exception is NGC4460 ($v_{\mathrm{LG}}=542$~km~s$^{-1}$) with $S\mathrm{_{p}}= 41$~mJy and $S\mathrm{_{int}}= 4.30$~Jy~km~s$^{-1}$ \citep{kovac2009}. We investigated the HIJASS data cube (by eye) and identify NGC4460 at the end of negative bandpass sidelobes emerging from the strong double source NGC4490/NGC4485. NGC4460 has reduced peak brightness ($S\mathrm{_{p}} \simeq 31$~mJy~beam$\mathrm{^{-1}}$) due to the negative bandpass sidelobes, which prevents this \HI source to be listed in our catalogue. We conclude that negative bandpass sidelobes may cause a deficit in completeness for sources with $16 \leq S\mathrm{_{p}} \leq 45$~mJy and $w_{50} \leq 100$~km~s$^{-1}$ (see Figure~\ref{fig:comp_one}), as their peak flux densities may lie below the detection threshold. However, this is a small effect, given the little area affected by negative bandpass sidelobes.

\section{The HI catalogue}
\label{sec:cat}

\subsection{Parametrization}
\label{sec:parametrization}

The sources found using {\small DUCHAMP} are parameterized with an interactive script utilising the {\small MIRIAD} software package \citep{sault1995}. The parametrization is necessary as {\small DUCHAMP} gives initial values -- dealing with double detections and obtaining reliable \HI parameters requires user input. Our method is similar to \citet{lang2003} and \citet{koribalski2004} and this will be briefly described in the following section.

A spatially integrated spectrum is generated from the original HIJASS cube, using the {\small DUCHAMP} output parameters for each source (RA, Dec., velocity, velocity widths). The spectral extent of the \HI source (velocity range) can be adjusted as necessary. An integrated \HI intensity map over this velocity range is generated using the {\small MIRIAD} task {\small MOMENT}. Given a specified fitting area (in relation to the source size), {\small IMFIT} is used to determine the central position of the \HI source by fitting a two-dimensional Gaussian to the integrated \HI intensity map. The obtained central coordinates are used to generate a new spatially integrated spectrum weighted according to the extent of the \HI source (see below). In general, we fit and subtract a first-order baseline fit to the spectrum (higher order baseline fits are labelled in the catalogue). The \HI parameters such as peak and integrated flux, the velocity widths, the clipped rms noise and the systemic velocity are measured using {\small MBSPECT}. In the catalogue we quote the integrated flux measured with the robust moments, the 20 and 50~per~cent velocity widths calculated from the width maximiser and the systemic velocity as the center of the 20~per~cent velocity width (see \citealt{meyer2004} for definitions).

An \HI source can be unresolved or extended with respect to the gridded beam size. To identify point and extended sources, we compare the integrated flux measurements obtained from (\textit{i}) spectra where every pixel is weighted according to the expected beam response (beam-weighted as for point sources) to (\textit{ii}) spectra where the flux is directly summed (as for extended sources). The considered area is given by a box surrounding the central position, e.g. of size $5 \times 5$~pixel$^2$ or $7 \times 7$~pixel$^2$ that correspond to $20' \times 20'$ and $28' \times 28'$, respectively. A potentially extended source is identified due to an increase in the integrated flux from a beam-weighted spectrum to a summed spectrum in a bigger box. This increase can also occur due to confusion with a nearby \HI source which might be included in the bigger box - such sources are flagged as double sources (d) in the catalogue and are parameterized by generating a beam-weighted spectrum as for point sources. Truly extended sources (e), which are indicated by an increase in the integrated flux, are confirmed by investigating the residual map of the Gaussian fit. An extended source will show flux in the residual map whereas a point source is well described by the Gaussian fit.

\subsection{The HI catalogue}
\label{sec:catalogue}

The {\small DUCHAMP} detection list in the smoothed and reconstructed HIJASS megacube contains 140 candidate galaxies with systemic velocities between 300 and 1900~km~s$^{-1}$. After the parametrization process, 4~galaxies have systemic velocities below 300~km~s$^{-1}$. In addition, we identified 11~\HI sources with systemic velocities between 150 and 300~km~s$^{-1}$ that are included in the sample as they are obviously real galaxies as opposed to Galactic emission or high-velocity clouds. However, \HI sources with systemic velocities below 300~km~s$^{-1}$ are not included in Section~\ref{sec:res}. Furthermore we identified several double, triple and quad detections that are separated during the parametrization process leading to a total of 173~\HI sources that are parameterized from the original HIJASS megacube. We exclude 7 detections from the presented catalogue that are unconfirmed in GBT follow-up observations (see Section~\ref{sec:reliability2}). Therefore a total of 166~\HI sources are listed in our catalogue in Section~\ref{sec:catalogue}.

The parameterized \HI sources are cross-correlated with galaxies listed in the NASA/IPAC Extragalactic Database (NED; as of September 2012) and the Sloan Digital Sky Survey (SDSS DR8). We only consider galaxies with cross-identifications in NED and galaxies with spectroscopic redshifts with confidence levels $z_{conf}>0.95$ in SDSS as possible counterparts. We associate an \HI source with a catalogued galaxy, if this galaxy is within the gridded beam size (or within the extent of the \HI intensity contours for large sources) and within the velocity range of the \HI source. We find 54~\HI sources that have multiple candidate counterparts. These are potentially confused \HI sources and may be interacting systems. To determine the most likely optical counterpart (listed in the first row of the catalogue), the \HI position uncertainty, the heliocentric velocity, the optical galaxy size (in relation to the \HI mass and size) and the morphology are considered. 

We find only one out of the 166 catalogued \HI sources that does not have an associate optical counterpart listed in NED/SDSS or visible on Palomar Digitized Sky Survey (DSS) images. This candidate galaxy/\HI cloud -- HIJASS~J1219+46 -- is discussed in Section~\ref{sec:new}.

Table~\ref{tab:A_cat} lists the \HI sources and their optical counterparts as follows: column~(A1) gives the HIJASS name; columns~(A2)$-$(A4) list the central \HI position (RA [J2000], Dec. [J2000]) and its uncertainty $\sigma\mathrm{_{pos}}$ (all uncertainties are discussed in Section~\ref{sec:uncertainties}); column~(A5) contains the systemic velocity (km~s$^{-1}$) with its uncertainty $\sigma\mathrm{_{v}}$; column~(A6) states `ID' for \HI detections that have an optical identification and `PUO' for the previously uncatalogued gas-rich object; columns~(A7)$-$(A10) list the details of the optical counterparts, i.e. name, RA (J2000), Dec. (J2000) and heliocentric velocity (km~s$^{-1}$); columns~(A11) and (A12) give the angular separation (arcmin) and velocity offset (km~s$^{-1}$) from the \HI parameters. The most likely optical counterpart is given in the first row and galaxies that lie within the gridded beam or within the extent of the \HI intensity contours (for large sources) and might confuse the \HI content are quoted in the subsequent rows. 

The measured \HI properties of the HIJASS detections are given in Table~\ref{tab:B_cat}. The columns are as follows: column~(B1) gives the HIJASS name; columns~(B2)-(B4) gives the measured and corrected peak flux densities ($S\mathrm{_p}$) with its uncertainty (in Jy); columns~(B5)-(B7) list the measured and corrected integrated flux ($S\mathrm{_{int}}$) with its uncertainty (in Jy~km~s$^{-1}$); columns~(B8) and (B9) contain the velocity widths at 50 and 20~per~cent of the peak flux density with their uncertainties (in km~s$^{-1}$); column~(B10) gives the clipped rms measured in {\small MBSPECT}; column~(B11) lists the distance to the galaxy (in Mpc) as obtained from the Extragalactic Distance Database (EDD; \citealt{tully2009} -- only distances obtained from Color-Magnitude Diagrams, Tip of the Red Giant Branch and other high quality methods are considered) or calculated from the local group velocity (see \citealt{koribalski2004} for HIPASS) -- we adopt $H_0=70$~km~s$^{-1}$~Mpc$^{-1}$; for galaxies within the cluster definition by \citet{tully1996} we adopt a distance of 17.1~Mpc; column~(B11) gives the logarithm of the \HI mass and column~(B12) contains flags for 78 extended (e) and 54 confused (c) \HI sources, 38 \HI souces which are part of {\small DUCHAMP} double detections (d), 10 \HI sources without a central position from a position fit (f) and therefore without major/minor axis measurements, 10 \HI sources lying nearby negative bandpass sidelobes (b), 10 newly detected \HI sources in the 21-cm emission line (n), 15 \HI sources with systemic velocities below 300~km~S$^{-1}$ (v), 34 sources where independent distance measurements are available (i) and numbers to indicate higher order baseline fits. The \HI mass in column~(B11) is calculated from the integrated flux ($S\mathrm{_{int}}$ and $S\mathrm{_{int}(corr)}$ for extended sources respectively) according to $M\mathrm{_{HI}}=2.356 \times 10^5 \times D^2 \times S\mathrm{_{int}}$ [M$_{\odot}$] with distances as given in column~(10). We note that \citet{kovac2009} also determine adequate distances in the local group frame, neglecting peculiar velocities (using \citealt{yahil1977}).

\begin{figure*}
\includegraphics[width=17cm]{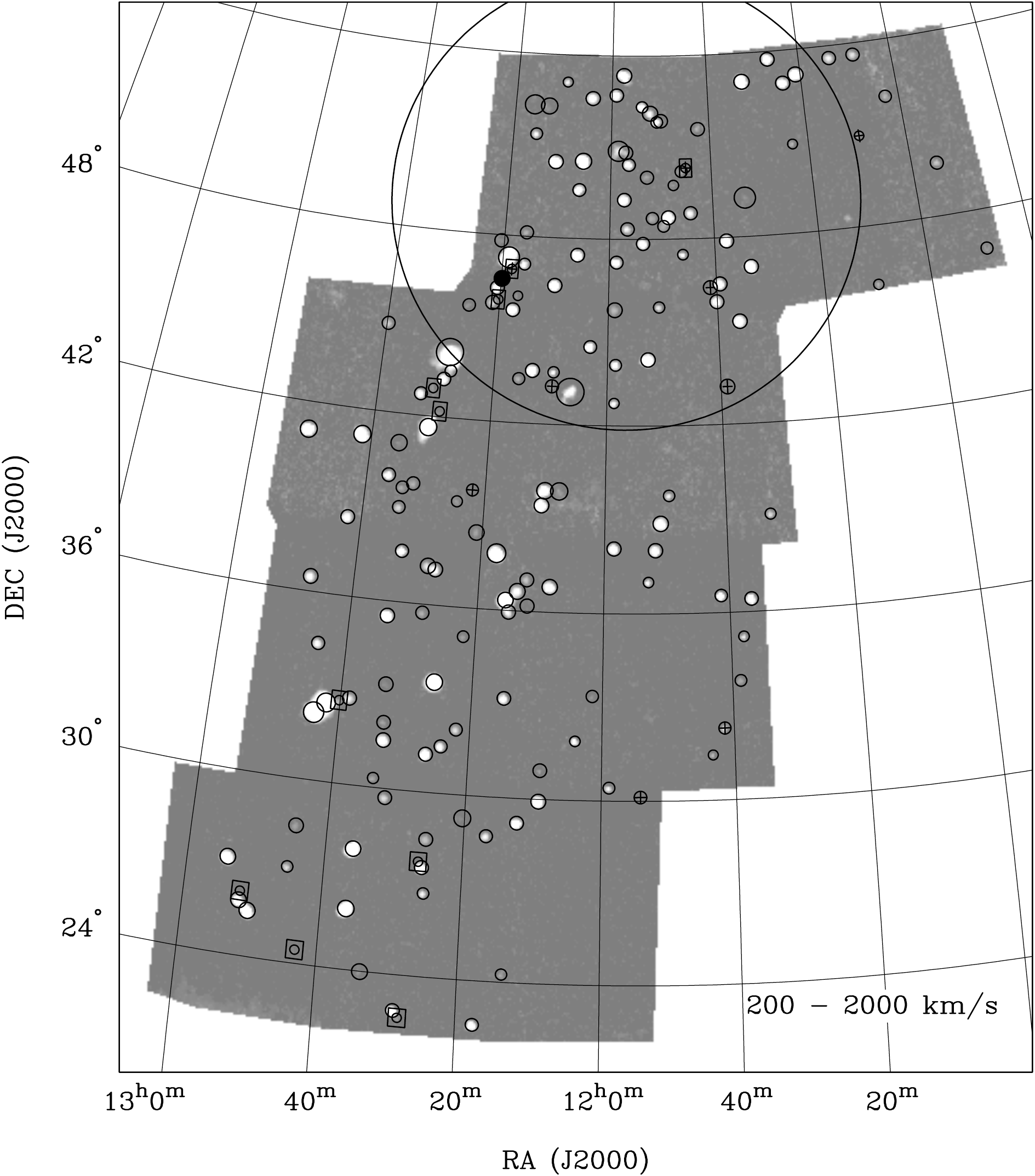}
\caption{\textbf{Velocity integrated \HI distribution of the Ursa Major region as obtained from the HIJASS data (grey scale).} The large circle (with radius 7.5~deg) drawn in the northern part shows the projected position of the Ursa Major cluster as defined by \citet{tully1996}. Overlaid are symbols to indicate the projected positions and sizes of \HI detections as presented in this catalogue - open circles indicate \HI sources with an associate optical counterpart whereas the filled circle marks the candidate galaxy/\HI cloud, HIJASS~J1219+46. This source and the \HI detections marked with crosses are newly detected in the 21-cm emission line. Circles surrounded by boxes indicate \HI detections for which no central position and no extent (major and minor axis) is measurable.}
\label{fig:skymap}
\end{figure*}

We show a velocity integrated \HI map with symbols marking the projected positions of all parameterised \HI sources in this catalogue in Figure~\ref{fig:skymap}. \HI sources with an associated optical counterpart are marked with open ellipses whereas the previously uncatalogued \HI source is marked with a filled ellipse (black). Furthermore previously catalogued sources that are newly detected in \HI are marked with crosses. The size of the ellipses indicate the measured major axis -- ellipses surrounded by boxes mark \HI detections for which no central position and no extent (major and minor axis) is measurable (usually weak \HI sources or double detections for which we retain the position of the optical counterpart). The projected position of the Ursa Major cluster as defined by \citep{tully1996} is overlaid with a large circle.

\subsection{HI parameter uncertainties}
\label{sec:uncertainties}

In Table~\ref{tab:A_cat} and \ref{tab:B_cat} we give several uncertainties of measured \HI parameters which are calculated as follows: 

The position uncertainty is derived from the gridded beam size ($FWHM$) and the \HI signal-to-noise \citep{koribalski2004}: $$\sigma_{\mathrm{pos}}=\frac{FWHM}{S\mathrm{_{p}}/rms}$$ This individual position uncertainty is considered when determining the most likely optical counterpart (see Table~\ref{tab:A_cat}). 

The peak flux uncertainty strongly depends on the rms noise for weak sources and to a lesser degree on the peak flux (see \citealt{barnes2001}). The formula we use is analogous to the one in HIPASS \citep{koribalski2004}, which is as follows: $$\sigma_{\mathrm{peak}}^2=rms^2+(0.05S\mathrm{_{p}})^2$$ 

The uncertainty in integrated flux is calculated as discussed in \citet{koribalski2004} whereby $$\sigma_{\mathrm{int}}=4 \times (\frac{S\mathrm{_{p}}}{rms})^{-1} \times (S\mathrm{_p} \times S\mathrm{_{int}} \times R)^{0.5}$$ with R=18.1~km~s$^{-1}$, which is the velocity resolution of the HIJASS data.

The systemic velocity uncertainty is $$\sigma_{\mathrm{v}}=3 \times (\frac{S\mathrm{_{p}}}{\sigma\mathrm{_{peak}}})^{-1} \times (R \times 0.5(w\mathrm{_{20}}-w\mathrm{_{50}}))^{0.5} $$ as discussed in \citet{koribalski2004}. 

The uncertainties in the 50\% and 20\% velocity widths are: $$\sigma_{\mathrm{50}}=2 \times \sigma_{\mathrm{v}}$$ $$\sigma_{\mathrm{20}}=3 \times \sigma_{\mathrm{v}}$$ as discussed in \citet{fouque1990}.

\onecolumn
\begin{landscape}
\begin{longtable}{lllcrclllrrr}
\caption[The HIJASS detections and possible optical counterparts.]{The HIJASS detections and possible optical counterparts.} 
\label{tab:A_cat} \\
  
  \hline\hline\\[-2ex]
   & 
  \multicolumn{1}{c}{RA} & 
  \multicolumn{1}{c}{Dec.} & 
  \multicolumn{1}{c}{$\sigma_{\mathrm{pos}}$}&    
  \multicolumn{1}{c}{$v\mathrm{_{sys}} \pm \sigma_{\mathrm{v}}$} & 
  &  
  & 
  \multicolumn{1}{c}{RA} &
  \multicolumn{1}{c}{Dec.} & 
  \multicolumn{1}{c}{vel.} &
  \multicolumn{2}{c}{pos. \& vel. offsets} \\

  \multicolumn{1}{c}{HIJASS name} & 
  \multicolumn{1}{c}{(J2000.0)} & 
  \multicolumn{1}{c}{(J2000.0)} & 
  \multicolumn{1}{c}{arcmin} &  
  \multicolumn{1}{c}{km s$^{-1}$} & 
  \multicolumn{1}{c}{Class} & 
  \multicolumn{1}{c}{NED ID$^{a,b}$} & 
  \multicolumn{1}{c}{(J2000.0)} & 
  \multicolumn{1}{c}{(J2000.0)} & 
  \multicolumn{1}{c}{km s$^{-1}$} &
  \multicolumn{1}{c}{arcmin} & 
  \multicolumn{1}{c}{km s$^{-1}$} \\

  \multicolumn{1}{c}{(A1)} & 
  \multicolumn{1}{c}{(A2)} & 
  \multicolumn{1}{c}{(A3)} & 
  \multicolumn{1}{c}{(A4)} &    
  \multicolumn{1}{c}{(A5)}  &
  \multicolumn{1}{c}{(A6)} &  
  \multicolumn{1}{c}{(A7)} & 
  \multicolumn{1}{c}{(A8)} & 
  \multicolumn{1}{c}{(A9)}  &
  \multicolumn{1}{c}{(A10)} &
  \multicolumn{1}{c}{(A11)} &
  \multicolumn{1}{c}{(A12)} \\[0.5ex] \hline
  \\[-1.8ex]
  \endfirsthead
  
  \multicolumn{12}{c}{{\tablename} \thetable{} -- Continued} \\[0.5ex]
   \hline\hline\\[-2ex]
   & 
  \multicolumn{1}{c}{RA} & 
  \multicolumn{1}{c}{Dec.} & 
  \multicolumn{1}{c}{$\sigma_{\mathrm{pos}}$}&    
  \multicolumn{1}{c}{$v\mathrm{_{sys}} \pm \sigma_{\mathrm{v}}$} & 
  &  
  & 
  \multicolumn{1}{c}{RA} &
  \multicolumn{1}{c}{Dec.} & 
  \multicolumn{1}{c}{vel.} &
  \multicolumn{2}{c}{pos. \& vel. offsets} \\

  \multicolumn{1}{c}{HIJASS name} & 
  \multicolumn{1}{c}{(J2000.0)} & 
  \multicolumn{1}{c}{(J2000.0)} & 
  \multicolumn{1}{c}{arcmin} &  
  \multicolumn{1}{c}{km s$^{-1}$} & 
  \multicolumn{1}{c}{Class} & 
  \multicolumn{1}{c}{NED ID$^{a,b}$} & 
  \multicolumn{1}{c}{(J2000.0)} & 
  \multicolumn{1}{c}{(J2000.0)} & 
  \multicolumn{1}{c}{km s$^{-1}$} &
  \multicolumn{1}{c}{arcmin} & 
  \multicolumn{1}{c}{km s$^{-1}$} \\

  \multicolumn{1}{c}{(A1)} & 
  \multicolumn{1}{c}{(A2)} & 
  \multicolumn{1}{c}{(A3)} & 
  \multicolumn{1}{c}{(A4)} &    
  \multicolumn{1}{c}{(A5)}  &
  \multicolumn{1}{c}{(A6)} &  
  \multicolumn{1}{c}{(A7)} & 
  \multicolumn{1}{c}{(A8)} & 
  \multicolumn{1}{c}{(A9)}  &
  \multicolumn{1}{c}{(A10)} &
  \multicolumn{1}{c}{(A11)} &
  \multicolumn{1}{c}{(A12)} \\[0.5ex] \hline
  \\[-1.8ex]
   \endhead

HIJASS J1048+46 & 10:48:43.33 & 46:43:10.0 & 2.7 & 741 $\pm$6 & ID & UGC05917 & 10:48:53.73 & 46:43:14.6 & 741 & 1.8 & 0\\
HIJASS J1054+49 & 10:54:57.73 & 49:42:53.6 & 2.9 & 1347 $\pm$15 & ID & UGC06029 & 10:55:02.23 & 49:43:33.1 & 1363 & 1.0 & 16\\
\multicolumn{1}{c}{-} & \multicolumn{1}{c}{-} & \multicolumn{1}{c}{-} & \multicolumn{1}{c}{-} & \multicolumn{1}{c}{-} & \multicolumn{1}{c}{-} & J105502.69+494340.9 & 10:55:02.70 & 49:43:40.9 & 1357 & 1.1 & 10\\
HIJASS J1102+52 & 11:02:48.05 & 52:06:40.5 & 2.0 & 949 $\pm$7 & ID & UGC06113 & 11:02:47.85 & 52:06:48.8 & 951 & 0.1 & 2\\
HIJASS J1108+53 & 11:08:08.79 & 53:36:41.6 & 2.1 & 1226 $\pm$6 & ID & UGC06182 & 11:08:03.01 & 53:36:57.6 & 1238 & 0.9 & 12\\
\multicolumn{1}{c}{-} & \multicolumn{1}{c}{-} & \multicolumn{1}{c}{-} & \multicolumn{1}{c}{-} & \multicolumn{1}{c}{-} & \multicolumn{1}{c}{-} & J110801.86+533702.7 & 11:08:01.87 & 53:37:02.7 & 1176 & 1.1 & 50\\
\multicolumn{1}{c}{-} & \multicolumn{1}{c}{-} & \multicolumn{1}{c}{-} & \multicolumn{1}{c}{-} & \multicolumn{1}{c}{-} & \multicolumn{1}{c}{-} & J110819.94+533627.7 & 11:08:19.94 & 53:36:27.8 & 1087 & 1.7 & 139\\
HIJASS J1109+50 & 11:09:32.53 & 50:56:24.0 & 1.6 & 906 $\pm$7 & ID & UGC06202 & 11:09:36.50 & 50:55:36.3 & 920 & 1.0 & 14\\
\end{longtable}
\end{landscape}
\twocolumn

\onecolumn
\begin{landscape}
\begin{longtable}{lcccrrcrrrccc}
\caption[The \HI properties of all HIJASS detections]{The \HI properties of all HIJASS detections} 
\label{tab:B_cat} \\
  
  \hline\hline\\[-2ex]
   & 
  \multicolumn{1}{c}{$S\mathrm{_p}$} & 
  \multicolumn{1}{c}{$S\mathrm{_p}$(corr)} & 
  \multicolumn{1}{c}{$\sigma_{\mathrm{peak}}$} & 
  \multicolumn{1}{c}{$S\mathrm{_{int}}$} & 
  \multicolumn{1}{c}{$S\mathrm{_{int}}$(corr)} & 
  \multicolumn{1}{c}{$\sigma_{\mathrm{int}}$} & 
  \multicolumn{1}{c}{$w_{50} \pm \sigma_{\mathrm{50}}$} & 
  \multicolumn{1}{c}{$w_{20} \pm \sigma_{\mathrm{20}}$} &
  \multicolumn{1}{c}{rms} & 
  \multicolumn{1}{c}{Distance} & 
  \\
  \\
  
  \multicolumn{1}{c}{HIJASS name} & 
  \multicolumn{1}{c}{Jy} & 
  \multicolumn{1}{c}{Jy} & 
  \multicolumn{1}{c}{Jy} &   
  \multicolumn{1}{c}{Jy km s$^{-1}$} & 
  \multicolumn{1}{c}{Jy km s$^{-1}$} & 
  \multicolumn{1}{c}{Jy km s$^{-1}$} & 
  \multicolumn{1}{c}{km s$^{-1}$} & 
  \multicolumn{1}{r}{km s$^{-1}$} &   
  \multicolumn{1}{c}{Jy} & 
  \multicolumn{1}{c}{Mpc} & 
  \multicolumn{1}{c}{$log M\mathrm{_{HI}}$} &
  \multicolumn{1}{c}{flags$^{a}$} \\
  
  \multicolumn{1}{c}{(B1)} & 
  \multicolumn{1}{c}{(B2)} & 
  \multicolumn{1}{c}{(B3)} & 
  \multicolumn{1}{c}{(B4)}  &
  \multicolumn{1}{r}{(B5)} &  
  \multicolumn{1}{c}{(B6)} &  
  \multicolumn{1}{c}{(B7)} &  
  \multicolumn{1}{c}{(B8)} &  
  \multicolumn{1}{c}{(B9)} & 
  \multicolumn{1}{c}{(B10)} &  
  \multicolumn{1}{c}{(B11)} &  
  \multicolumn{1}{c}{(B12)} &  
  \multicolumn{1}{c}{(B13)} \\[0.5ex] \hline
  \\[-1.8ex]
  \endfirsthead

  \multicolumn{12}{c}{{\tablename} \thetable{} -- Continued} \\[0.5ex]
  \hline\hline\\[-2ex]
   & 
  \multicolumn{1}{c}{$S\mathrm{_p}$} & 
  \multicolumn{1}{c}{$S\mathrm{_p}$(corr)} & 
  \multicolumn{1}{c}{$\sigma_{\mathrm{peak}}$} & 
  \multicolumn{1}{c}{$S\mathrm{_{int}}$} & 
  \multicolumn{1}{c}{$S\mathrm{_{int}}$(corr)} & 
  \multicolumn{1}{c}{$\sigma_{\mathrm{int}}$} & 
  \multicolumn{1}{c}{$w_{50} \pm \sigma_{\mathrm{50}}$} & 
  \multicolumn{1}{c}{$w_{20} \pm \sigma_{\mathrm{20}}$} &
  \multicolumn{1}{c}{rms} & 
  \multicolumn{1}{c}{Distance} & 
  \multicolumn{1}{c}{$log M\mathrm{_{HI}}$} &
  \\
  
  \multicolumn{1}{c}{HIJASS name} & 
  \multicolumn{1}{c}{Jy} & 
  \multicolumn{1}{c}{Jy} & 
  \multicolumn{1}{c}{Jy} &   
  \multicolumn{1}{c}{Jy km s$^{-1}$} & 
  \multicolumn{1}{c}{Jy km s$^{-1}$} & 
  \multicolumn{1}{c}{Jy km s$^{-1}$} & 
  \multicolumn{1}{c}{km s$^{-1}$} & 
  \multicolumn{1}{r}{km s$^{-1}$} & 
  \multicolumn{1}{c}{Jy} & 
  \multicolumn{1}{c}{Mpc} & 
  & 
  \multicolumn{1}{c}{flags$^{a}$} \\
  
  \multicolumn{1}{c}{(B1)} & 
  \multicolumn{1}{c}{(B2)} & 
  \multicolumn{1}{c}{(B3)} & 
  \multicolumn{1}{c}{(B4)}  &
  \multicolumn{1}{r}{(B5)} &  
  \multicolumn{1}{c}{(B6)} &  
  \multicolumn{1}{c}{(B7)} &  
  \multicolumn{1}{c}{(B8)} &  
  \multicolumn{1}{c}{(B9)} & 
  \multicolumn{1}{c}{(B10)} &  
  \multicolumn{1}{c}{(B11)} &  
  \multicolumn{1}{c}{(B12)} &  
  \multicolumn{1}{c}{(B13)} \\[0.5ex] \hline
  \\[-1.8ex]
   \endhead

HIJASS J1048+46 & 0.059 & - & 0.012 & 4.22 & - & 1.73 & 98$\pm$12 & 110$\pm$18 & 0.012 & 11.1 & 8.09 &  \\
HIJASS J1054+49 & 0.101 & 0.091 & 0.021 & 10.35 & 11.54 & 3.83 & 124$\pm$30 & 173$\pm$45 & 0.020 & 20.0 & 9.04 & ec \\
HIJASS J1102+52 & 0.106 & - & 0.017 & 4.39 & - & 1.75 & 46$\pm$14 & 68$\pm$21 & 0.016 & 14.5 & 8.34 &  \\
HIJASS J1108+53 & 0.109 & 0.098 & 0.017 & 8.75 & 9.76 & 2.72 & 94$\pm$12 & 109$\pm$18 & 0.016 & 18.6 & 8.90 & ec3 \\
HIJASS J1109+50 & 0.074 & - & 0.010 & 6.91 & - & 1.48 & 125$\pm$14 & 155$\pm$21 & 0.009 & 13.9 & 8.50 & n \\
HIJASS J1110+46 & 0.069 & - & 0.011 & 6.54 & - & 1.66 & 120$\pm$12 & 135$\pm$18 & 0.010 & 20.7 & 8.82 &  \\
HIJASS J1113+53 & 0.269 & - & 0.017 & 10.76 & - & 1.08 & 44$\pm$4 & 61$\pm$6 & 0.010 & 14.4 & 8.72 & c \\
HIJASS J1120+53 & 0.560 & - & 0.031 & 49.6 & - & 2.08 & 101$\pm$4 & 125$\pm$6 & 0.013 & 17.1 & 9.53 & cd \\
\end{longtable}
\end{landscape}
\twocolumn

\onecolumn
\begin{figure}
\begin{tabular}{p{3cm} p{3.5cm} p{1.5cm} p{3cm} p{3.5cm}}

\multicolumn{2}{c}{HIJASS J1226+27 (IC3341$^{f}$)}
 & &
\multicolumn{2}{c}{HIJASS J1228+42 (MCG+07-26-011$^{fd}$)}\\
\mbox{\includegraphics[height=3cm]{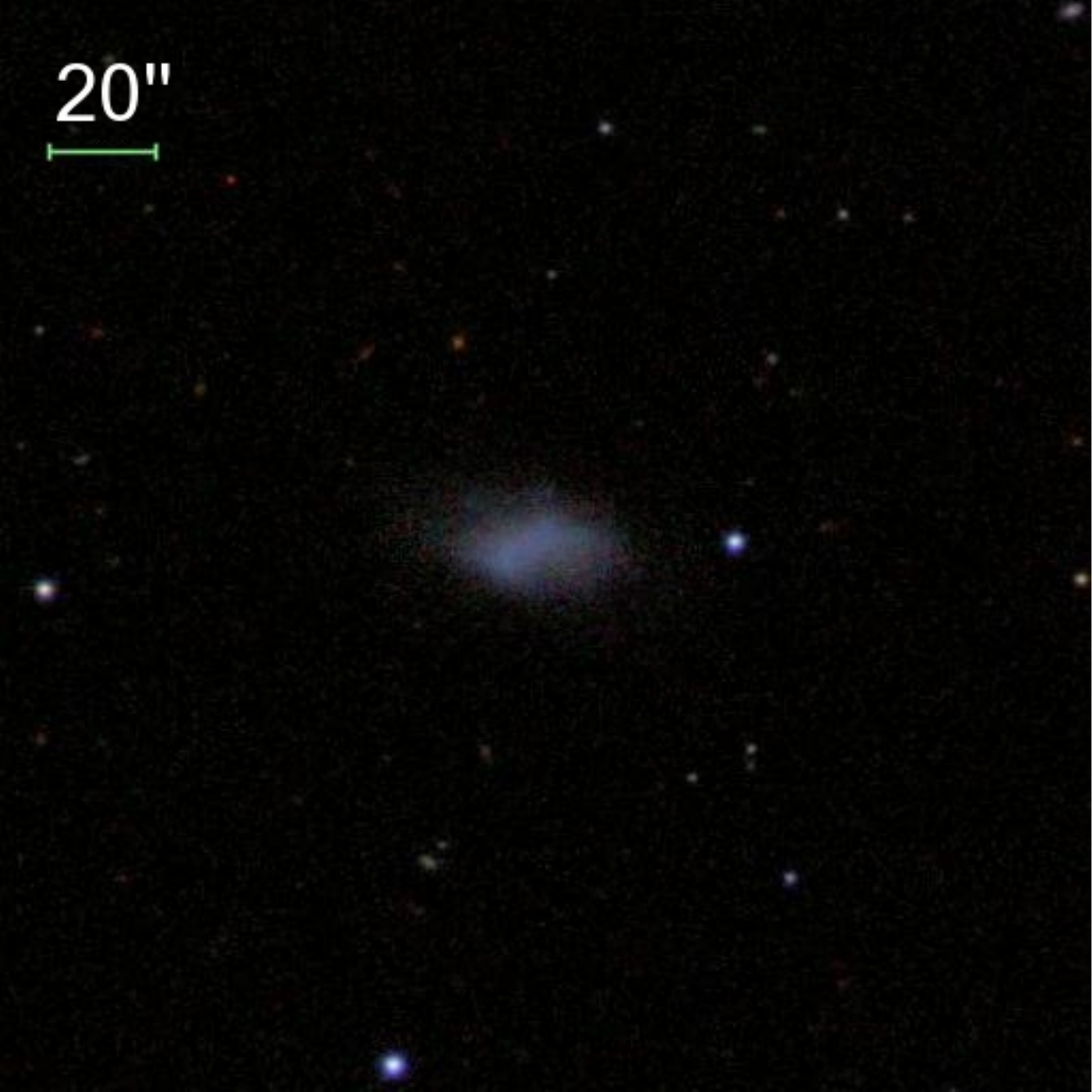}} & 
\mbox{\includegraphics[height=3cm]{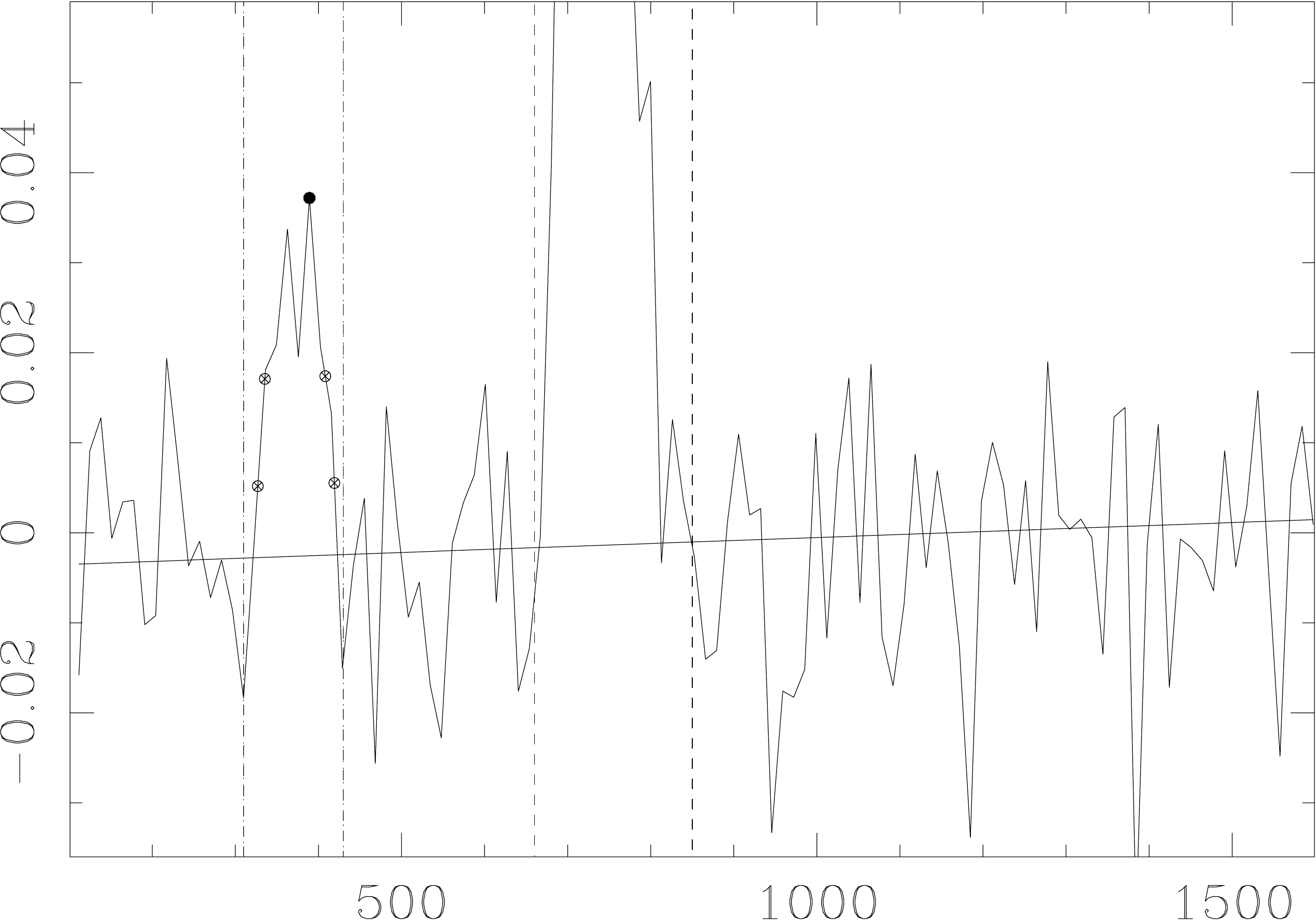}} & 
 & 
\mbox{\includegraphics[height=3cm]{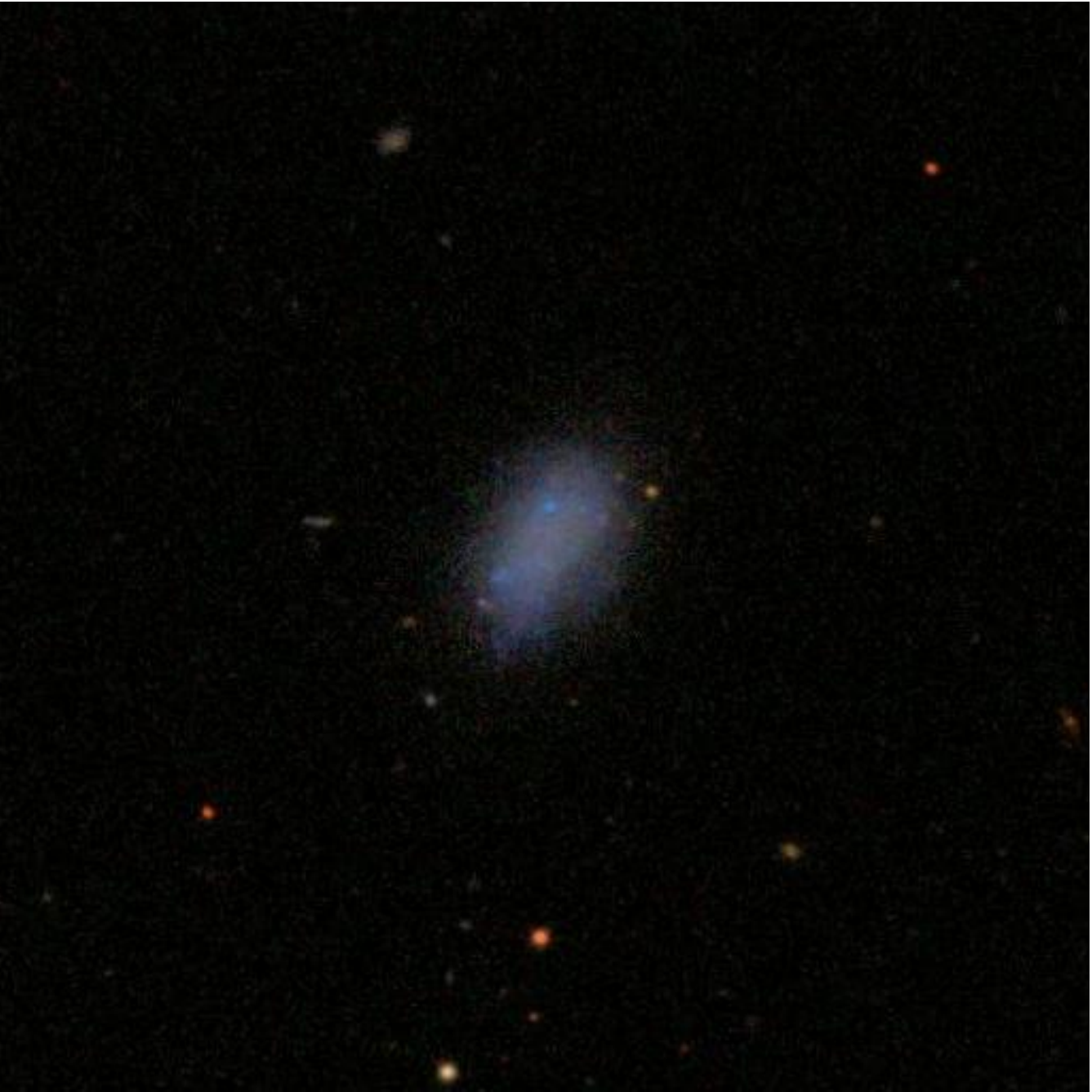}} & 
\mbox{\includegraphics[height=3cm]{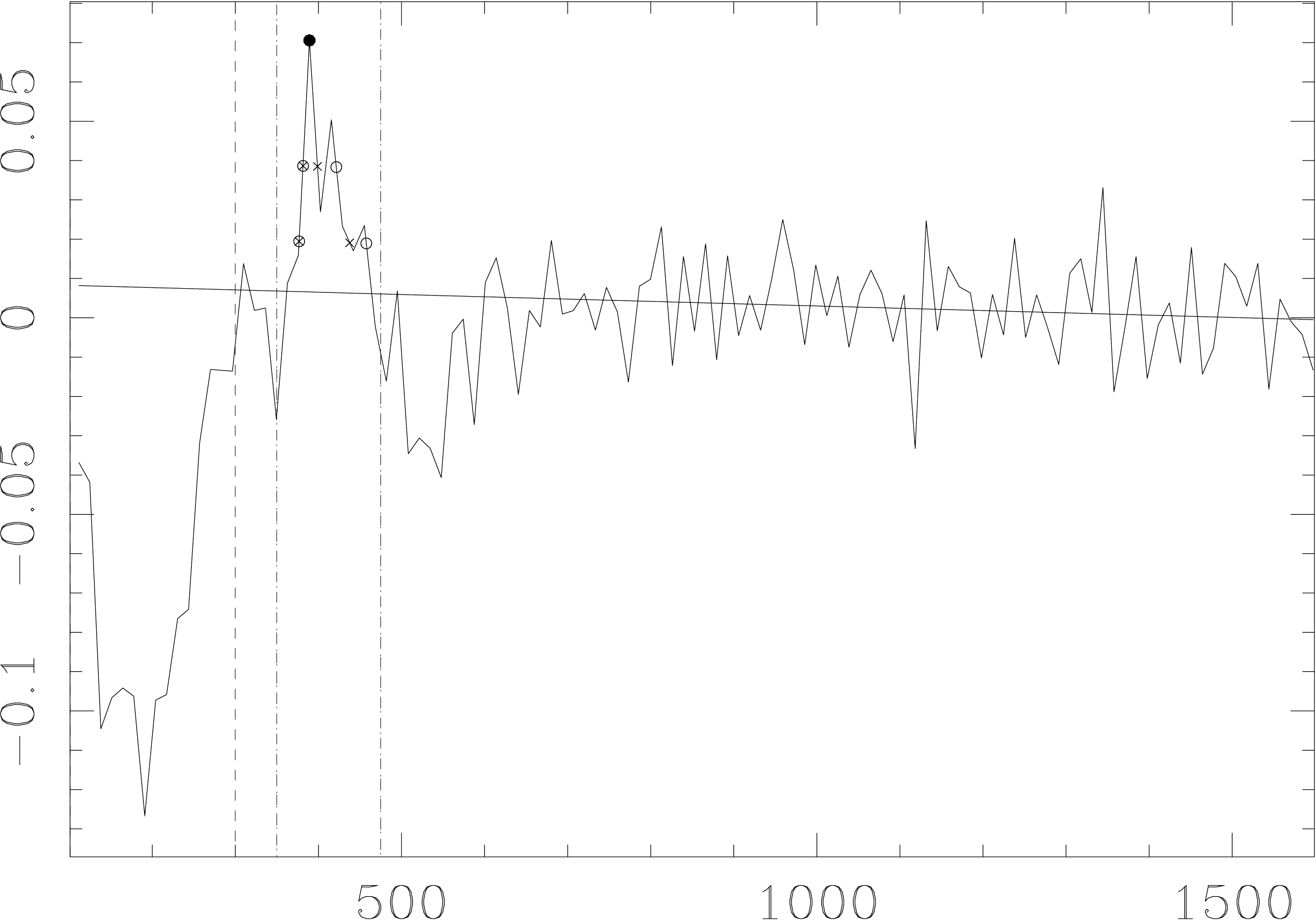}}\\
[7pt]

\multicolumn{2}{c}{HIJASS J1228+35 (UGC07605$^{e5i}$)}
 & &
\multicolumn{2}{c}{HIJASS J1235+41 (UGC07751)}\\
\mbox{\includegraphics[height=3cm]{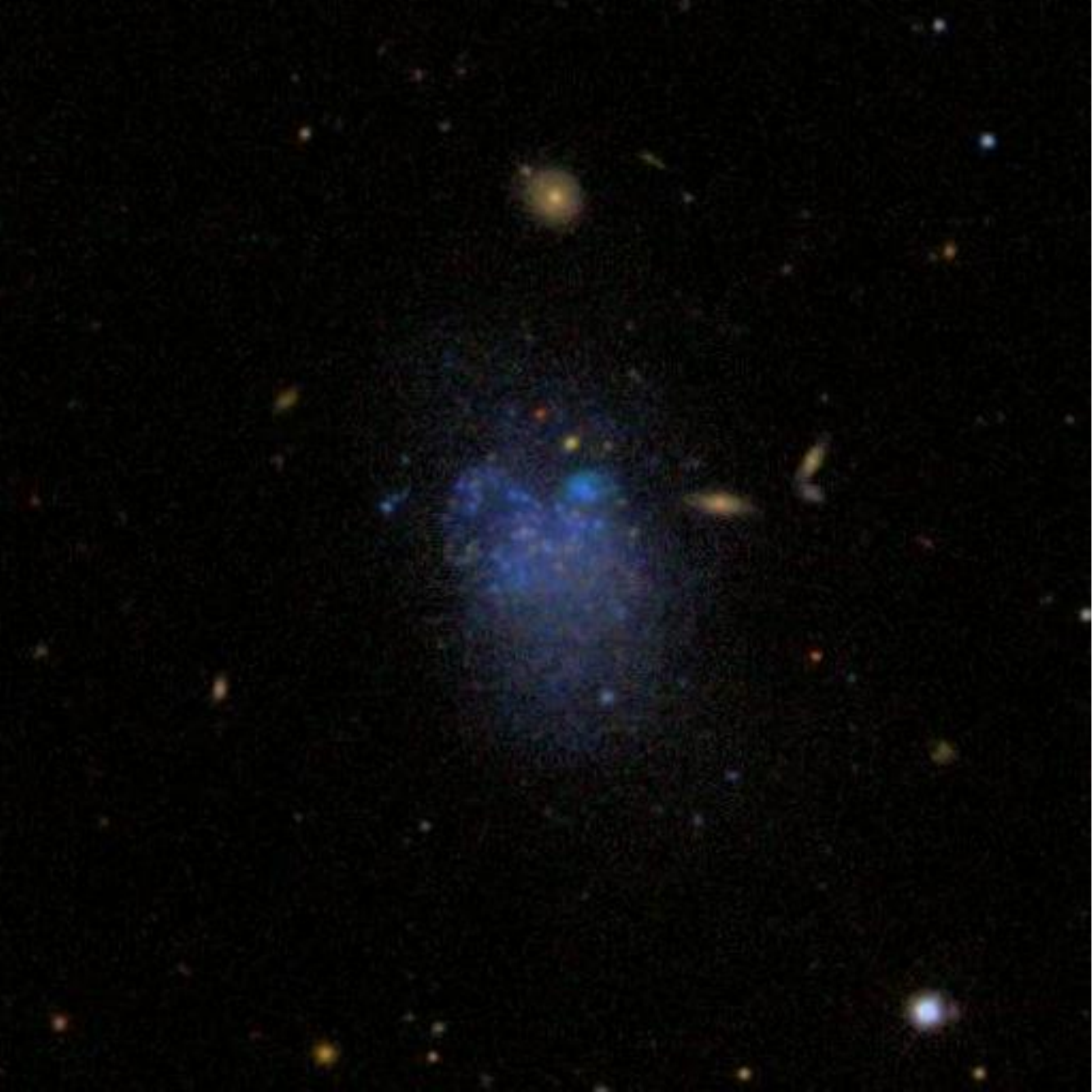}} & 
\mbox{\includegraphics[height=3cm]{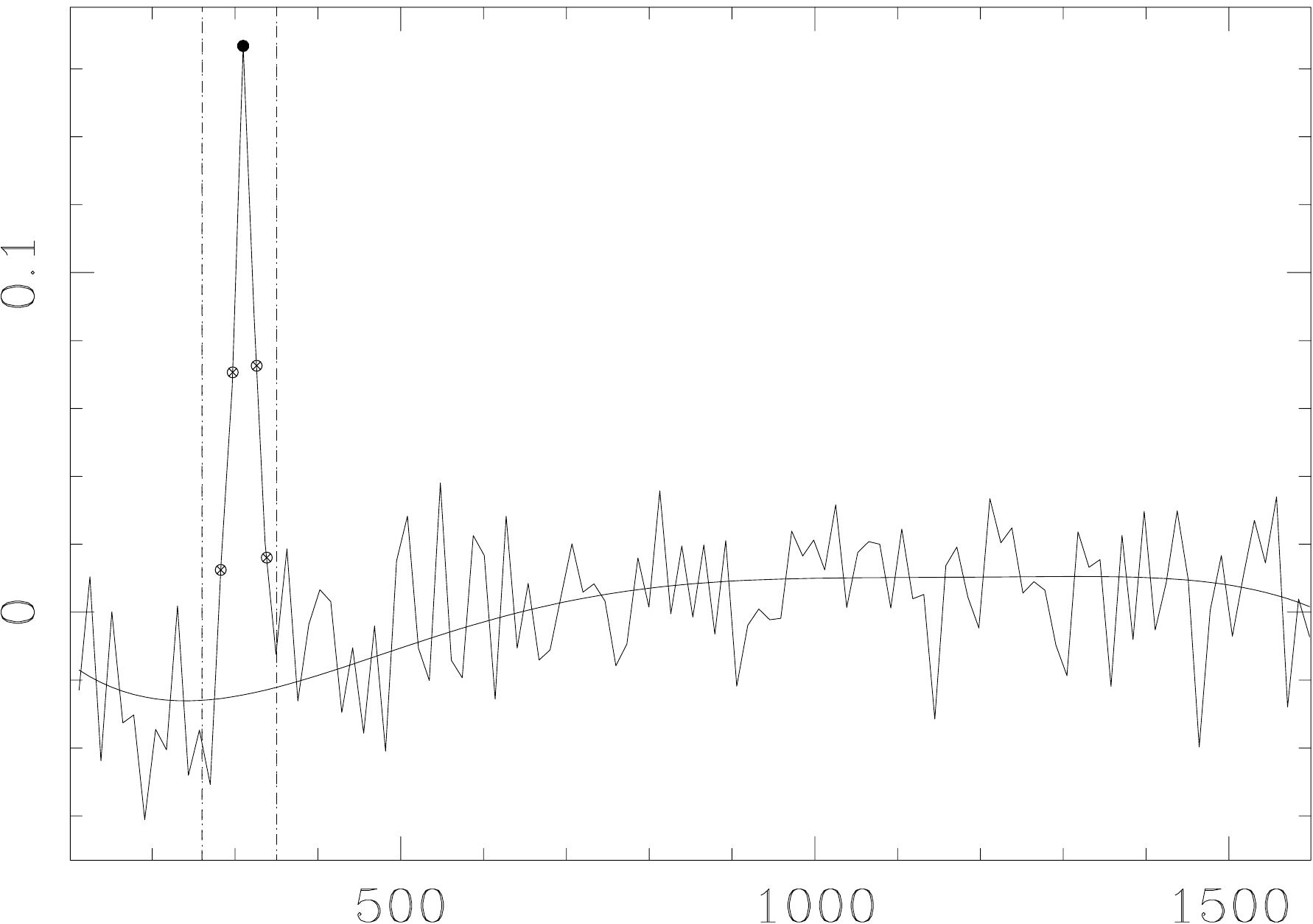}} & 
 & 
\mbox{\includegraphics[height=3cm]{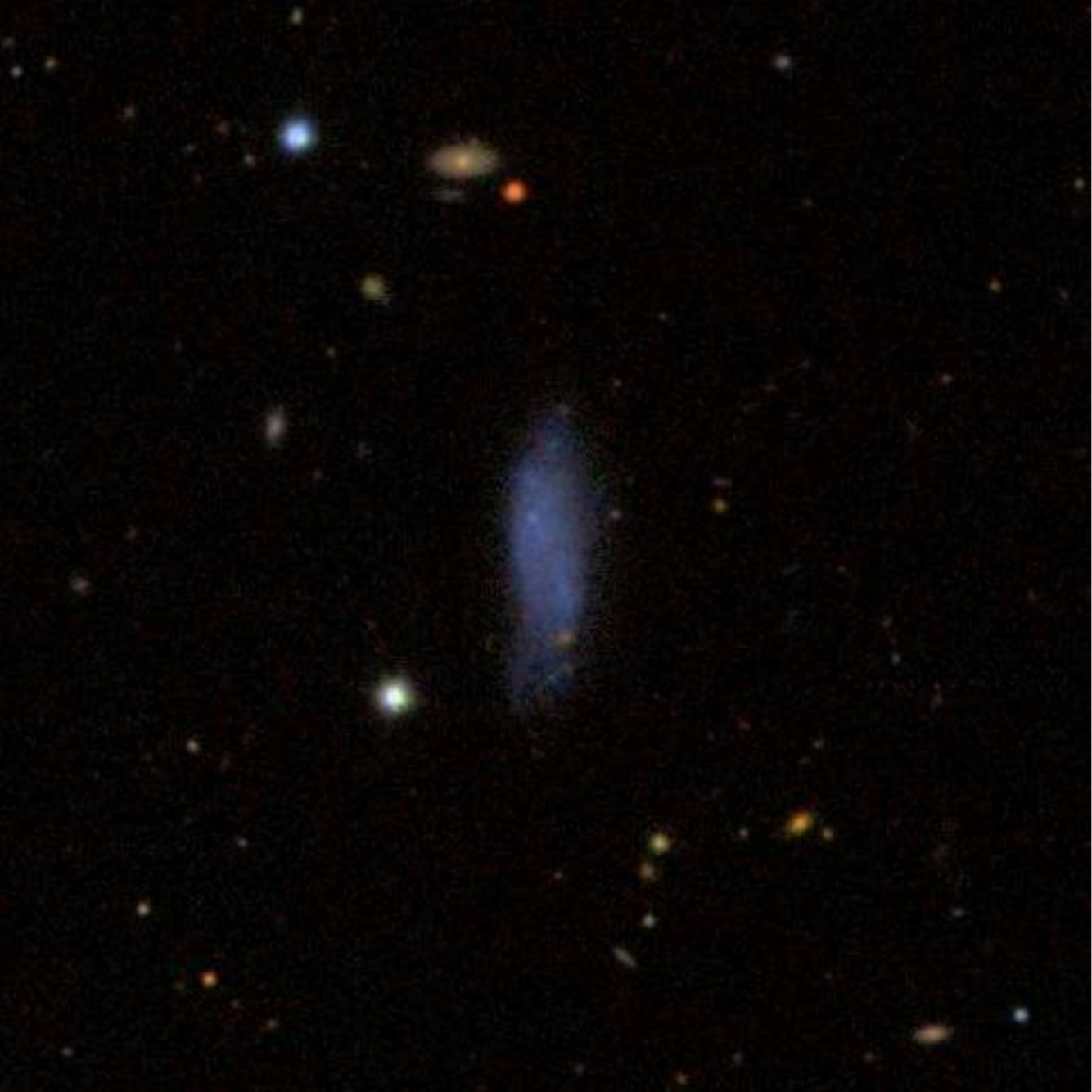}} & 
\mbox{\includegraphics[height=3cm]{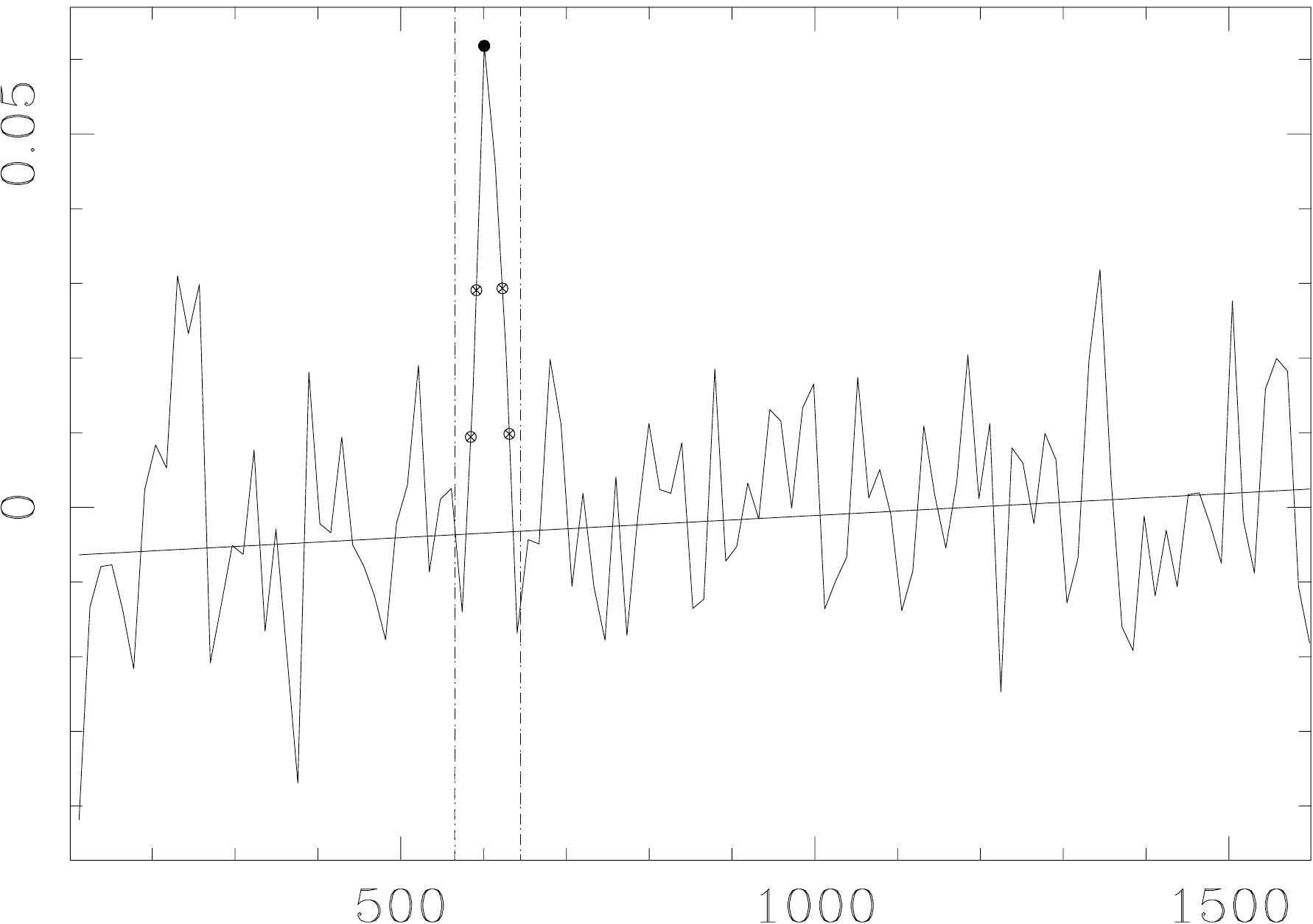}}\\
[7pt]

\multicolumn{2}{c}{HIJASS J1246+36 (UGC07949$^{ei}$)}
&&
\multicolumn{2}{c}{HIJASS J1228+22A (UGC07584$^{df}$)}\\
\mbox{\includegraphics[height=3cm]{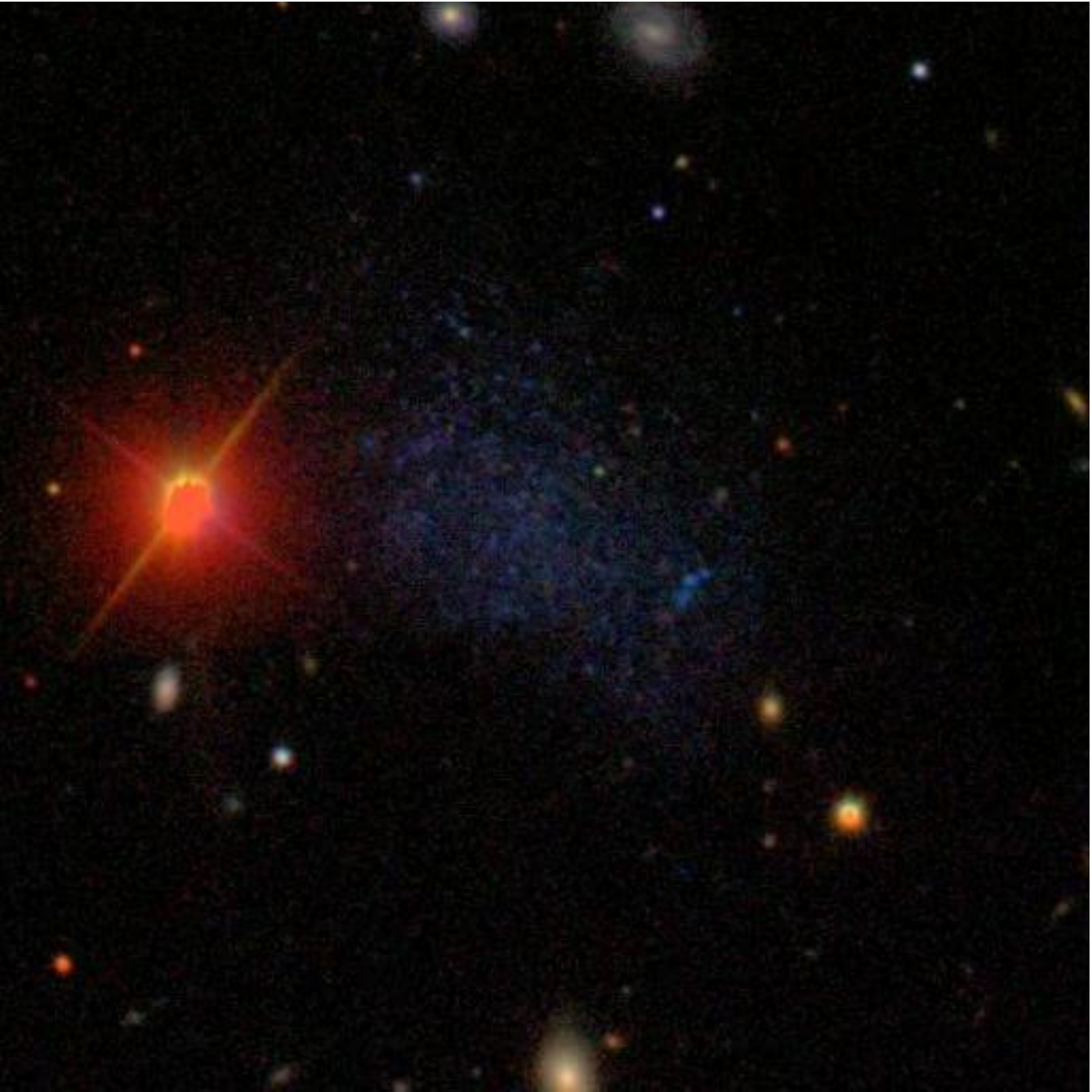}} & 
\mbox{\includegraphics[height=3cm]{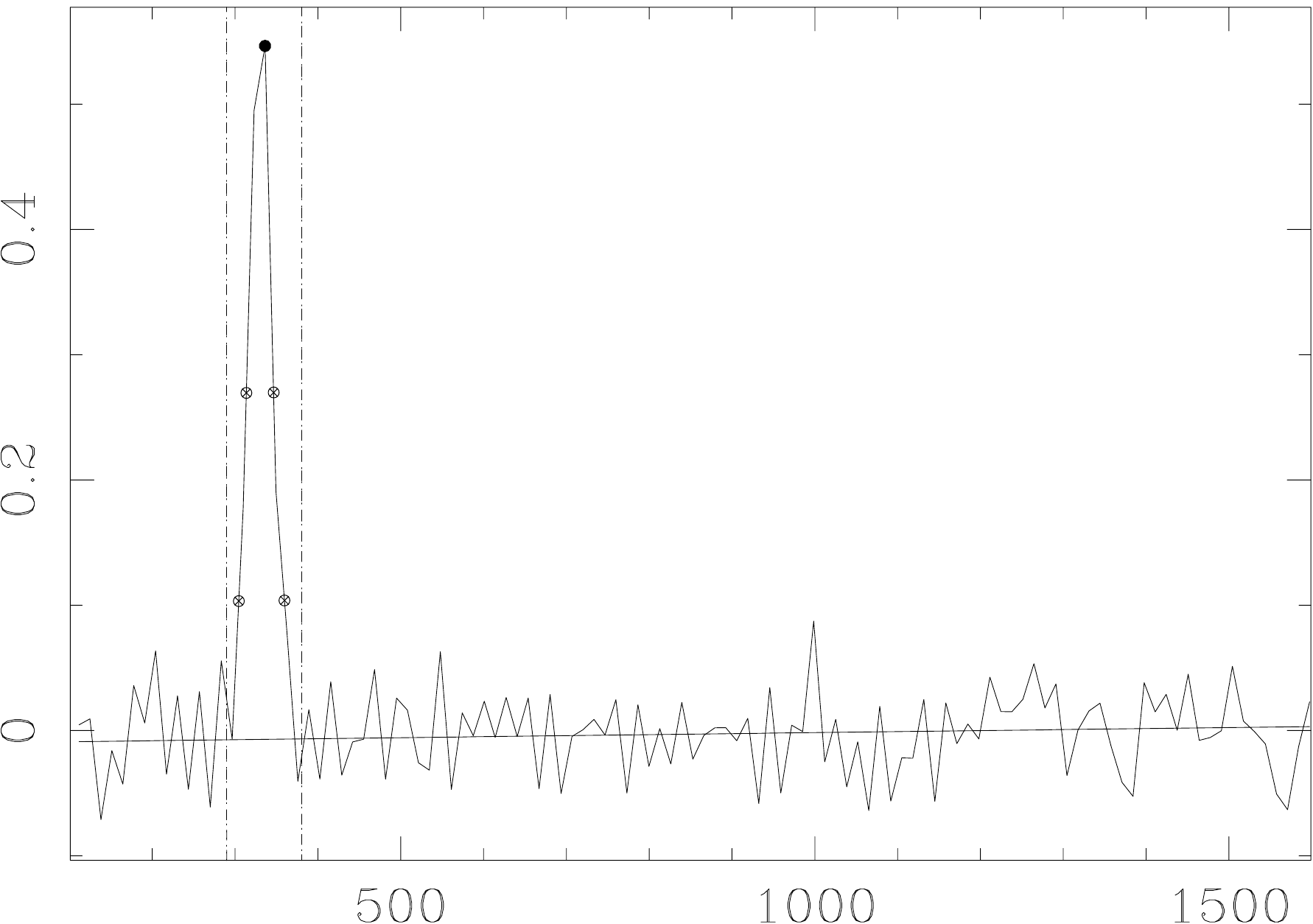}} & 
 & 
\mbox{\includegraphics[height=3cm]{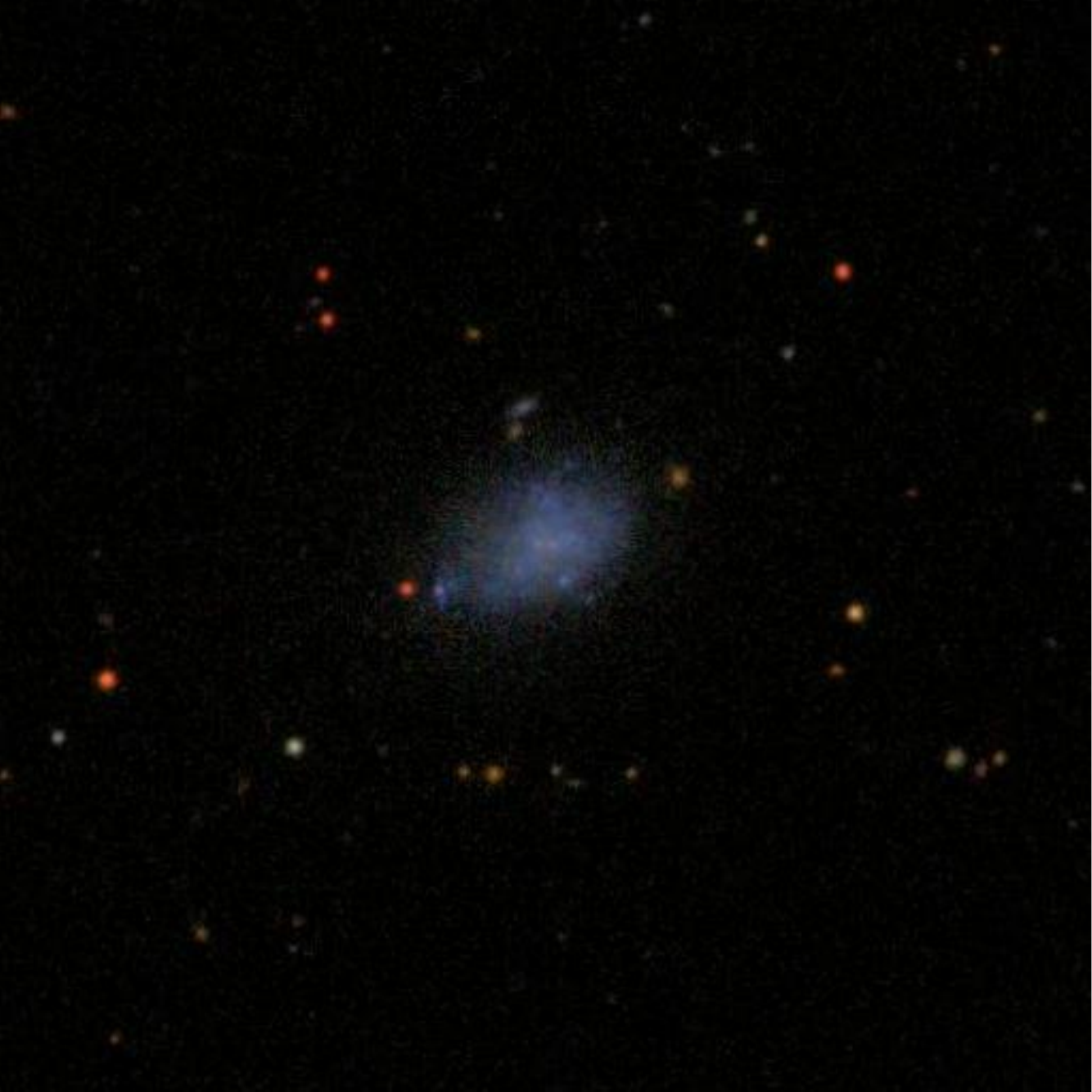}} & 
\mbox{\includegraphics[height=3cm]{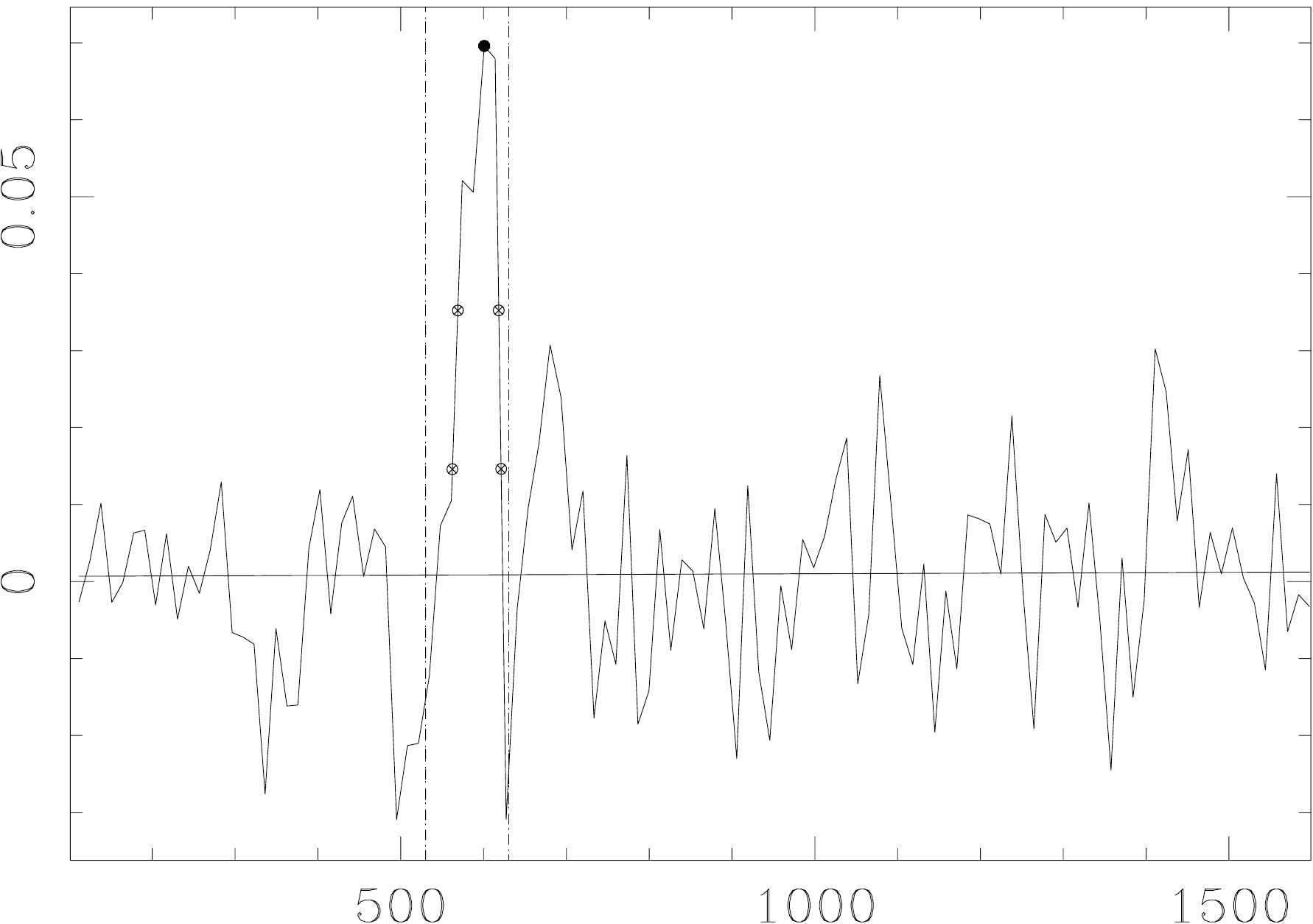}}\\
[7pt]

\multicolumn{2}{c}{HIJASS J1238+32 (UGCA292$^{ei}$)}
& &
\multicolumn{2}{c}{HIJASS J1220+38 (KUG1218+387)}\\
\mbox{\includegraphics[height=3cm]{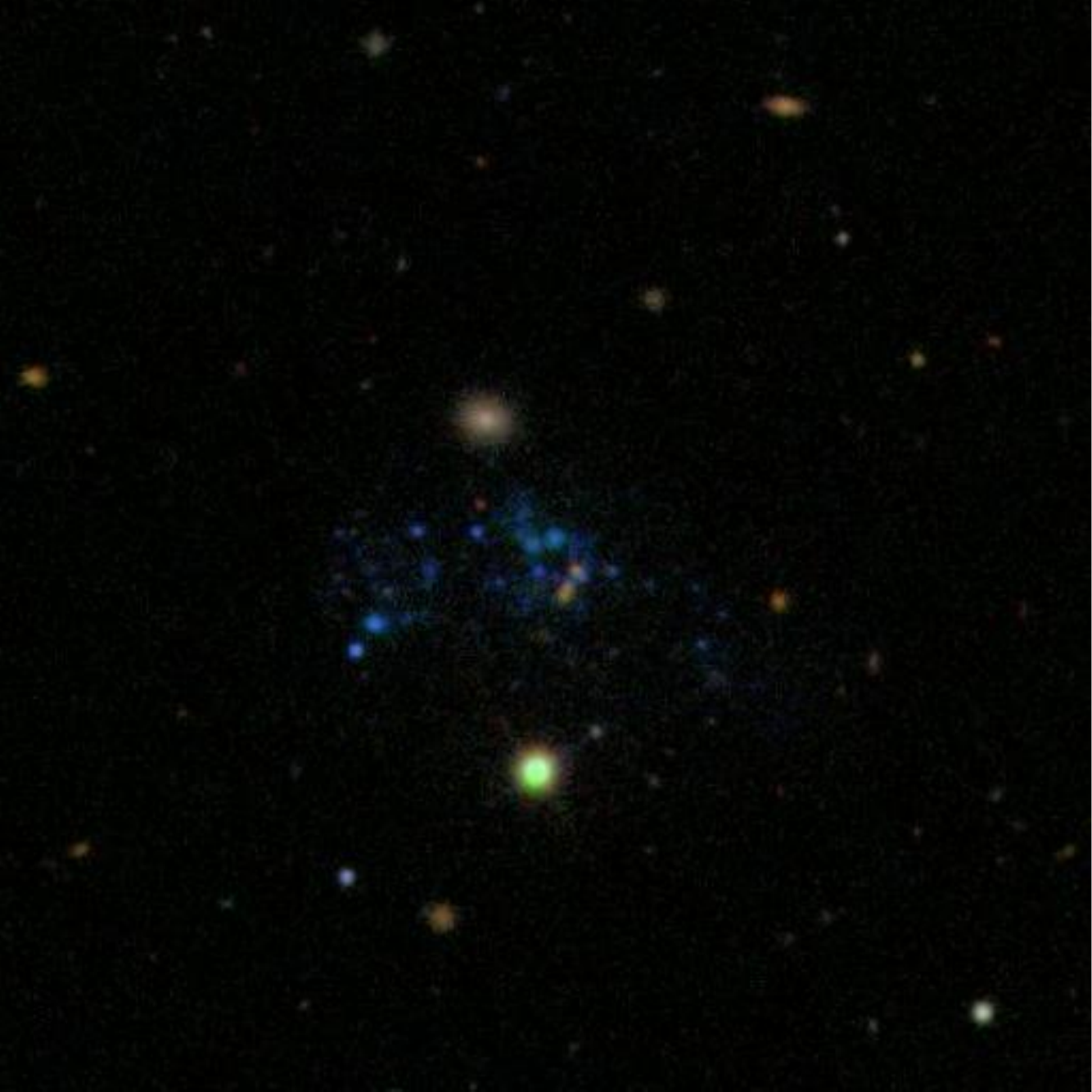}} & 
\mbox{\includegraphics[height=3cm]{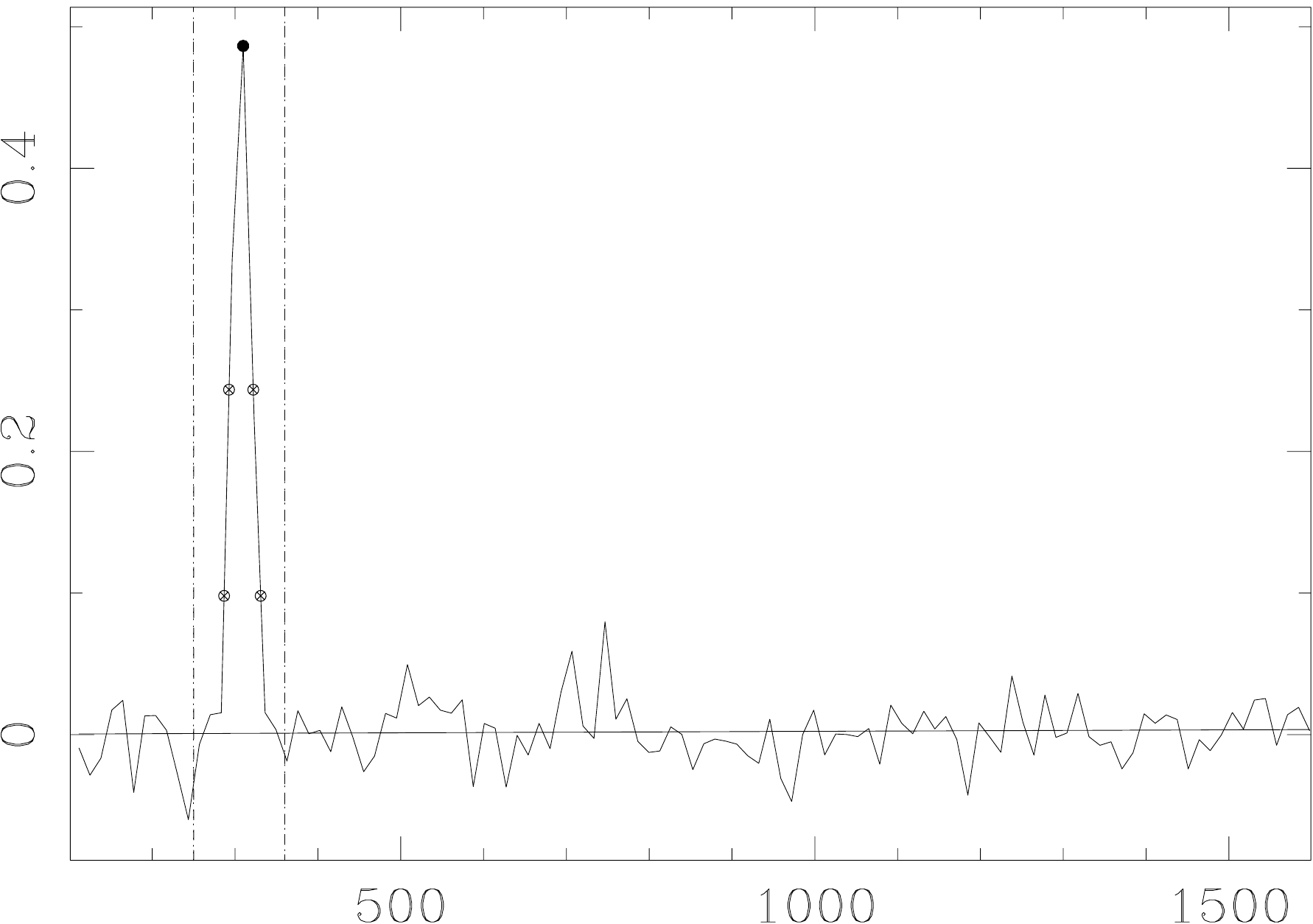}} & 
 & 
\mbox{\includegraphics[height=3cm]{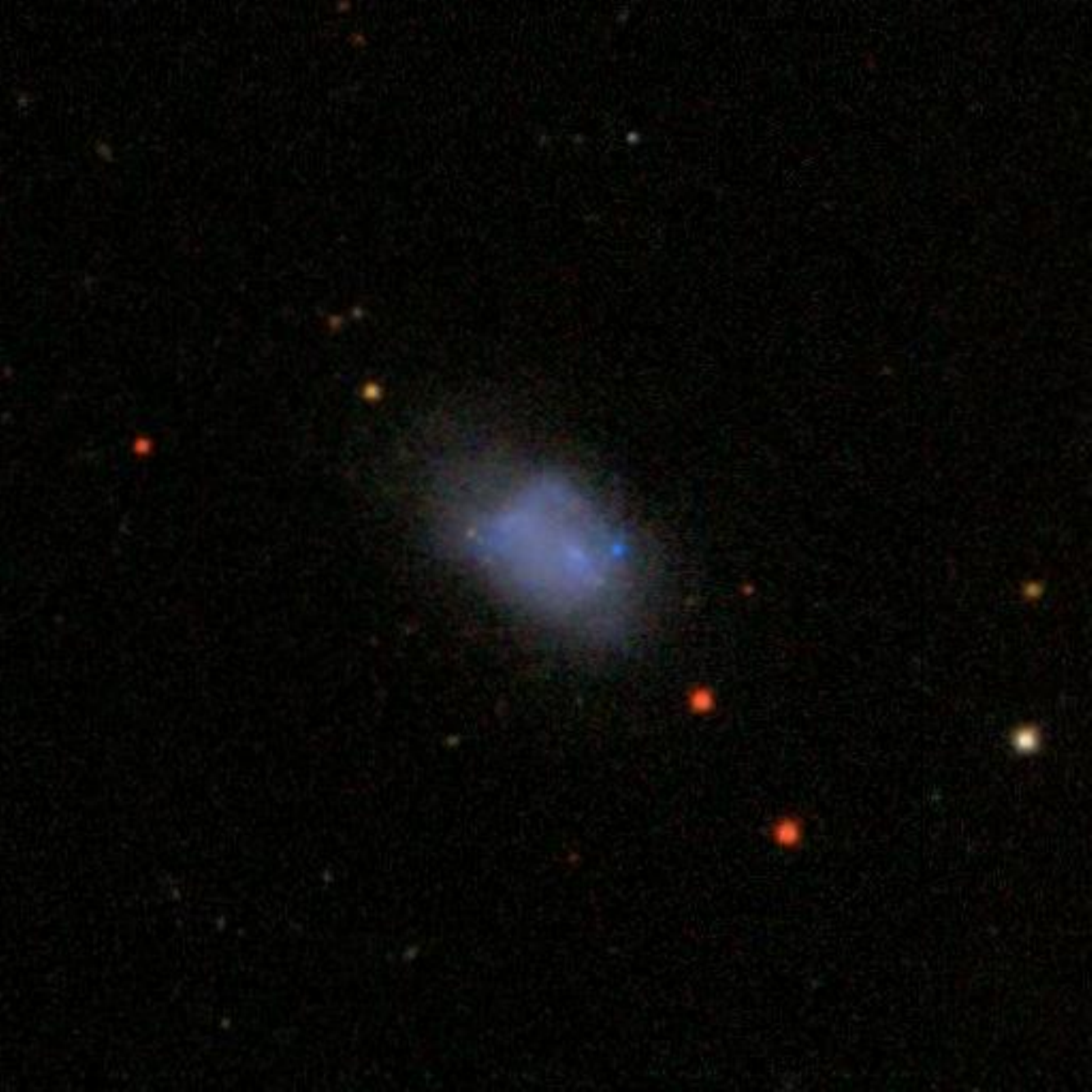}} & 
\mbox{\includegraphics[height=3cm]{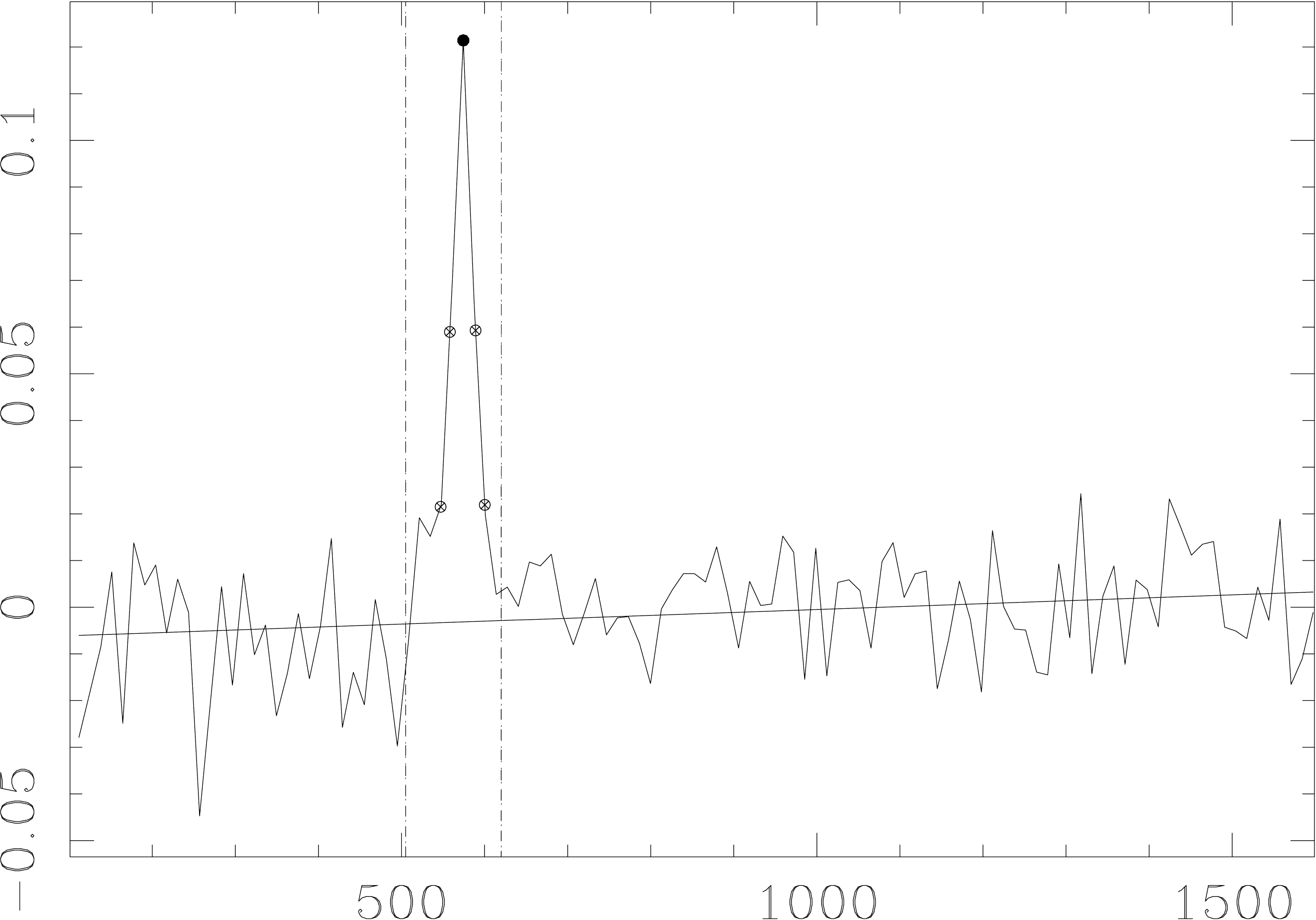}}\\
[7pt]

\multicolumn{2}{c}{HIJASS J1233+39 (MCG+07-26-024$^{d}$)}
 & &
\multicolumn{2}{c}{HIJASS J1201+33 (UGC07007)}\\
\mbox{\includegraphics[height=3cm]{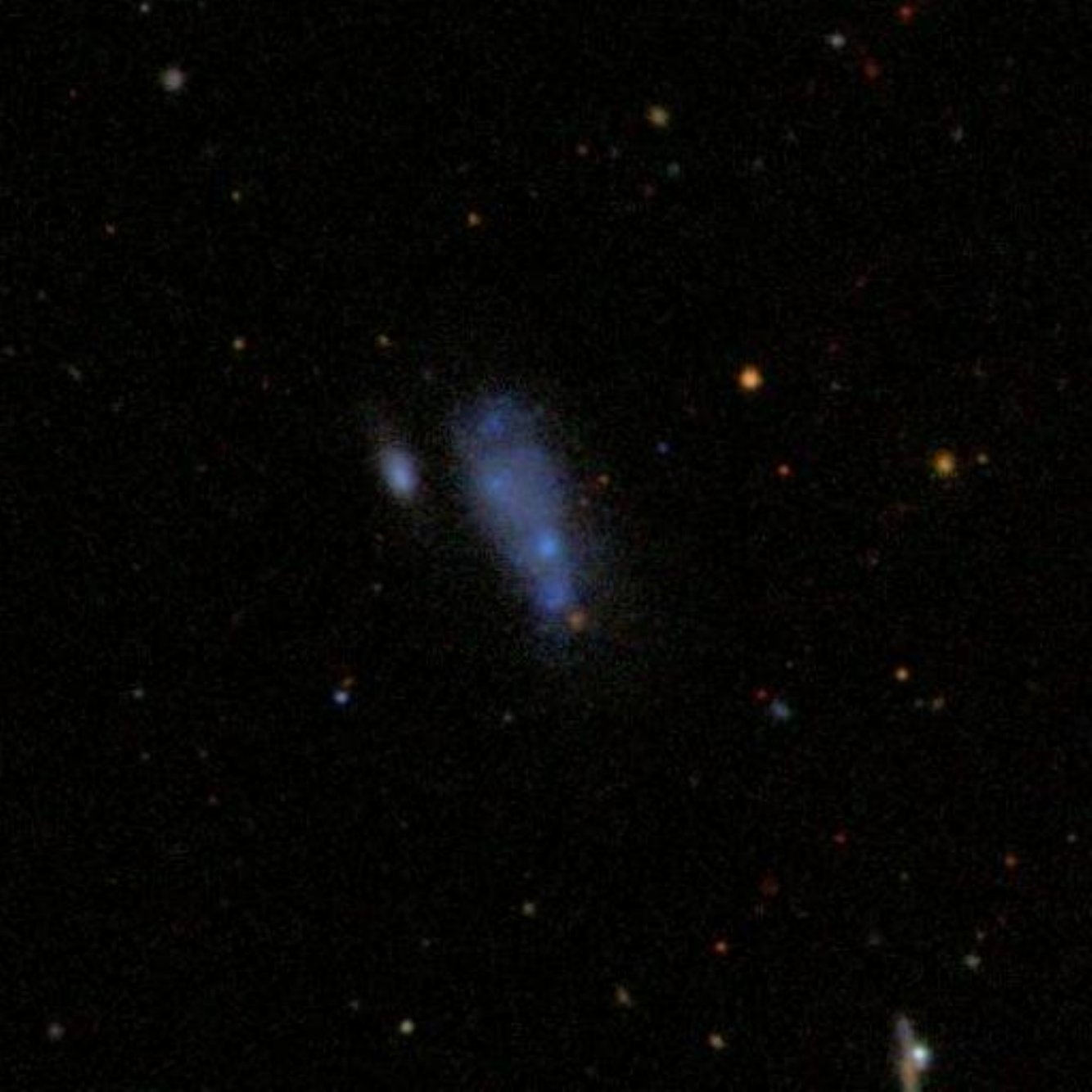}} & 
\mbox{\includegraphics[height=3cm]{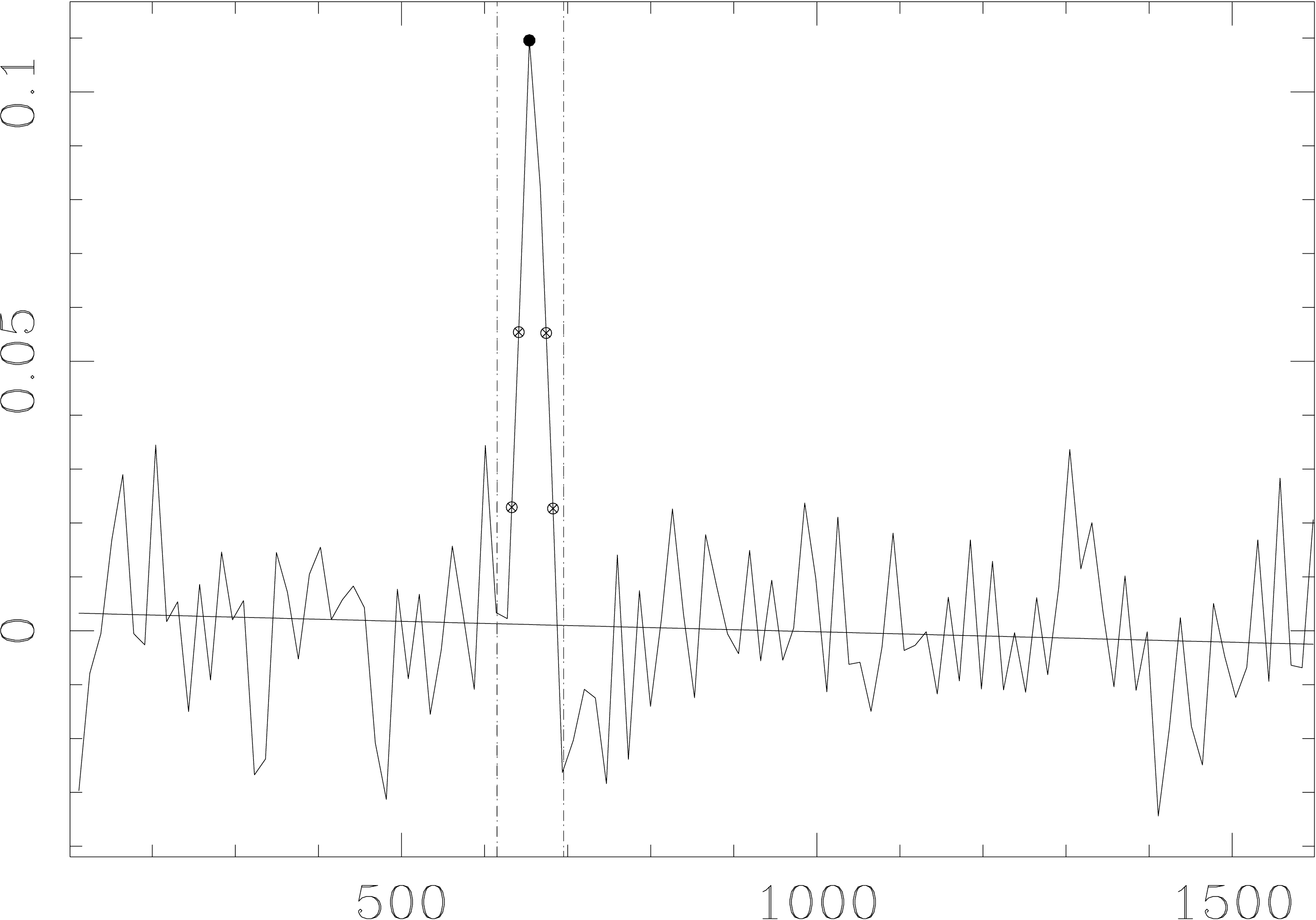}} & 
 & 
\mbox{\includegraphics[height=3cm]{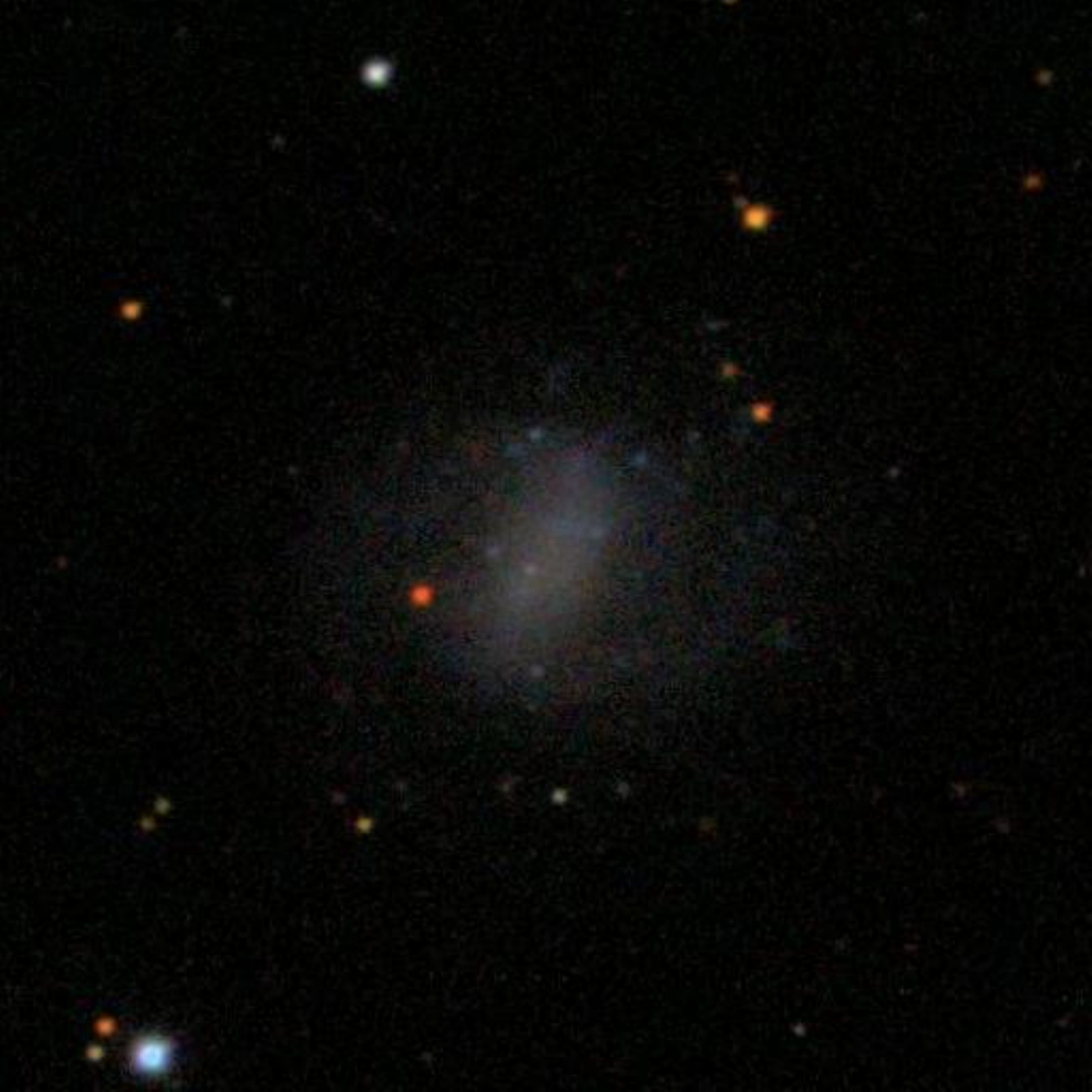}} & 
\mbox{\includegraphics[height=3cm]{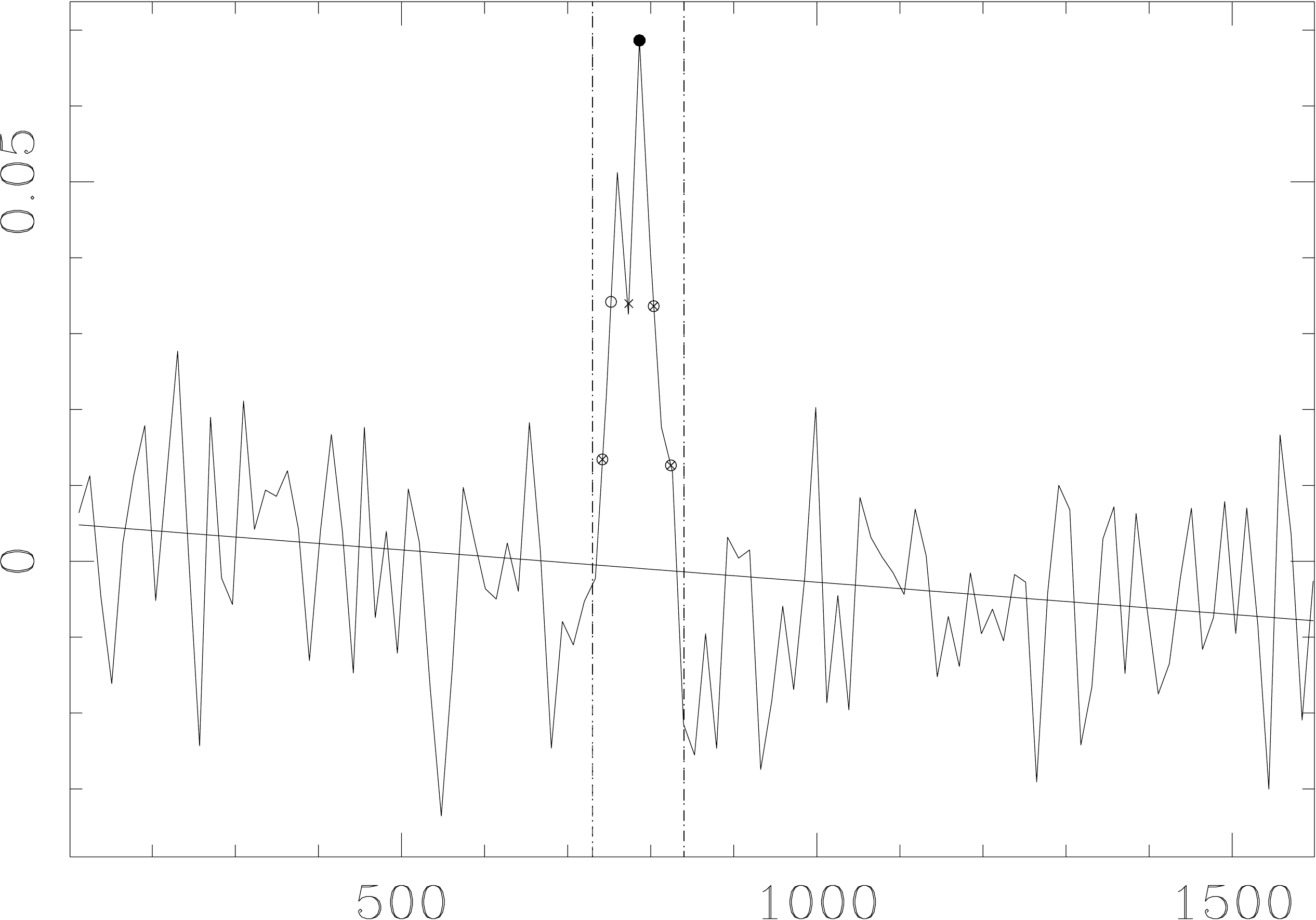}}\\
[7pt]

\end{tabular}
\caption{\textbf{Multi-colour SDSS images of the lowest \HI mass and previously catalogued galaxies with systemic velocities between 300 and 1900~km~s$^{-1}$ together with the HIJASS spectra.} The \HI masses increase from the left to the right and from the top to the bottom -- note that we do not show HIJASS~J1218+46 with an \HI mass between the ones of HIJASS~J1228+22A and HIJASS~J1238+32 as it is shown in Figure~\ref{fig:newlydetected}. The HIJASS name along with the most likely optical counterpart (in brackets) is given above each SDSS image and spectrum (including the flags as discussed in Section~\ref{sec:catalogue}). The SDSS images ($3.3' \times 3.3'$) are centered on the optical positions of the most likely counterparts. The \HI spectra have the optical velocities ($cz$ in km~s$^{-1}$) along the abscissas and the flux densities (in Jy) along the ordinates. The \HI line emission analysis is conducted within the velocity range as indicated by the dotted vertical lines. The baseline fits to the spectra are conducted within the displayed velocity range excluding the velocity ranges of \HI line emission analysis and of other \HI sources (e.g. NGC4393 at $\sim$650--850~km~s$^{-1}$ in the spectra of HIJASS~J1226+27). Note that we show the \HI spectra prior to baseline subtraction. Hence, the marked \HI peak flux as well as the 50 and 20~per~cent velocity widths slightly vary from the values given in the catalogue as they are measured after baseline subtraction.}
\label{fig:lowestHIMass}
\end{figure}
\twocolumn

\onecolumn
\setlength\fboxsep{0pt}
\setlength\fboxrule{1pt}

\begin{figure}
\begin{tabular}{p{3cm} p{3.5cm} p{1.5cm} p{3cm} p{3.5cm}}

\multicolumn{2}{c}{HIJASS J1219+46$^{dn}$}
& &
\multicolumn{2}{c}{HIJASS J1218+46 (SDSSJ121811.04+465501.2$^{fdnb}$)}\\
\fbox{\includegraphics[height=3cm]{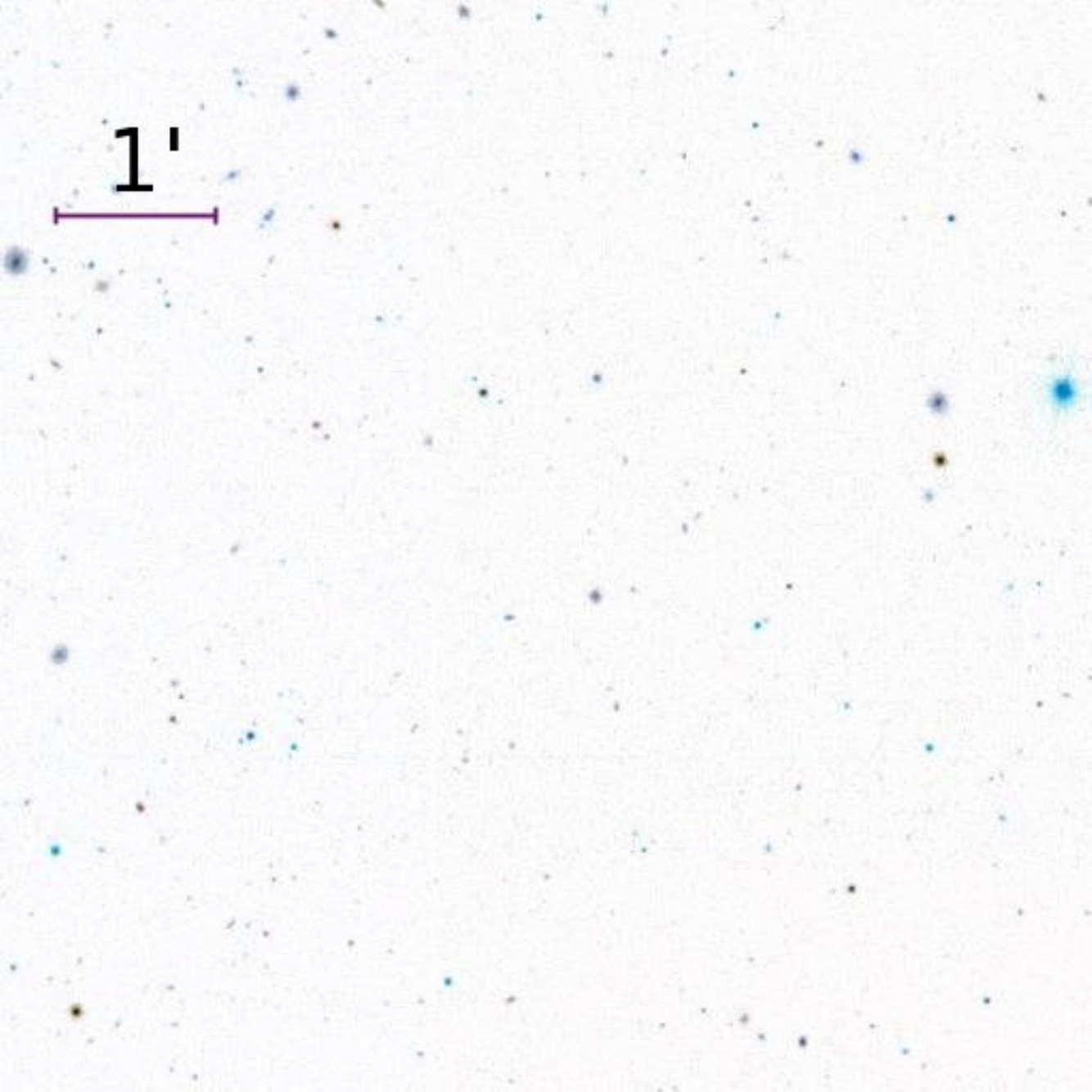}} & 
\mbox{\includegraphics[height=3.3cm]{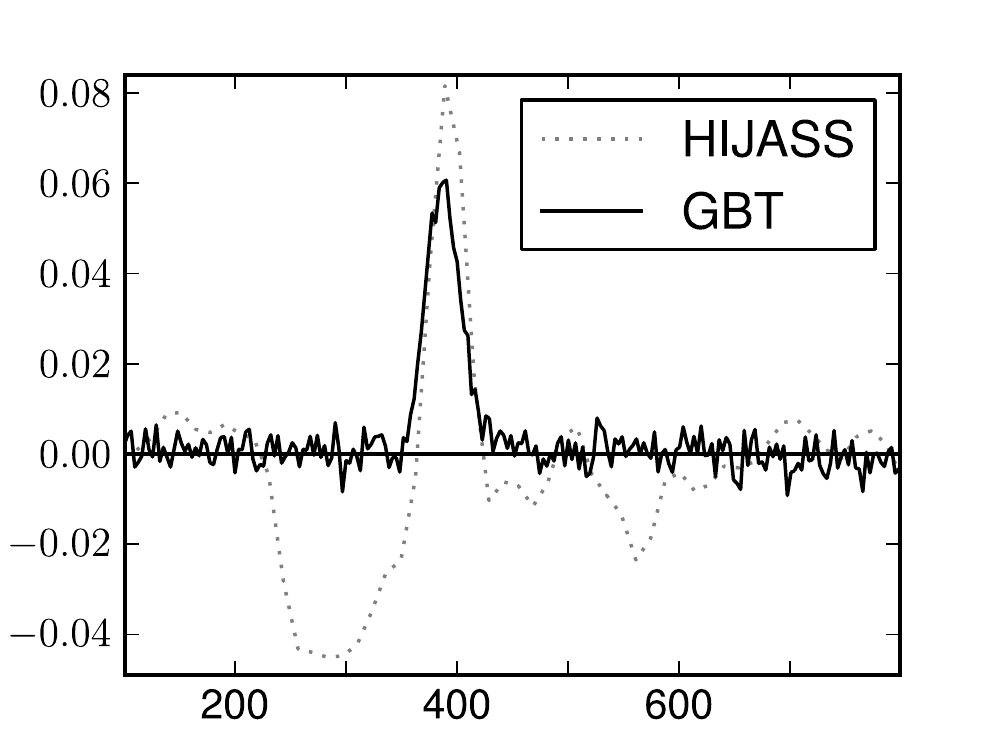}} &
&
\fbox{\includegraphics[height=3cm]{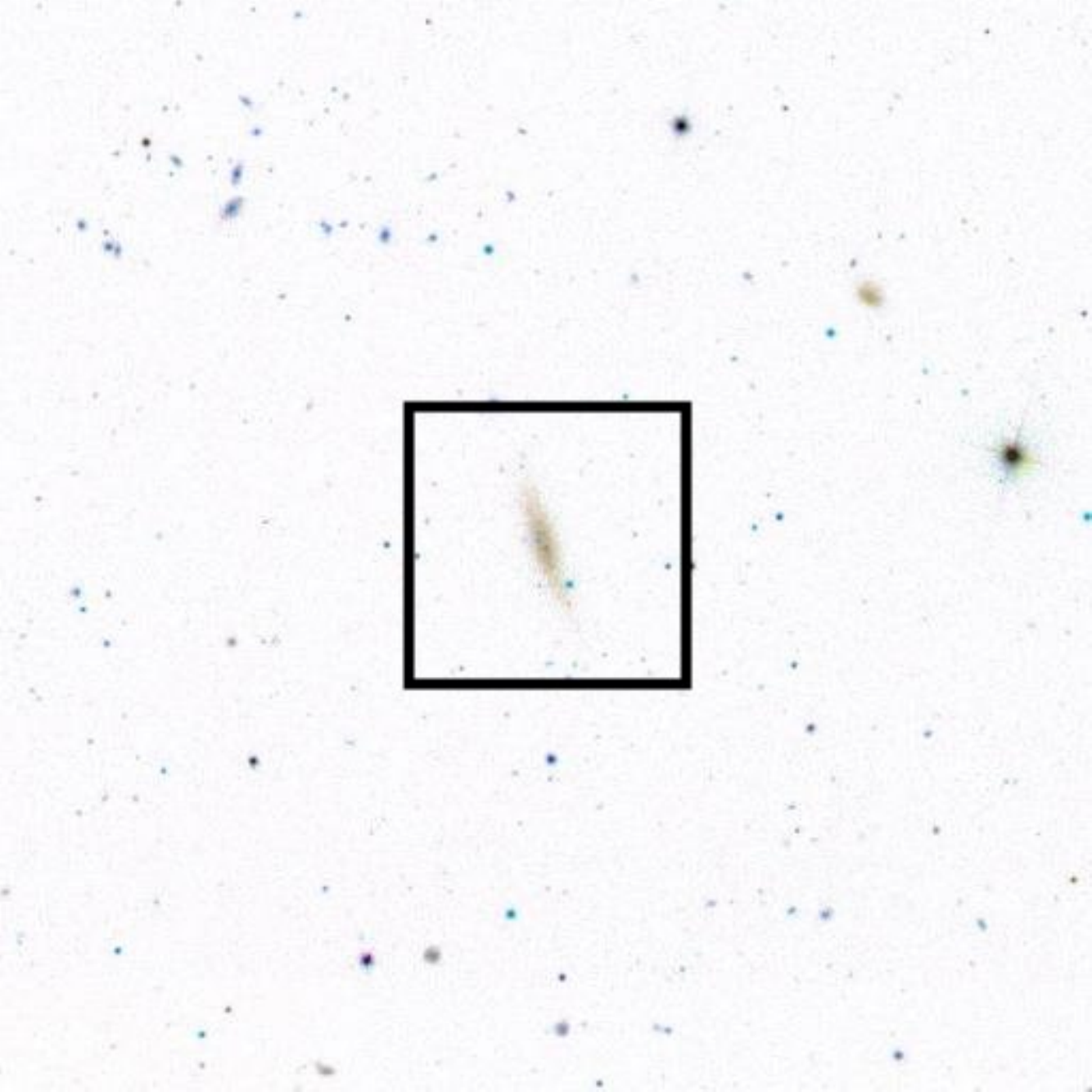}} & 
\mbox{\includegraphics[height=3.3cm]{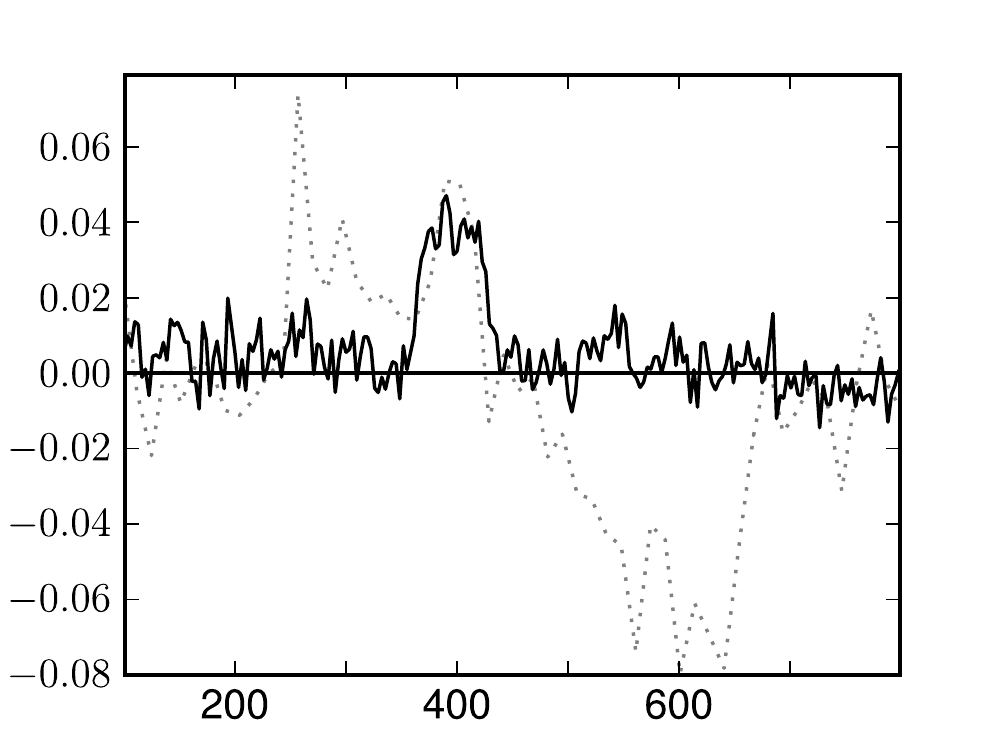}} \\
[12pt]

\multicolumn{2}{c}{HIJASS J1222+39 (SDSSJ122206.12+394452.4$^n$)}
 & &
\multicolumn{2}{c}{HIJASS J1209+43B (UGC07146$^{n}$)}\\
\fbox{\includegraphics[height=3cm]{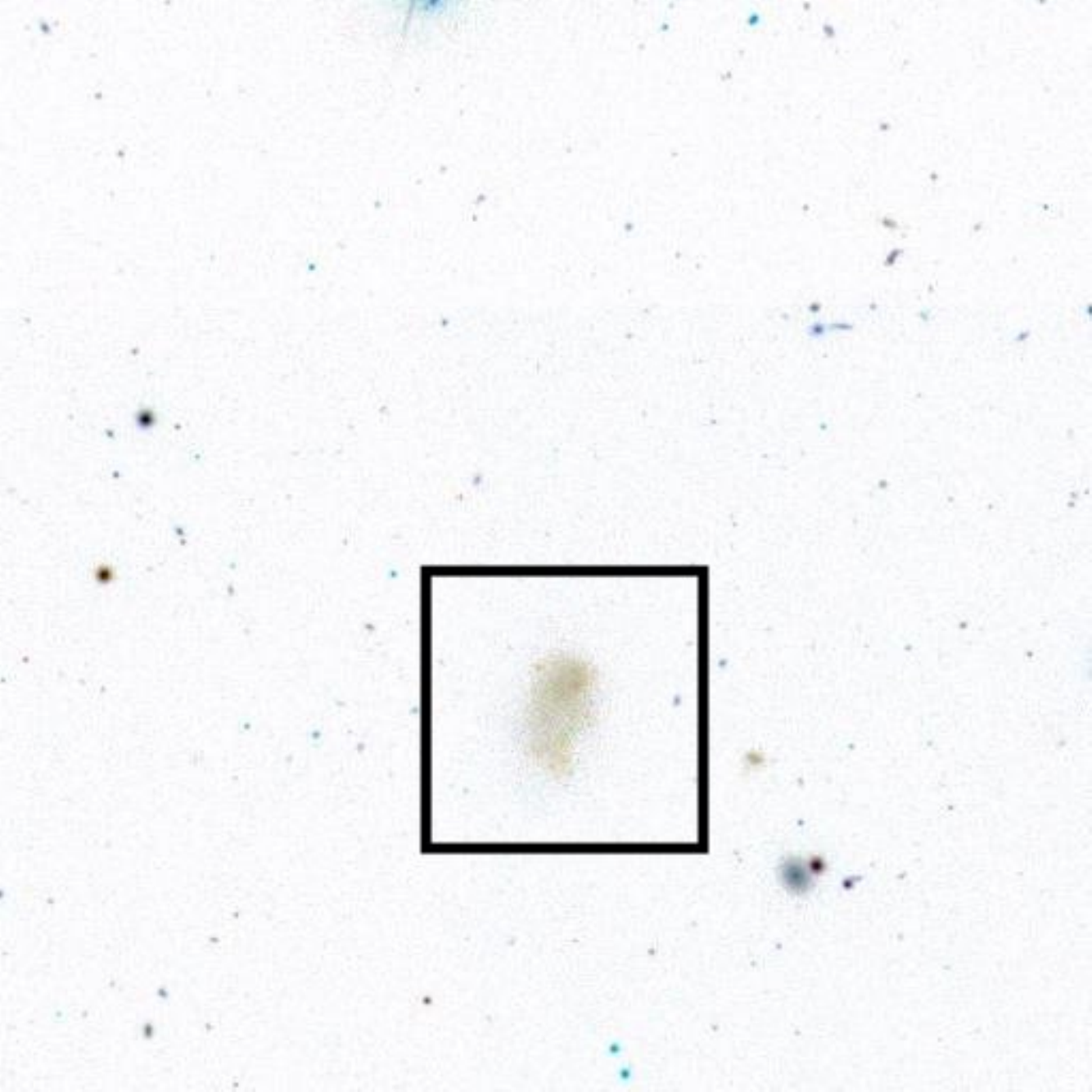}} & 
\mbox{\includegraphics[height=3.3cm]{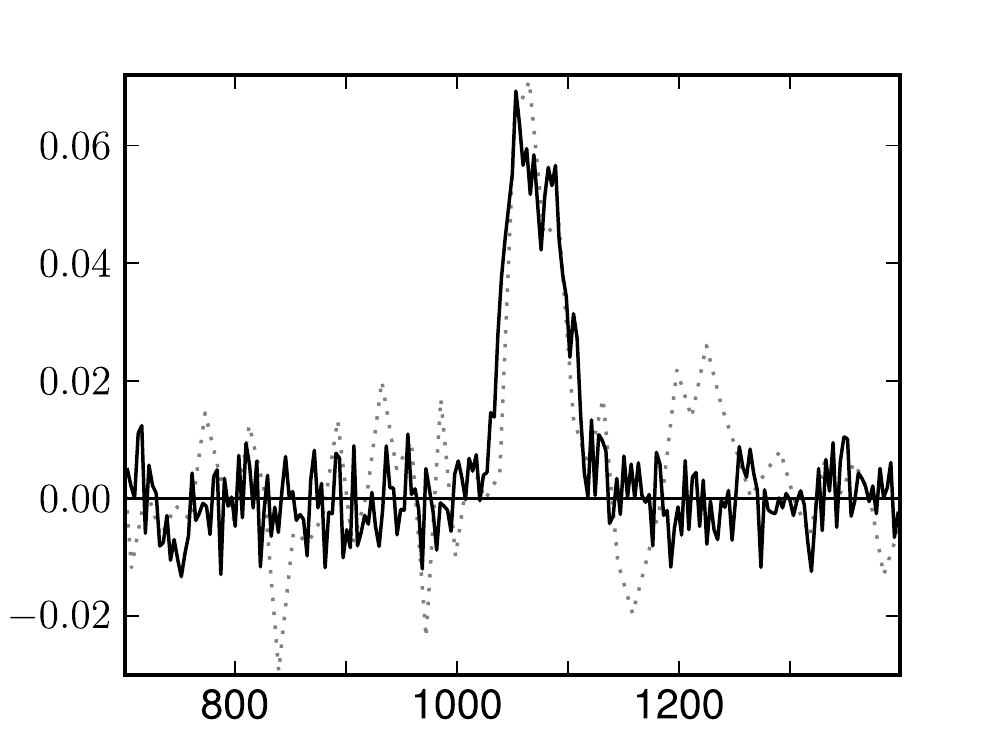}} & 
 & 
\fbox{\includegraphics[height=3cm]{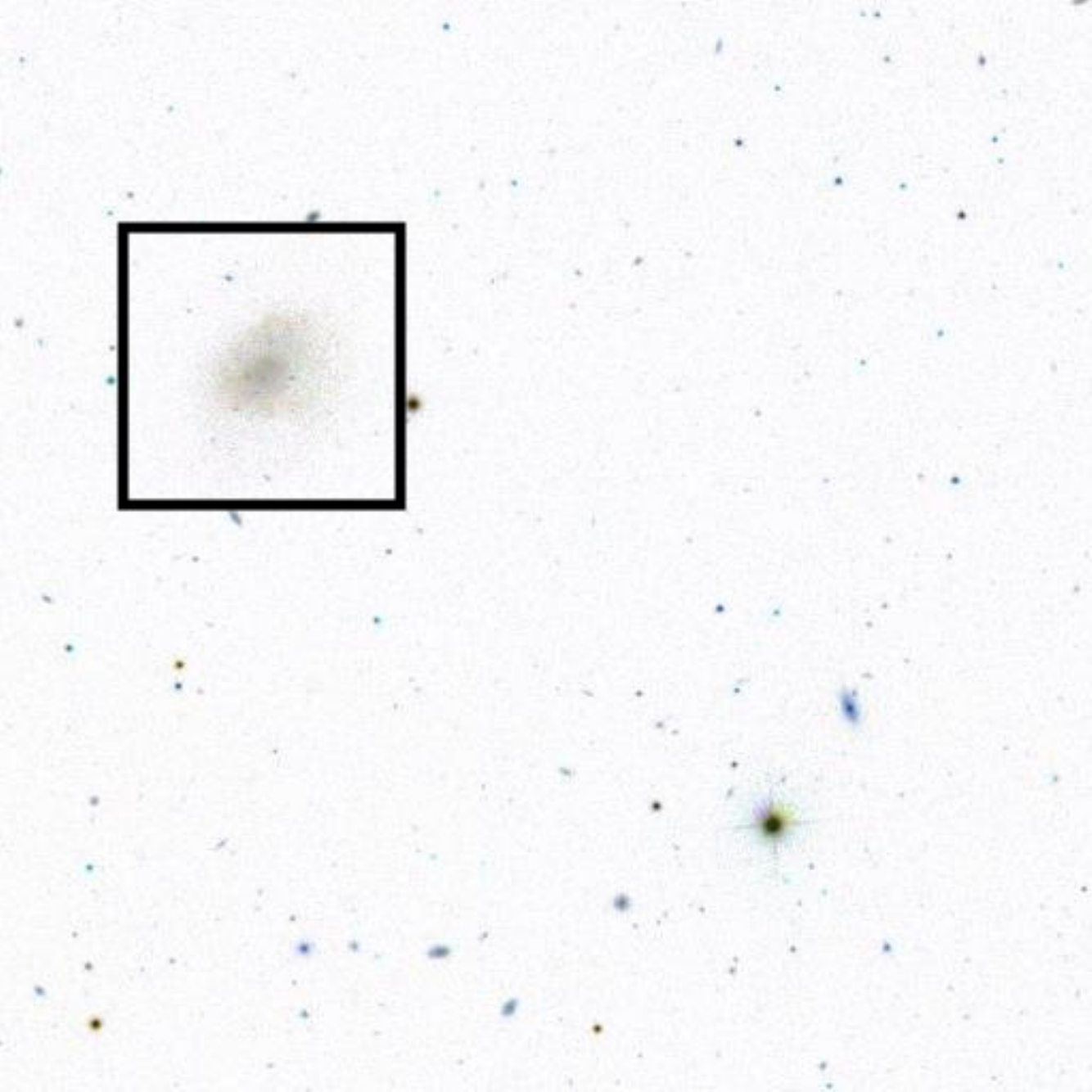}} & 
\mbox{\includegraphics[height=3.3cm]{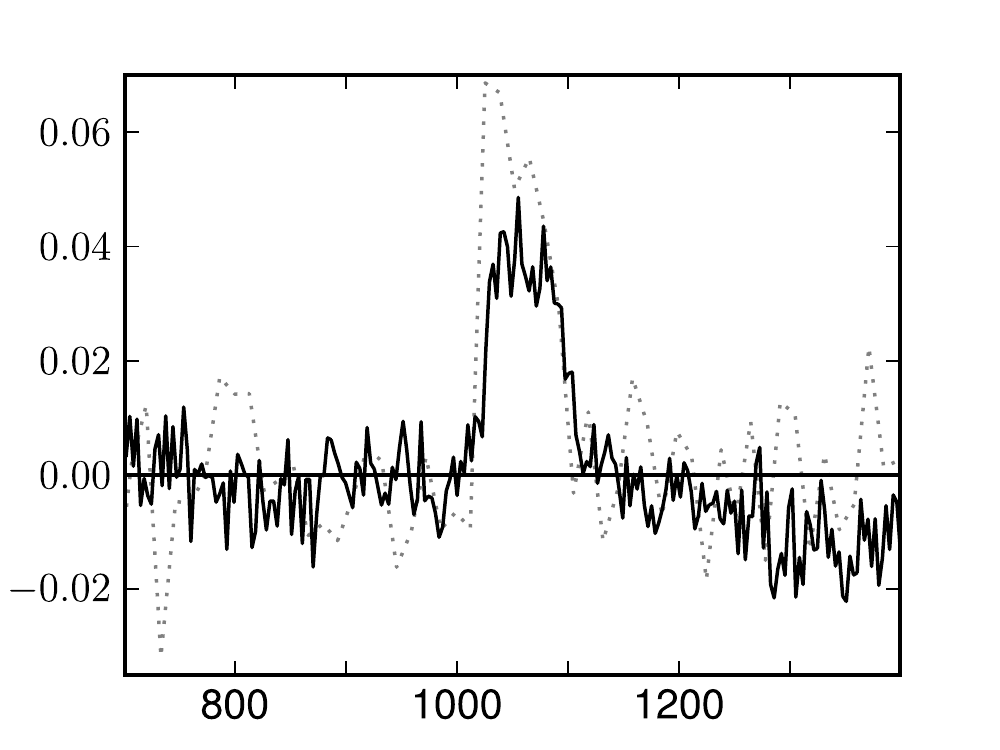}}\\
[12pt]

\multicolumn{2}{c}{HIJASS J1141+46 (CGCG242-075$^{dn}$)}
 & &
\multicolumn{2}{c}{HIJASS J1145+50 (SDSSJ114506.26+501802.4$^{fn}$)}\\
\fbox{\includegraphics[height=3cm]{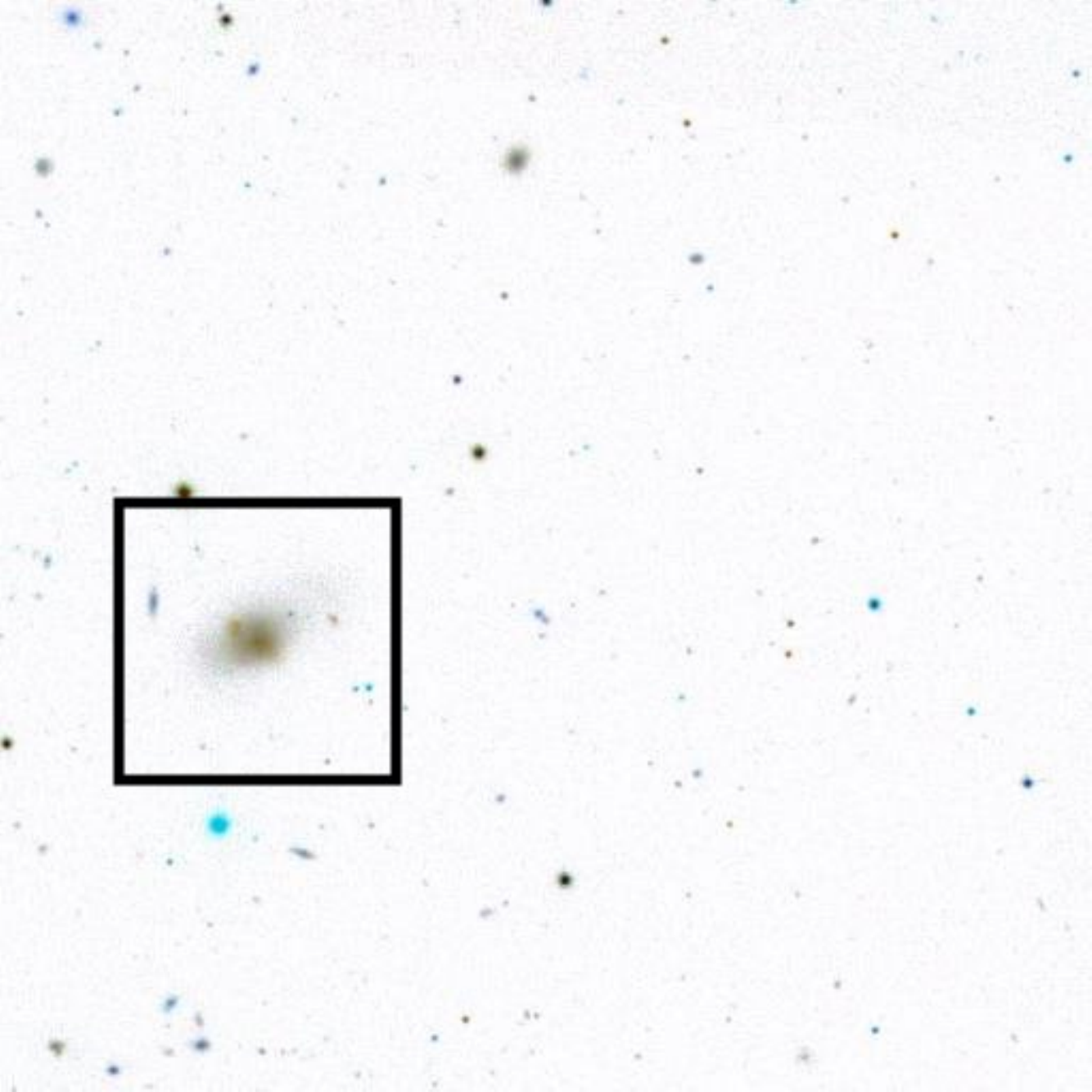}} & 
\mbox{\includegraphics[height=3.3cm]{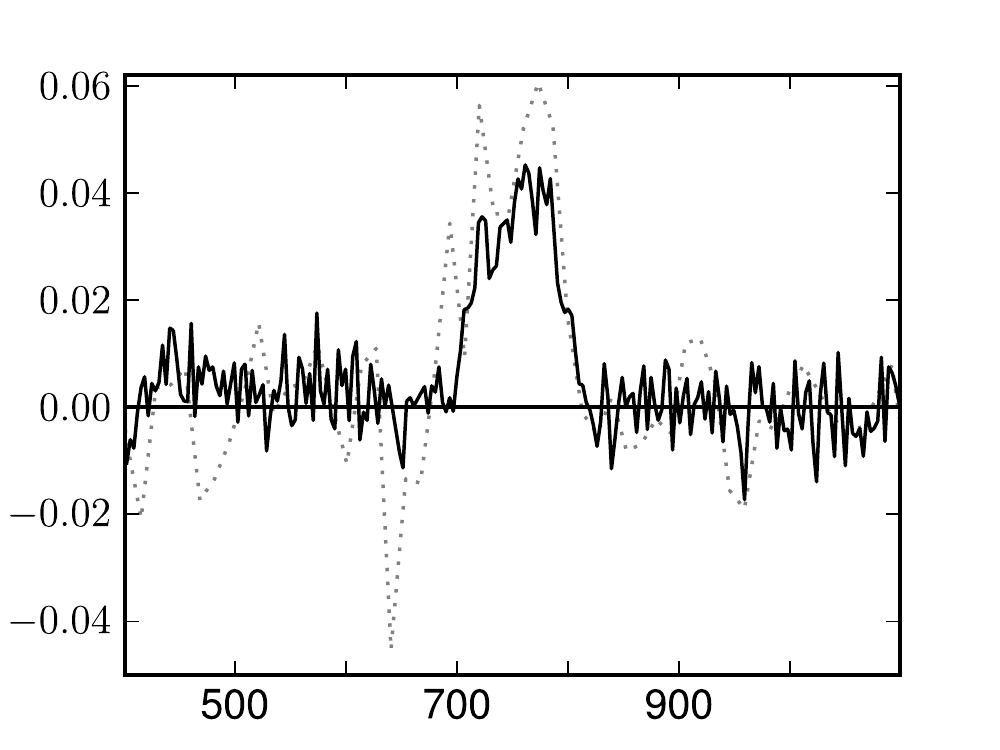}} & 
 & 
\fbox{\includegraphics[height=3cm]{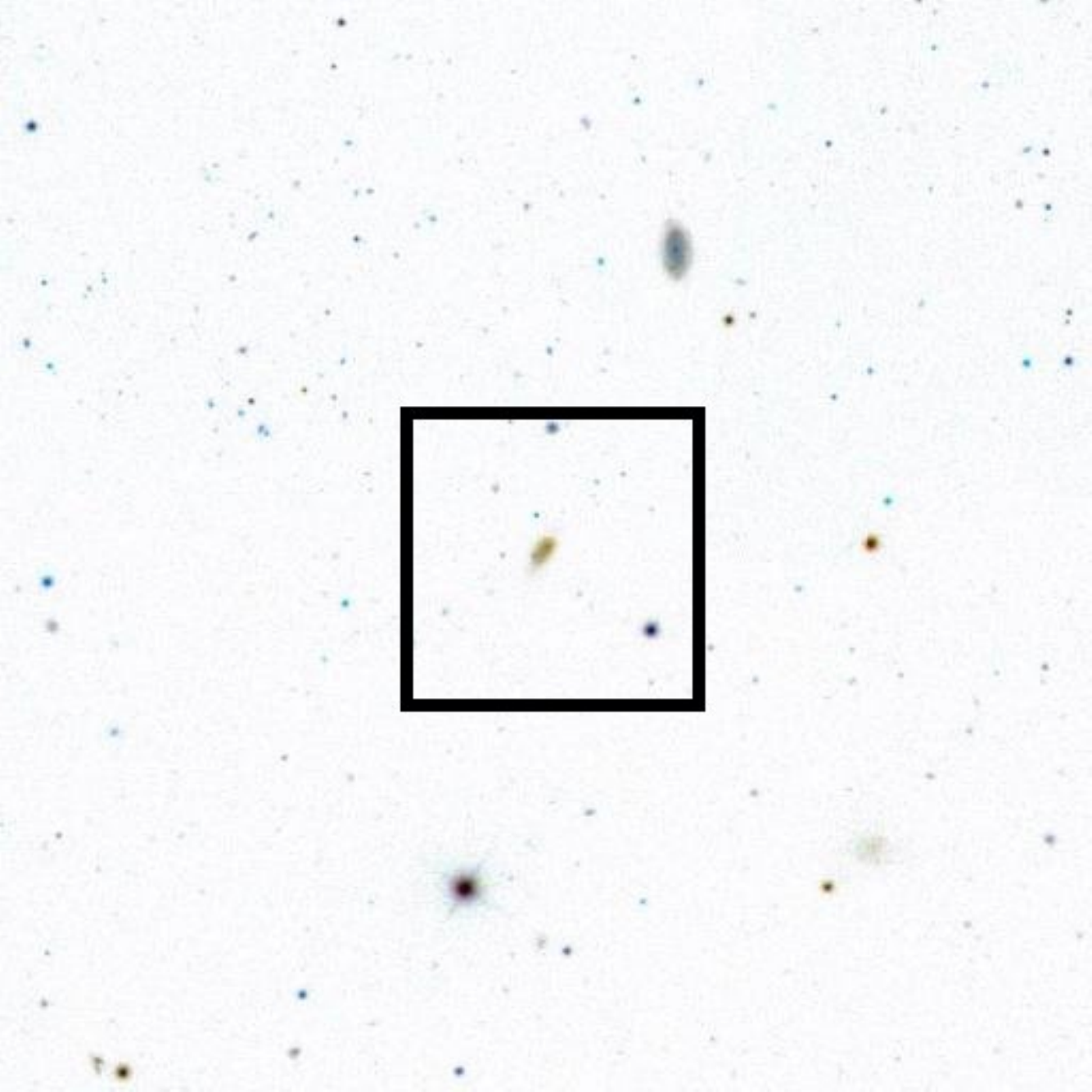}} & 
\mbox{\includegraphics[height=3.3cm]{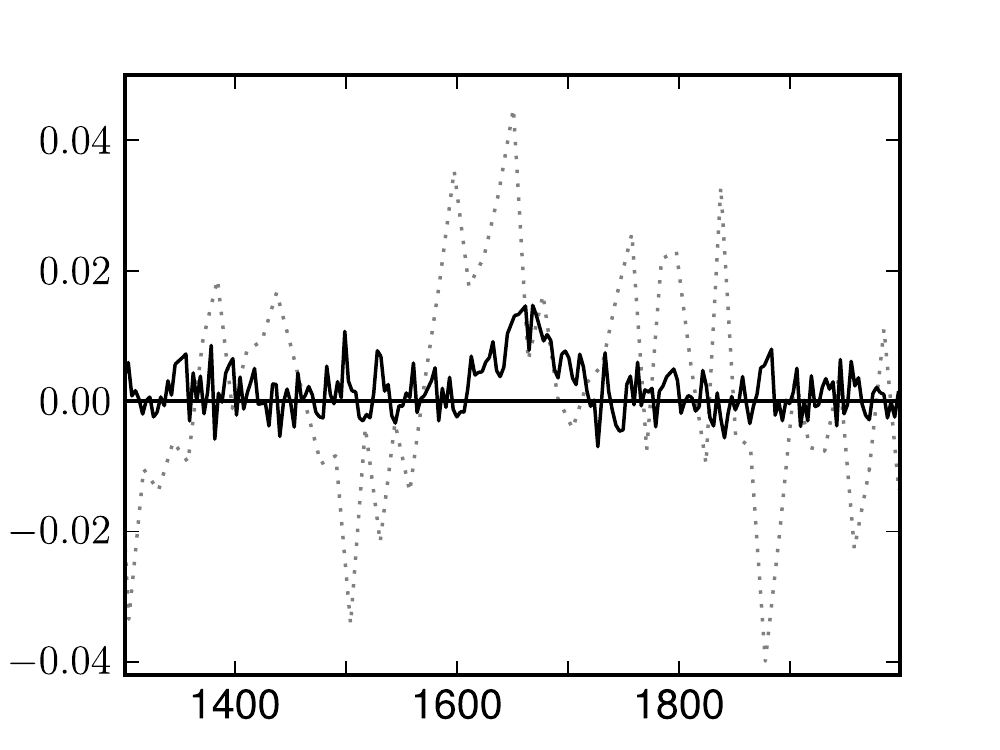}}\\
[12pt]

\multicolumn{2}{c}{HIJASS J1138+43 (UGC06611$^{cn}$)}
&&
\multicolumn{2}{c}{HIJASS J1154+30 (SDSSJ115404.42+300634.6$^{n}$)}\\
\fbox{\includegraphics[height=3cm]{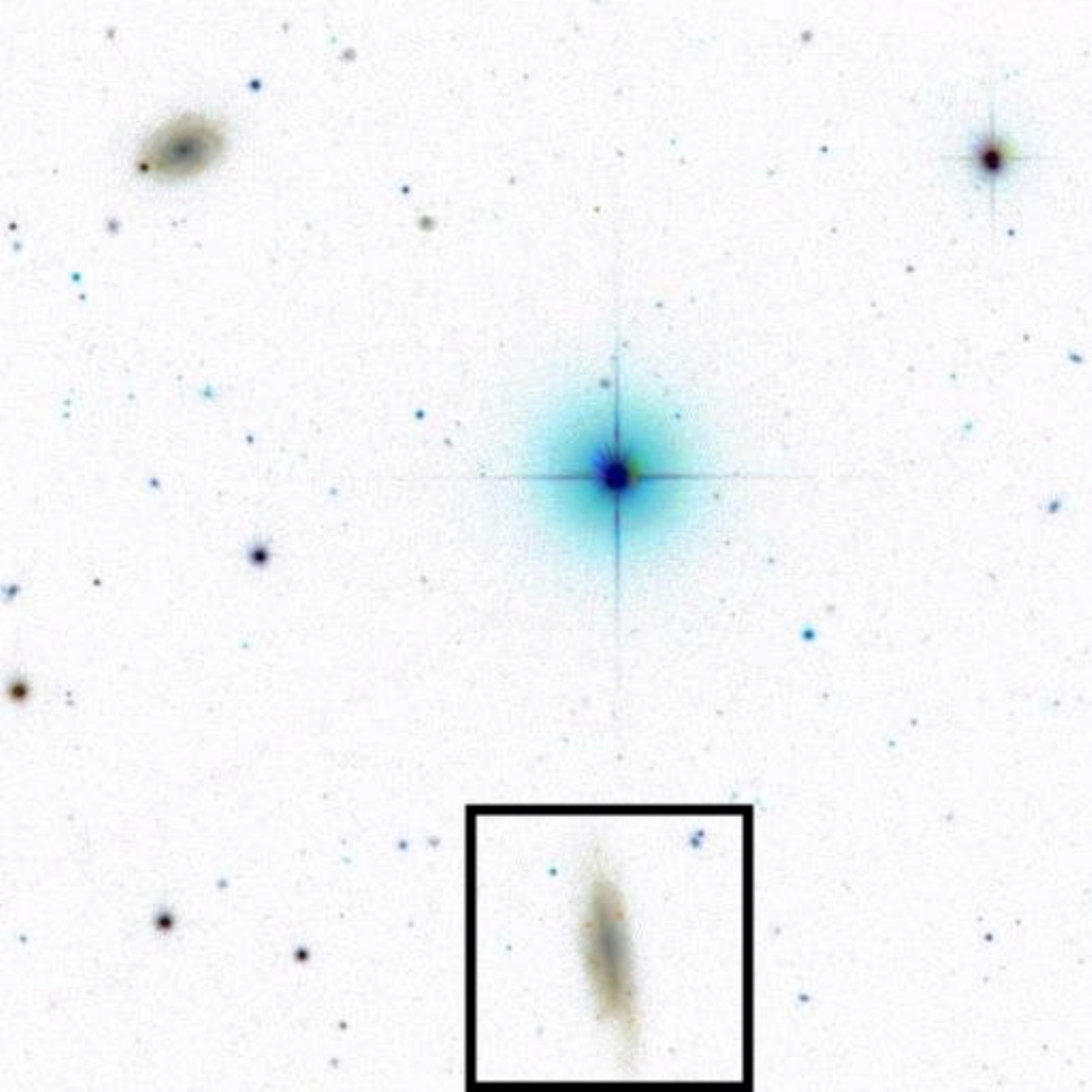}} & 
\mbox{\includegraphics[height=3.3cm]{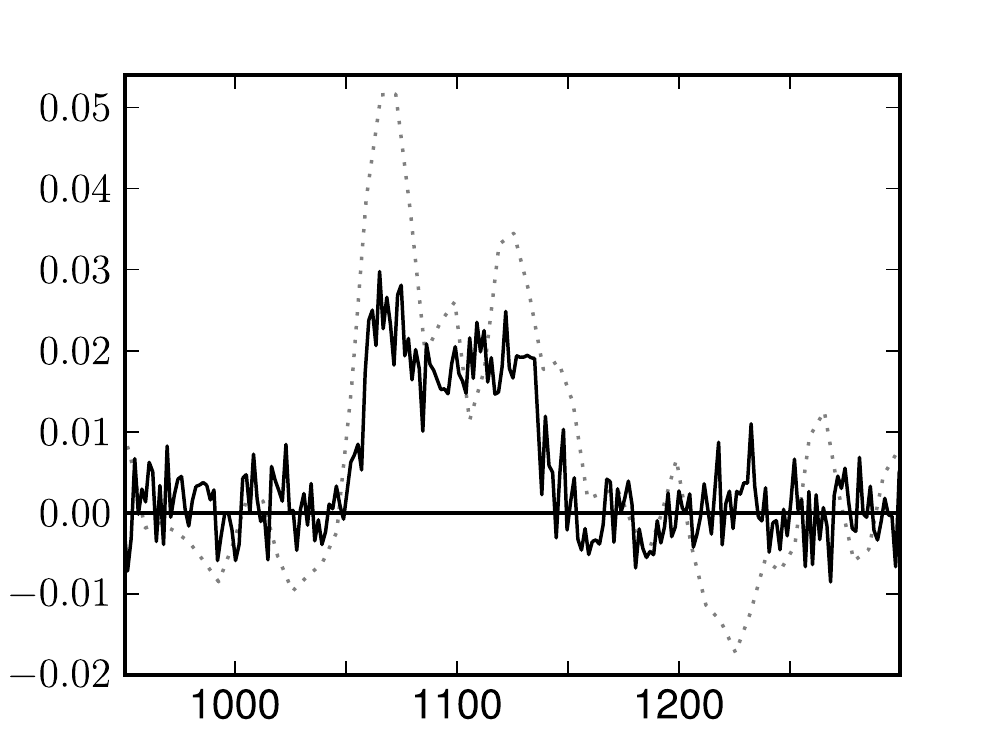}} & 
 & 
\fbox{\includegraphics[height=3cm]{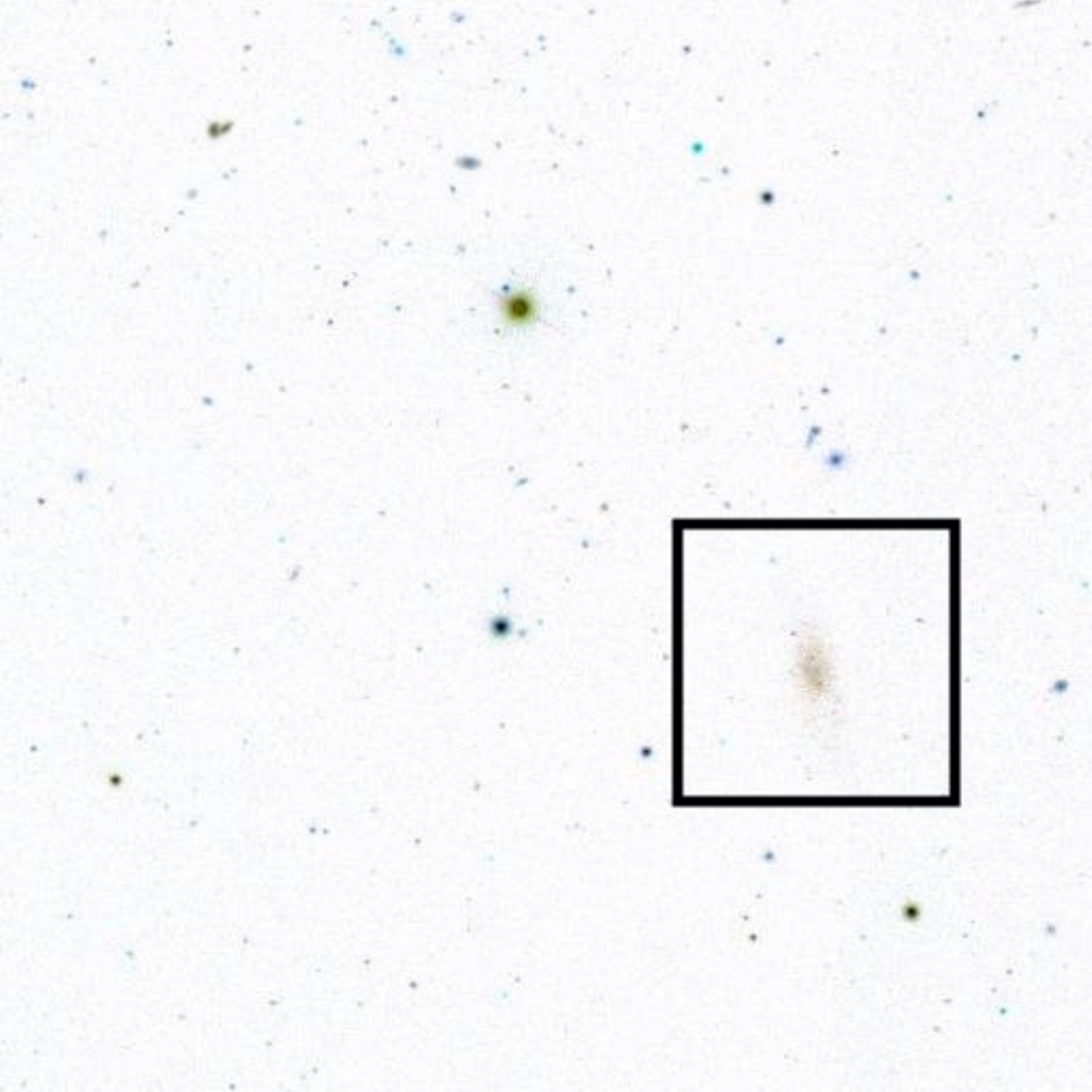}} & 
\mbox{\includegraphics[height=3.3cm]{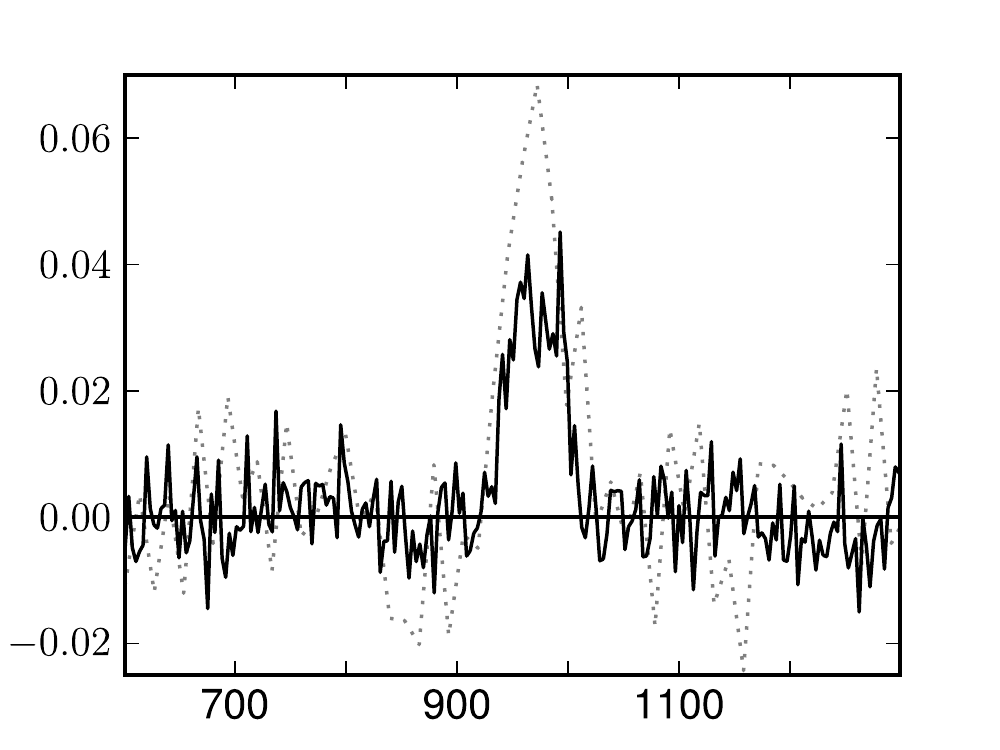}}\\
[12pt]

\multicolumn{2}{c}{HIJASS J1109+50 (UGC06202$^{n}$)}
&&
\multicolumn{2}{c}{HIJASS J1141+32 (MRK0746$^{ecn}$)}\\
\fbox{\includegraphics[height=3cm]{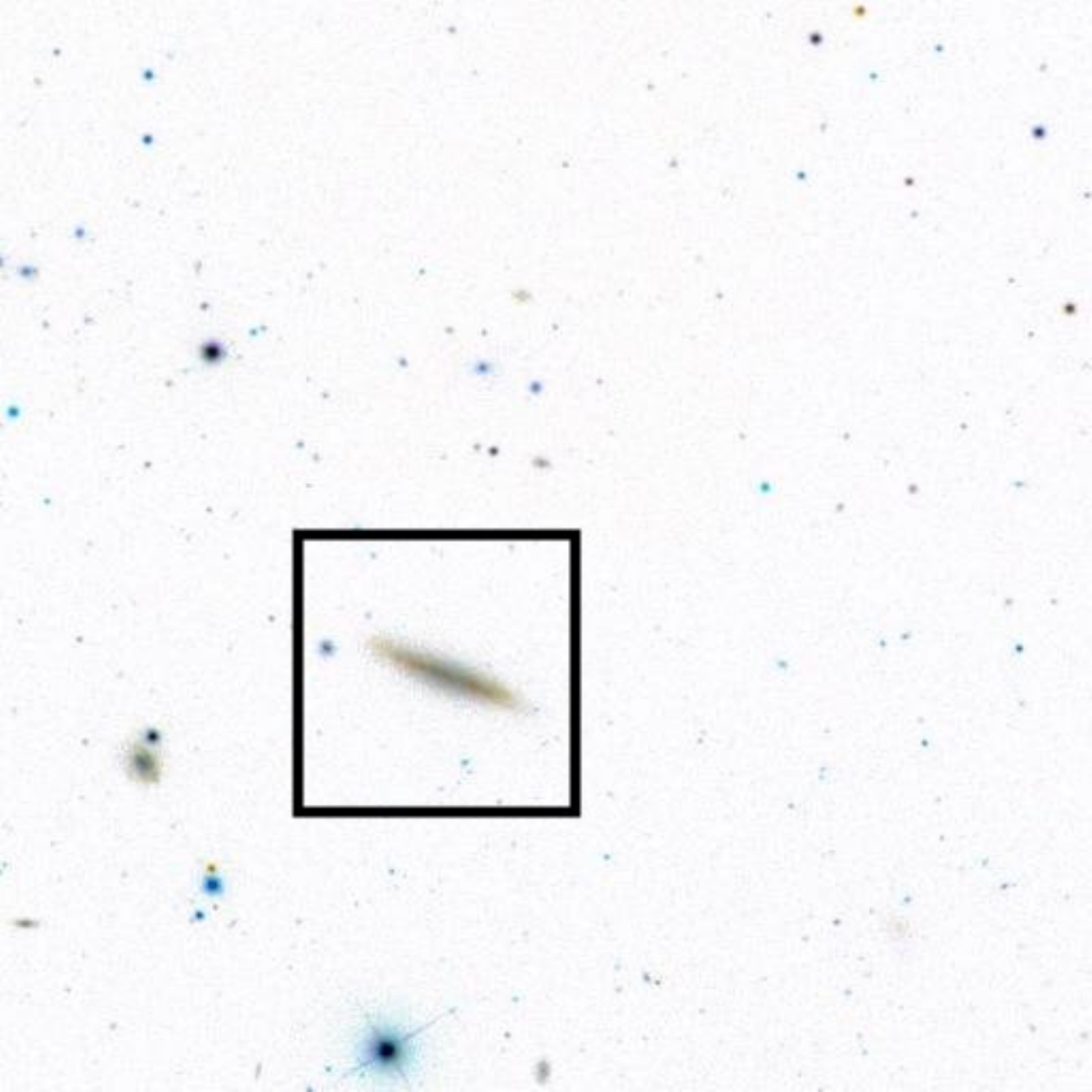}} & 
\mbox{\includegraphics[height=3.3cm]{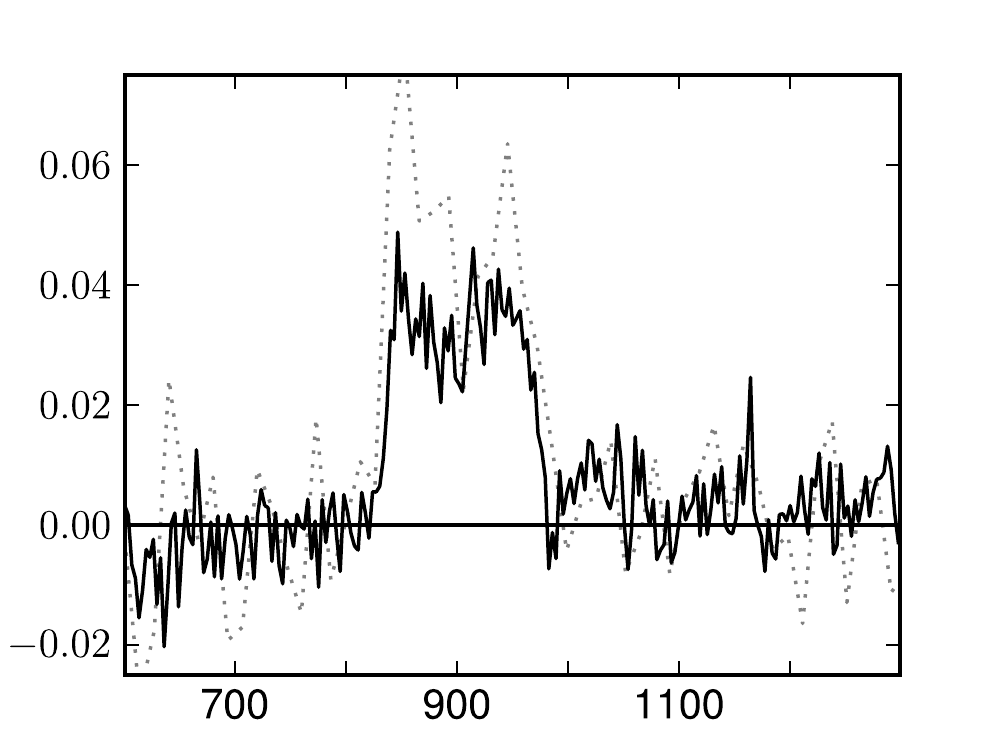}} & 
 & 
\fbox{\includegraphics[height=3cm]{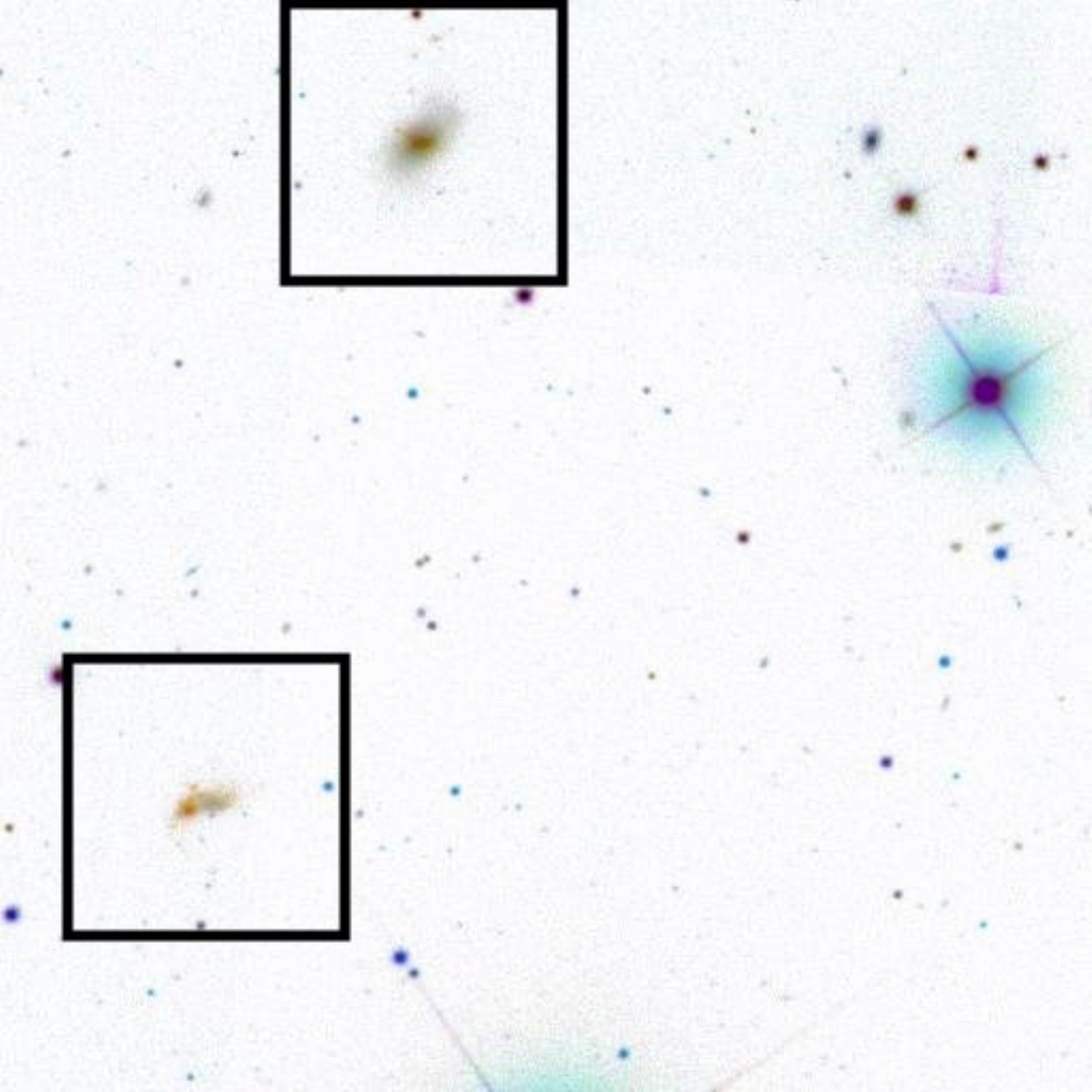}} & 
\mbox{\includegraphics[height=3.3cm]{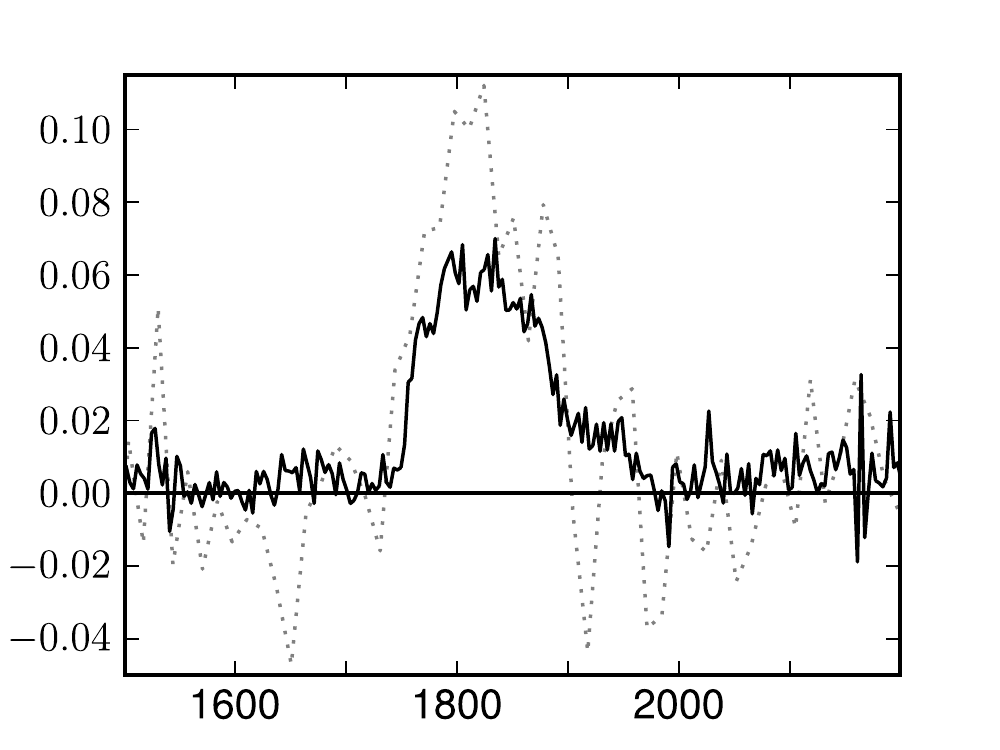}}\\
[12pt]

\end{tabular} 
\caption{\textbf{The first time detections in the 21-cm emission, including the candidate galaxy/HI cloud, HIJASS~J1219.} We show the inverted multi-colour SDSS images ($\sim 7' \times 7'$) together with the HIJASS and GBT spectra ordered with increasing \HI mass (similar to Figure~\ref{fig:lowestHIMass}). (\textit{left}) The SDSS images are centered on the \HI positions. Where available, the optical counterparts that match in position and velocity are marked with boxes. There is no obvious optical counterpart apparent in the SDSS image of HIJASS~J1219+46. (\textit{right}) For comparison, we show the GBT spectrum on top of the HIJASS spectrum (after baseline subtraction). The spectra have the optical velocities ($cz$ in km~s$^{-1}$) along the abscissas and the flux densities (in Jy) along the ordinates. The velocity range of the \HI line emissions as well as the negative troughs in the HIJASS spectra of HIJASS~J1219+46 (220-360~km~s$^{-1}$ dotted/grey) and HIJASS~J1218+46 (430-700~km~s$^{-1}$ dotted/grey) are excluded from the baseline fits. HIJASS~J1141+32 is further discussed in Section~\ref{sec:interacting}. Note that we show the spectra of HIJASS~J1219+46 and HIJASS~J1138+43 3-point Hanning smoothed.}
\label{fig:newlydetected}
\end{figure}
\twocolumn

\subsection{Candidate galaxy/HI cloud}
\label{sec:new}

The presented catalogue of 166 \HI sources in the Ursa Major region contains 162~\HI detections, which are clearly associated with previously catalogued galaxies as listed in the NASA/IPAC Extragalactic Database (NED -- with cross-identfications) or the Sloan Digital Sky Survey (SDSS DR8 -- redshift confidence level $z_{conf}>0.95$). Another three HIJASS detections -- namely HIJASS~J1226+27, HIJASS~J1243+24 and HIJASS~J1233+24 -- can be matched up with \HI sources listed in the ALFALFA 40~per~cent catalogue \citep{haynes2011} and have been assigned to IC3341, KUG1240+247 and MAPS-NGP~O$\_$378$\_$0206985 respectively. Note that HIJASS~J1233+24 was first catalogued in \citet{lang2003} using the same HIJASS data set and is therefore listed in NED (without cross-identifications). Therefore, only one \HI detection -- HIJASS~J1219+46 -- does not have a previously catalogued optical counterpart with known velocity listed in NED, SDSS or ALFALFA (HIJASS~J1219+46 lies outside the ALFALFA declination range).

To identify a potential optical counterpart of the candidate galaxy/\HI cloud we investigate SDSS and Digitized Sky Survey (DSS) images. We compare the images centered on the central position of HIJASS~J1219+46 to the ones of the lowest \HI mass and previously catalogued galaxies in the sample (see Figure~\ref{fig:lowestHIMass}). One might expect that the candidate dwarf galaxy has a similar optical counterpart as the known galaxies within the same mass bin - however, as it has not been previously catalogued one might also expect it to have lower surface brightness and thus be less easily visible. All of the lowest \HI mass galaxies show faint patches of nebulosity in the multi-colour SDSS images (Figure~\ref{fig:lowestHIMass}). Furthermore, we show HIJASS sources that are newly detected in the 21-cm emission line in Figure~\ref{fig:newlydetected}. These SDSS images are centred on the \HI position and therefore show typical offsets from the optical counterparts. For clarification we invert the colours of the multi-colour SDSS images in Figure~\ref{fig:newlydetected} (as opposed to Figure~\ref{fig:lowestHIMass}). We do not find a similar galaxy in the images for HIJASS~J1219+46 within the search radius (the position uncertainty is 1.2~arcmin as given in Table~\ref{tab:A_cat}). There are multiple galaxies listed in SDSS with no redshift measurements, which match in position to the candidate galaxy/\HI cloud. To determine the optical counterpart(s) and/or confirm the \HI cloud accurate position is required from follow-up observations. It is also quite possible that deeper optical imaging is required to determine any optical counterparts.

\begin{figure}
\mbox{\includegraphics[width=8.2cm]{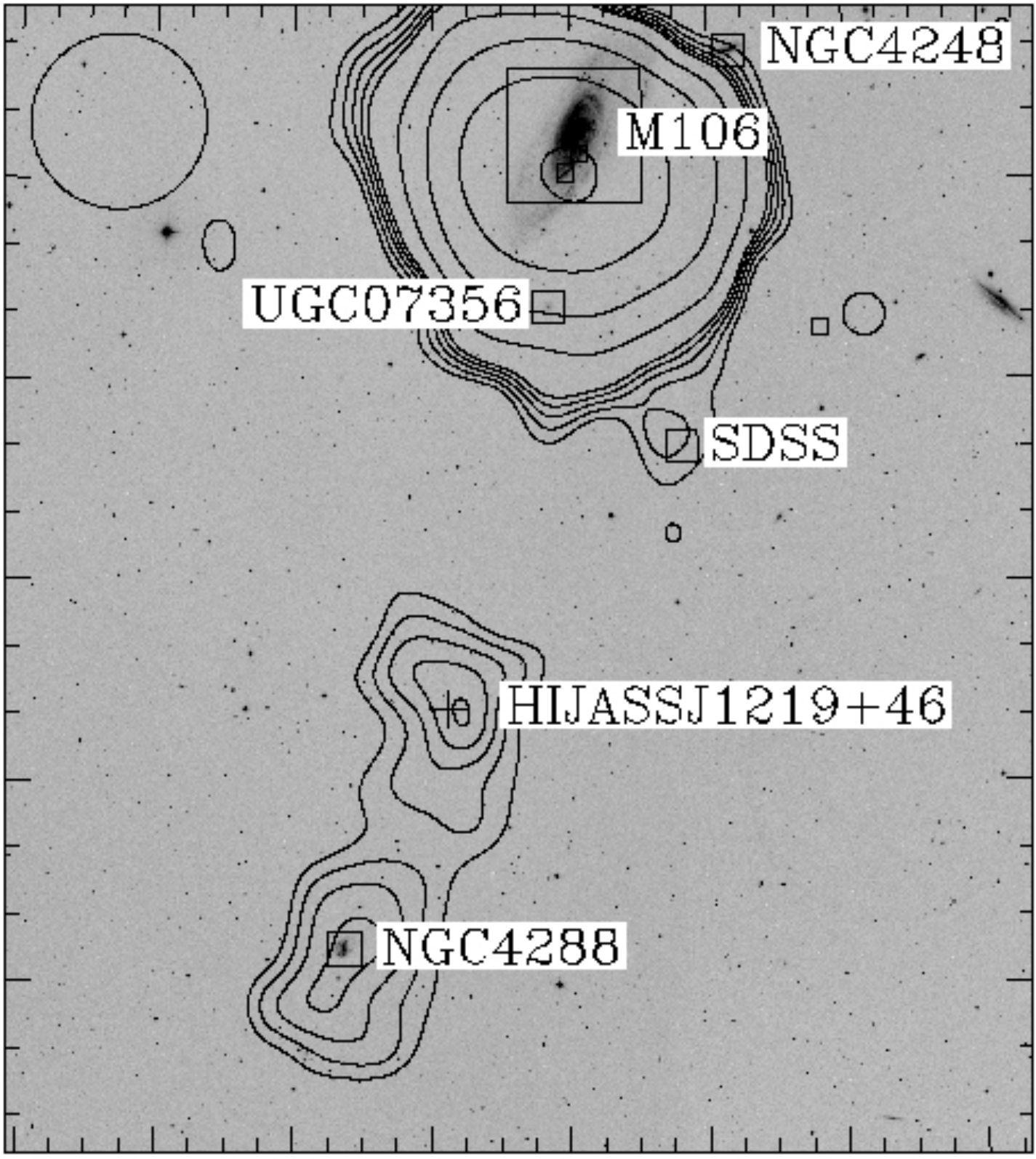}}
\caption{\textbf{Candidate galaxy/\HI cloud found in the HIJASS data -- HIJASS~J1219+46 (indicated by a cross and labelled in the Figure).} We show a DSS blue image (1.4~deg $\times$ 1.4~deg) with overlaid \HI intensity contours at 1, 1.5, 2, 2.5, 3, 6, 12, 24 and 48~Jy~beam$\mathrm{^{-1}}$~km~s$^{-1}$ integrated over the velocity range from 360 to 430~km~s$^{-1}$. The gridded beam is displayed in the top left corner (FWHM 13.1~arcmin). Galaxies residing in this region are marked with boxes and some are labelled in the Figure.}
\label{fig:cloud}
\end{figure}

In Figure~\ref{fig:cloud} we show a 1.4~deg~$\times$~1.4~deg DSS B-band image of the candidate galaxy/\HI cloud and its surroundings with overlaid \HI intensity contours integrated over the velocity range from 360 to 430~km~s$^{-1}$. HIJASS~J1219+46 is a confused source in the {\small DUCHAMP} output with HIJASS~J1220+46/NGC4288. The angular separation to NGC4288 is 0.3~deg, which corresponds to a distance of $\sim$40~kpc assuming that the region lies at a distance of 7.6~Mpc. To the north of HIJASS~J1219+46 lies HIJASS~J1218+46 (SDSSJ121811.04+465501.2) and HIJASS~J1218+47/M106 (including NGC4248, UGC07356 and two SDSS galaxies). The angular separation to M106 is 0.7~deg, which corresponds to $\sim$93~kpc. Although, HIJASS~J1219+46 lies nearby negative bandpass sidelobes emerging from HIJASS~J1218+47/M106 the Green Bank Telescope follow-up observations confirm the \HI properties measured in HIJASS as given in Table~\ref{tab:GBT_cat} (the GBT and HIJASS spectra are shown in Figure~\ref{fig:newlydetected} -- the negative bandpass sidelobes are apparent at 220 to 360~km~s$^{-1}$ in the HIJASS spectrum). The nature of HIJASS~J1219+46 could be tidally stripped gas or a dwarf companion.

\citet{wilcots1996} observed NGC4288 using the Very Large Array (D-configuration; channel width of 5.2~km~s$^{-1}$ and a rms nois of 1.0~mJy~beam$^{-1}$) finding a gas cloud to the south of NGC4288 (also see \citealt{kovac2009}). Furthermore, the gas cloud shows a tidally disrupted \HI stream that extends well beyond the optical disk of NGC4288 \citep{wilcots1996} towards HIJASS~J1219+46. However, HIJASS~J1219+46 lies just outside the area covered by \citet{wilcots1996} and \citet{kovac2009}.

\subsection{Comparison of measured HI parameters with values in the literature and follow-up observations}
\label{sec:literature}

\begin{figure}
\mbox{\includegraphics[width=8.5cm]{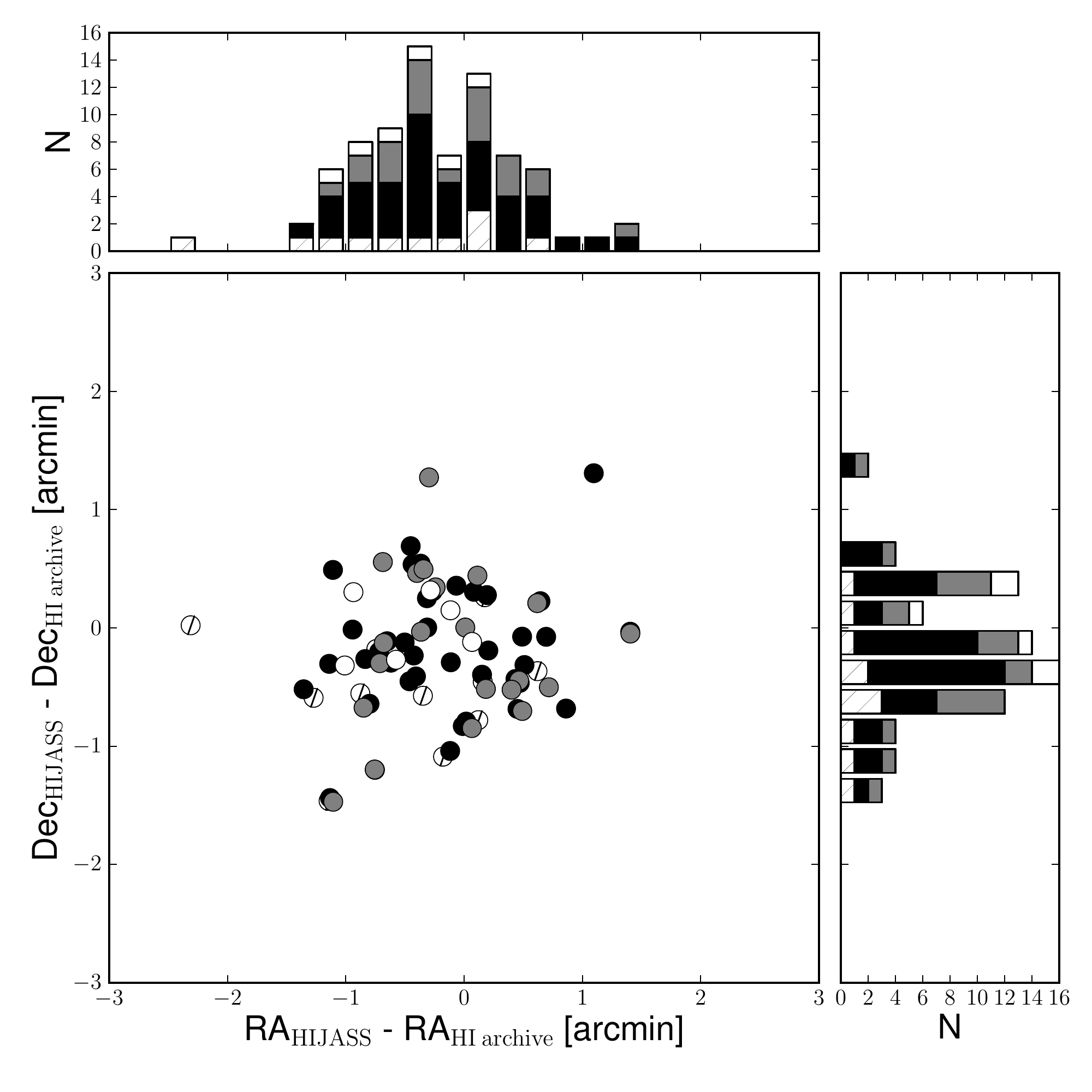}}
\mbox{\includegraphics[width=9cm]{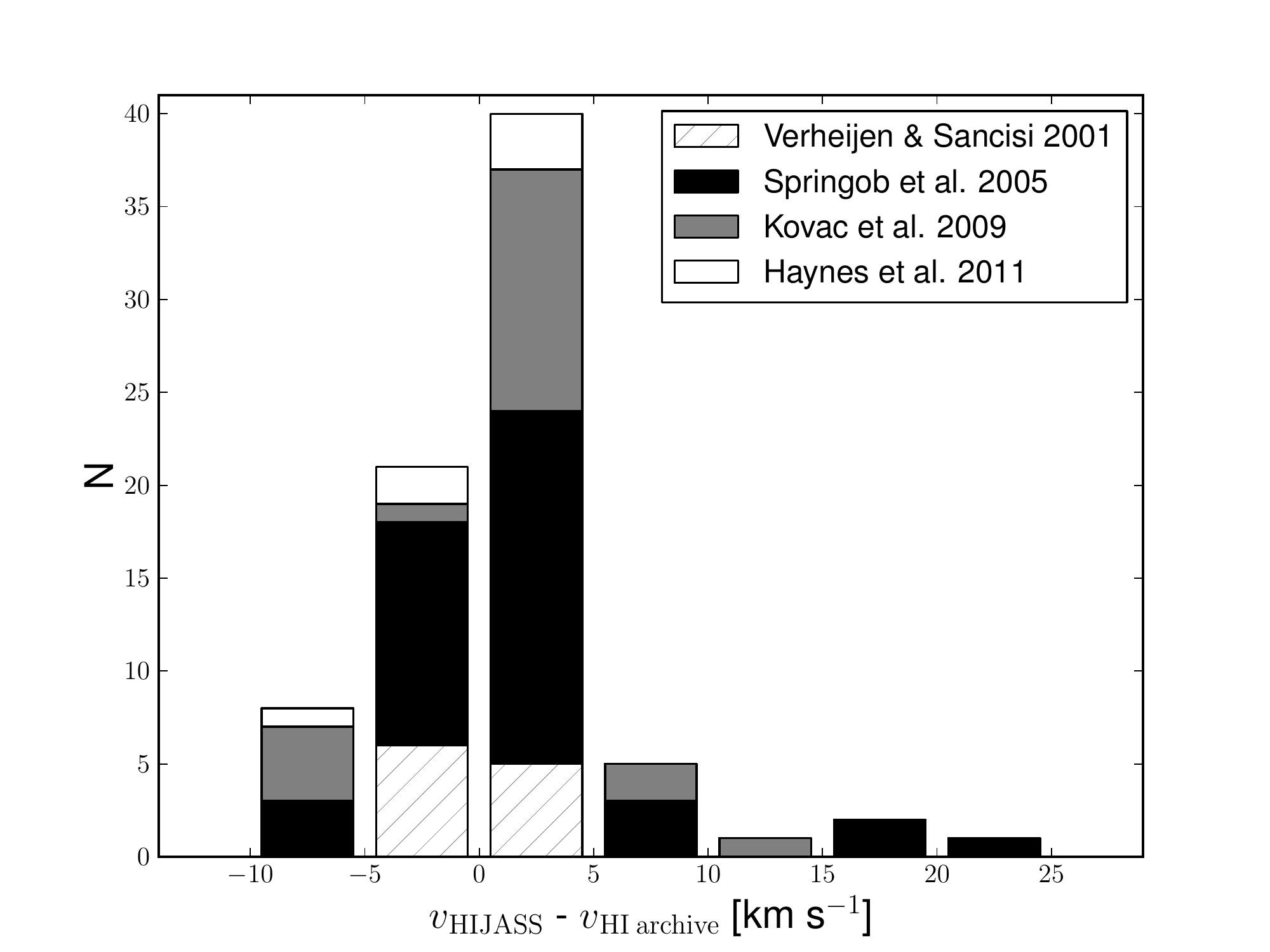}}
\caption{The differences between \HI parameters as listed in this paper and previously obtained \HI parameters (the used literature is noted in the key) for \HI sources with only \textit{one} associate optical counterpart within the gridded beam size and the velocity range of the \HI source. (\textit{Top}) the position offsets; (\textit{bottom}) the systemic velocity offsets -- the comparison to \citet{kovac2009} uses local group velocities as discussed in \citet{yahil1977}.}
\label{fig:offsetsPOS}
\end{figure}

We find 165~\HI sources with one or more associate optical counterparts as listed in NED and SDSS in the HIJASS data. For \HI detections with \textit{one, clear} associate counterpart, i.e. \HI sources that are not confused (c), not double detections (d) and not in the vicinity of negative bandpass (b) sidelobes (as labelled in this catalogue in Section~\ref{sec:catalogue}), we show the position and velocity offsets from our measured \HI parameters to previously obtained \HI parameters in Figure~\ref{fig:offsetsPOS}. We compare our results to recent literature values: (\textit{i}) \citet{verheijen2001} extensively studied 43 spiral galaxies in the Ursa Major cluster using WSRT, (\textit{ii}) \citet{springob2005} have compiled a homogenous sample of \HI properties of more than 9000 galaxies from single dish observations, (\textit{iii}) \citet{kovac2009} have conducted a blind \HI survey in the Canes Venatici region using the WSRT and (\textit{iv}) the ongoing ALFALFA survey using the Arecibo covers a strip in the southern part of the region ($+24^{\circ}\leq\delta\leq+28^{\circ}$) \citep{haynes2011}. The congruent sample of clear \HI sources in this study and the literature includes 78 galaxies. The central positions and velocities presented in this paper are generally consistent with the literature values within the errors (see Figure~\ref{fig:offsetsPOS}).

\begin{figure}
\includegraphics[width=9cm]{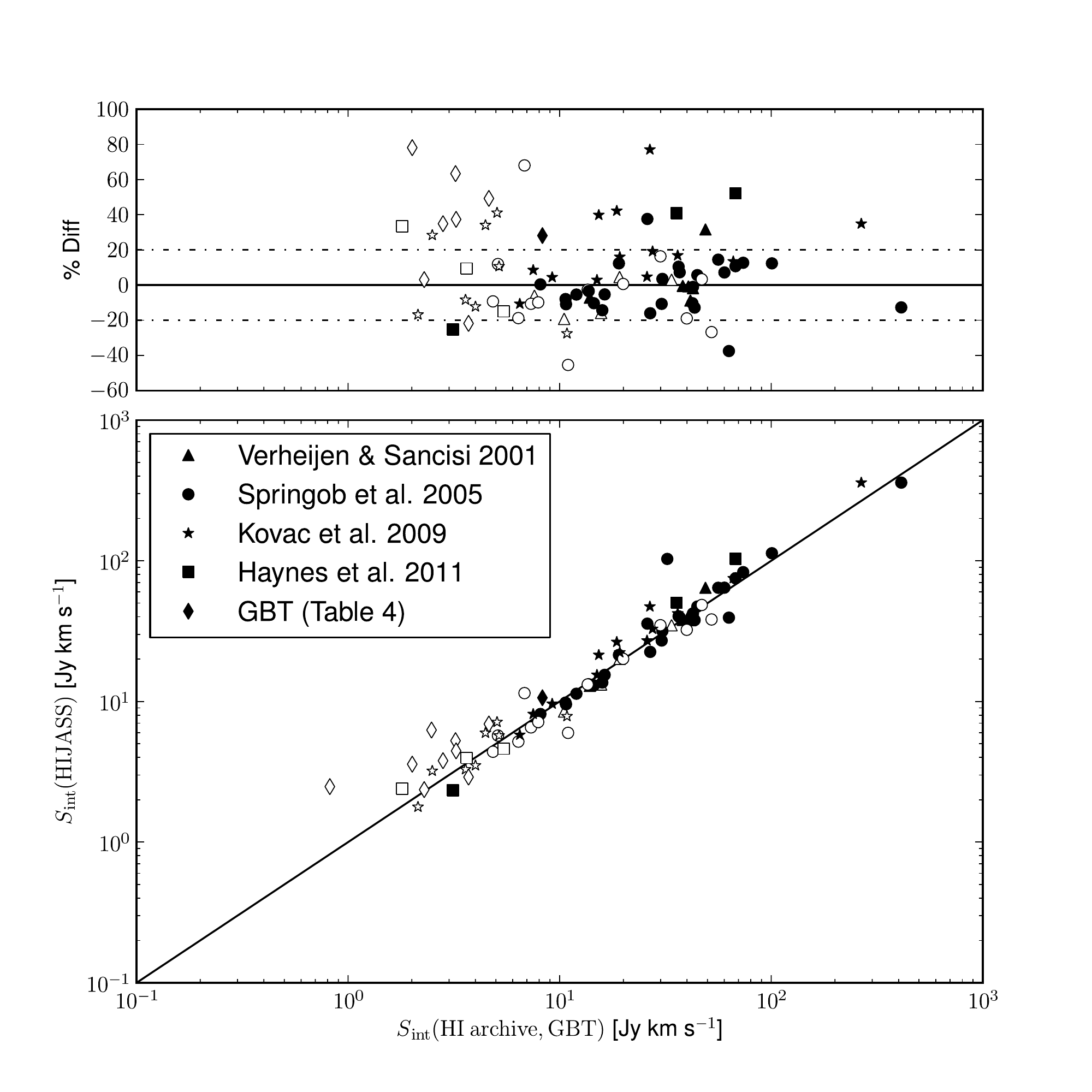}
\caption{\textbf{Comparison of integrated fluxes measured in HIJASS with values in the literature and measured with the GBT (follow-up observations):} (\textit{top}) deviation of values measured in HIJASS compared to the ones in the literature and obtained with the GBT; (\textit{bottom}) the integrated flux values from the literature and the GBT plotted against the ones measured in HIJASS. The used literature is noted in the key. The light coloured markers represent point sources; the dark ones are for extended sources as listed in the catalogue (Table~\ref{tab:B_cat}). The solid lines mark the unity relationship and the dashed lines (\textit{top panel}) limit the 20~per~cent difference in the integrated flux measurements.}
\label{fig:offsetsSINT}
\end{figure}

In Figure~\ref{fig:offsetsSINT} we compare the integrated fluxes measured in HIJASS to recent literature values as well as the GBT follow-up observations listed in Table~\ref{tab:GBT_cat}. We show a comparison of the percentage difference of integrated fluxes measured in HIJASS to the ones in the literature and with the GBT (\textit{top}) as well as the comparison of the absolute flux values for the sample (\textit{bottom}). The percent difference is calculated as follows: $$ \mathrm{\% Diff} = \frac{S_{int} \mathrm{(HIJASS)} - S_{int} \mathrm{(HI \: archive,GBT)}}{S_{int} \mathrm{(HI \: archive,GBT)}} \times 100 $$

Some integrated fluxes differ by more than 20$\%$, in particular in comparison to \citet{kovac2009}. This may be explained as they use an interferometric dataset (obtained with WSRT) which is not sensitive to diffuse \HI emission compared to a single-dish observation.

There are some data points that lie outside the displayed range in Figure~\ref{fig:offsetsSINT} (\textit{top}): HIJASS~J1254+27/NGC4789A with Diff~=~219$\%$ in comparison with \citet{springob2005}. We find a wide range of literature values for this dwarf irregular galaxy. NGC4789A has a large \HI to optical radius ratio \citep{hoffman1996}, therefore some telescopes may resolve out significant flux. There are two outliers in comparison with the GBT follow-up observations -- HIJASS~J1218+46/SDSSJ121811.04+465501.2 (Diff~=~152$\%$) and HIJASS~J1145+50/SDSSJ114506.26+501802.4 (Diff~=~202$\%$). See Figure~\ref{fig:newlydetected} for the comparison of the GBT and HIJASS spectra. HIJASS~J1218+46 is a double detection with HIJASS~J1218+47 including M106, UGC07356, NGC4248 and two SDSS galaxies (also see Figure~\ref{fig:cloud}). Given that the angular separation between SDSSJ121811.04+465501.2 and M106 is 24.5~arcmin and the angular separation between SDSSJ121811.04+465501.2 and UGC07356 is 14.4~arcmin, these galaxies may contribute a significant amount of flux to the HIJASS detection (FWHM 13.1~arcmin) in comparison to the GBT observation (FWHM 9~arcmin). Regarding SDSSJ114506.26+501802.4, the mean signal-to-noise ratio in the HIJASS data is only 1.5 -- therefore the rms noise contributes significantly to the measured integrated \HI flux in comparison to the GBT data. In general, this literature comparison shows that \HI properties of galaxies measured with different telescopes and obtained with different parametrization procedures lead to consistent results within the errors. However, for \HI detections near the detection limit, the noise can cause fluctuations in the \HI measurements.

\begin{table*}
\begin{center}
\caption{The \HI properties of newly detected \HI sources as measured with the GBT.}
\begin{tabular}{lrccccrrr}

  \hline\hline\\[-2ex]
  &
  \multicolumn{1}{c}{$v\mathrm{_{sys}} \pm \sigma_{\mathrm{v}}$} & 
  \multicolumn{1}{c}{$S\mathrm{_p}$} & 
  \multicolumn{1}{c}{$\sigma_{\mathrm{peak}}$} &
  \multicolumn{1}{c}{$S\mathrm{_{int}}$} & 
  \multicolumn{1}{c}{$\sigma_{\mathrm{int}}$} & 
  \multicolumn{1}{c}{$w_{50} \pm \sigma_{\mathrm{50}}$} & 
  \multicolumn{1}{c}{$w_{20} \pm \sigma_{\mathrm{20}}$} &
  \multicolumn{1}{c}{rms} \\
    
  \multicolumn{1}{c}{HIJASS name} &
  \multicolumn{1}{c}{km s$^{-1}$} &
  \multicolumn{1}{c}{Jy} & 
  \multicolumn{1}{c}{Jy} &
  \multicolumn{1}{c}{Jy km s$^{-1}$} & 
  \multicolumn{1}{c}{Jy km s$^{-1}$} & 
  \multicolumn{1}{c}{km s$^{-1}$} & 
  \multicolumn{1}{c}{km s$^{-1}$} & 
  \multicolumn{1}{c}{Jy} \\ 

  \multicolumn{1}{c}{(A1)} & 
  \multicolumn{1}{c}{(A5)} &    
  \multicolumn{1}{c}{(B2)} & 
  \multicolumn{1}{c}{(B4)} & 
  \multicolumn{1}{c}{(B5)} & 
  \multicolumn{1}{c}{(B7)}  &
  \multicolumn{1}{c}{(B8)} &  
  \multicolumn{1}{c}{(B9)} &  
  \multicolumn{1}{c}{(B10)} \\[0.5ex] \hline
  \\[-1.8ex]
   
HIJASS J1109+50  &  905 $\pm$ 2 & 0.049 & 0.006 & 4.63 & 0.36 & 132 $\pm$ 4 & 146 $\pm$ 6 & 0.005  \\
HIJASS J1138+43  &  1190 $\pm$ 2 & 0.030 & 0.004 & 3.22 & 0.28 & 155 $\pm$ 4 & 171 $\pm$ 6 & 0.004  \\
HIJASS J1141+46  &  755 $\pm$ 3 & 0.045 & 0.006 & 3.24 & 0.33 & 75 $\pm$ 6 & 105 $\pm$ 9 & 0.005  \\
HIJASS J1141+32  &  1852 $\pm$ 4 & 0.070 & 0.008 & 8.29 & 0.55 & 123 $\pm$ 8 & 198 $\pm$ 12 & 0.007  \\
HIJASS J1145+50  &  1663 $\pm$ 8 & 0.015 & 0.005 & 0.82 & 0.27 & 68 $\pm$ 16 & 105 $\pm$ 24 & 0.005  \\
HIJASS J1154+30  &  972 $\pm$ 2 & 0.045 & 0.007 & 2.01 & 0.31 & 60 $\pm$ 4 & 72 $\pm$ 6 & 0.006  \\
HIJASS J1209+43B  &  1061 $\pm$ 3 & 0.049 & 0.007 & 2.81 & 0.38 & 69 $\pm$ 6 & 90 $\pm$ 9 & 0.007  \\
HIJASS J1218+46  &  398 $\pm$ 2 & 0.047 & 0.006 & 2.48 & 0.30 & 62 $\pm$ 4 & 75 $\pm$ 6 & 0.006  \\
HIJASS J1219+46  & 390 $\pm$ 1 & 0.061 & 0.004 & 2.29 & 0.14 & 36 $\pm$ 2 & 56 $\pm$ 3 & 0.003 \\
HIJASS J1222+39  &  1071 $\pm$ 2 & 0.069 & 0.006 & 3.71 & 0.28 & 59 $\pm$ 4 & 81 $\pm$ 6 & 0.005  \\

\hline
\end{tabular}
\label{tab:GBT_cat}
\end{center}
\end{table*}

\section{Discussion and results}
\label{sec:res}

\subsection{Velocity Distribution}
\label{sec:velocity}

\begin{figure}
\includegraphics[width=8.5cm]{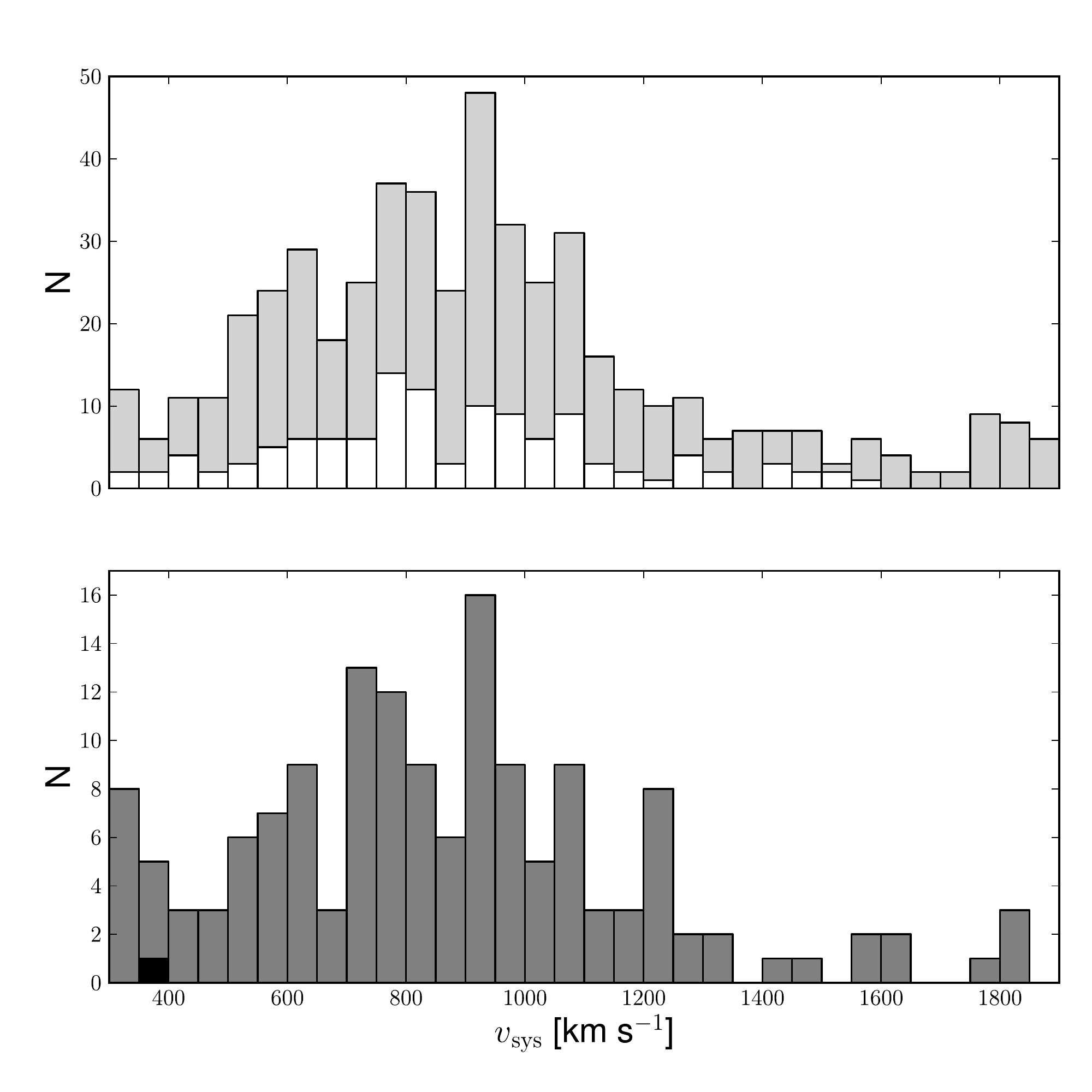}
\caption{Velocity distribution of galaxies residing in the Ursa Major region: (\textit{top}) galaxies as listed in SDSS (white) and additional galaxies listed in NED (grey); (\textit{bottom}) \HI selected galaxies as catalogued in this paper with an associate optical counterpart coloured in grey whereas the nearby candidate galaxy/\HI cloud is shown in black.}
\label{fig:hist_vel}
\end{figure}

The velocity distribution of galaxies residing in the Ursa Major region is shown in Figure~\ref{fig:hist_vel}. The top panel shows the velocity distribution of galaxies as listed in SDSS (white) with spectroscopic redshifts with confidence levels $z_{conf}>0.95$ and additional galaxies listed in NED (grey) with cross-identifications. To avoid duplicates we cross-correlate the galaxies listed in NED and SDSS within 1~arcmin and 50~km~s$^{-1}$ of the central position and velocity. The bottom panel shows the velocity distribution of the \HI selected galaxies as catalogued in this paper with systemic velocities between 300 and 1900~km~s$^{-1}$. The \HI detections are colour-coded according to \HI sources with associate optical counterparts (in grey) and the previously uncatalogued candidate galaxy/\HI cloud (in black). 

One can see the large scale structure of the region - the overdensity appearing at $\sim$950~km~s$^{-1}$ in the optical and (less clearly) in the \HI selected sample indicates the presence of the Ursa Major cluster. 
There are 212~galaxies listed in SDSS and NED within the Ursa Major cluster definition by \citet{tully1996}. In this paper we present 51~\HI detections within the cluster definition by \citet{tully1996} that contain 96~previously catalogued galaxies. The high number of confused detections in our data is due to a high density of galaxies in the cluster. We will discuss cluster/group membership, substructures and HIJASS non-detections in a forthcoming paper but the number of high-probability cluster members is likely to increase significantly from the 79 cluster members identified in \citet{tully1996}.

There are many groups nearby the cluster, which overlap and make the cluster difficult to define. In paricular foreground galaxy groups with velocities $v < 700$~km~s$^{-1}$ lead to an asymmetric galaxy distribution in the optical and \HI selected sample. The number of galaxies with velocities $v > 1300$~km~s$^{-1}$ is lower than the number of galaxies residing in the foreground. This may be a selction effect or could indicate the filamentary structure: the Ursa Major cluster (at 17.1~Mpc) is connected with the Virgo cluster (at $\sim$16~Mpc \citealt{mei2007}) via the galaxies and galaxy groups lying at the low velocity end.

Figure~\ref{fig:hist_vel} also shows one nearby candidate galaxy/\HI cloud. We find no candidate galaxies/\HI clouds in the Ursa Major cluster, where one might expect to find some e.g. due to previously uncatalogued gas-rich dwarf irregular galaxies or the presence of stripped \HI gas. Given that our \HI mass limit at the distance of the Ursa Major cluster is 2.5$\times 10^8$~M$_{\odot}$ (assuming a velocity width of 100~km~s$^{-1}$) we cannot rule out candidate galaxies/\HI clouds below the \HI mass limit.

\subsection{HI Mass Distribution}

\begin{figure}
\includegraphics[width=9cm]{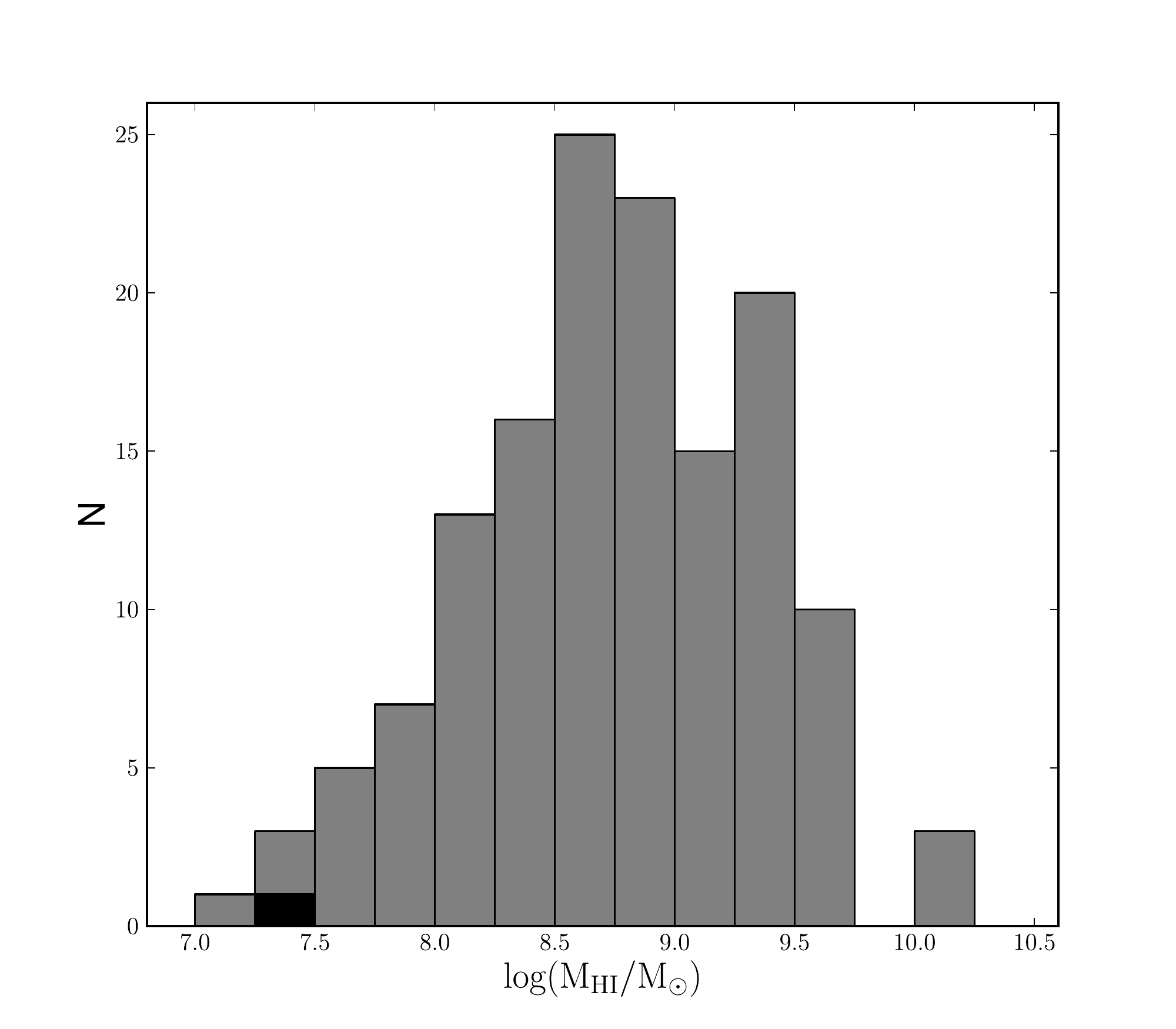}
\caption{The \HI mass histogram of the catalogued sources in this paper with systemic velocities between 300 and 1900~km~s$^{-1}$. Grey shadings correspond to \HI sources with an associate optical counterpart whereas the candidate galaxy/\HI cloud is shown in black.}
\label{fig:hist_mass}
\end{figure}

Figure~\ref{fig:hist_mass} shows the \HI mass distribution of 151~\HI sources as listed in this catalogue (excluded are 15 galaxies with central velocities below 300~km~s$^{-1}$). \HI sources that have been previously detected and are listed in NED/SDSS are shown in grey whereas the candidate galaxy/\HI cloud is presented in black. The latter inhabits the low \HI mass end and is likely to be a dwarf galaxy, i.e. of the same galaxy population as the previously catalogued galaxies within the same mass bin. \citet{taylor2005} note that `dark' (optically invisible) \HI galaxies are unlikely to be detected in our studied mass range between $10^{7}$ and $10^{10.5}$~M$_{\odot}$ as the \HI gas is normally associated with star formation or is destroyed. \citet{meurer2006} also find that dormant (non-star-forming) galaxies with \HI masses above $3 \times 10^7$~M$_{\odot}$ are very rare. However, high resolution follow-up observations are required to determine the nature of HIJASS~J1219+46. We discussed the candidate galaxy/\HI cloud in more detail in Section~\ref{sec:new}. We note that we did not find a numerous low-mass population of gas-rich dwarf galaxies to solve the dwarf deficiency problem as discussed in \citet{trentham2001}. 

We can place an upper limit on the \HI mass for the galaxies listed in NED/SDSS within the Ursa Major region, which we did not detect in HIJASS. The survey sensitivity is $2.5\times10^8$ M$_{\odot}$ (assuming a velocity width of 100~km~s$^{-1}$ and a distance of 17.1~Mpc). A lower \HI mass limit is valid for galaxies lying in the foreground groups and filamentary structure (with $v < 700$~km~s$^{-1}$ and distances $D < 17.1$~Mpc), wheras for background galaxies ($v > 1200$~km~s$^{-1}$) with distances $D > 17.1$~Mpc a slightly higher \HI mass limit is appropriate (as $M\mathrm{_{HI}} = 8.38 \times 10^5 \times D^2$).

\subsection{Candidates for galaxy-galaxy interactions}
\label{sec:interacting}

In this paper, there are 54~\HI detections (33~per~cent) which have two or more optical counterparts as listed in NED/SDSS within the gridded beam size and within the velocity range of the \HI detection (see \HI detections labelled with `c' in the catalogue in Section~\ref{sec:catalogue}). We can not clearly associate the \HI content with only one optical counterpart. However, these \HI detections with multiple optical counterparts are intriguing objects as galaxy-galaxy interactions might take place due to the small separation between the galaxies (e.g. 10~arcmin corresponds to 50~kpc at a distance of 17.1~Mpc). 

\begin{figure*}
\begin{tabular}{p{8cm} p{8cm}}
\includegraphics[width=8cm]{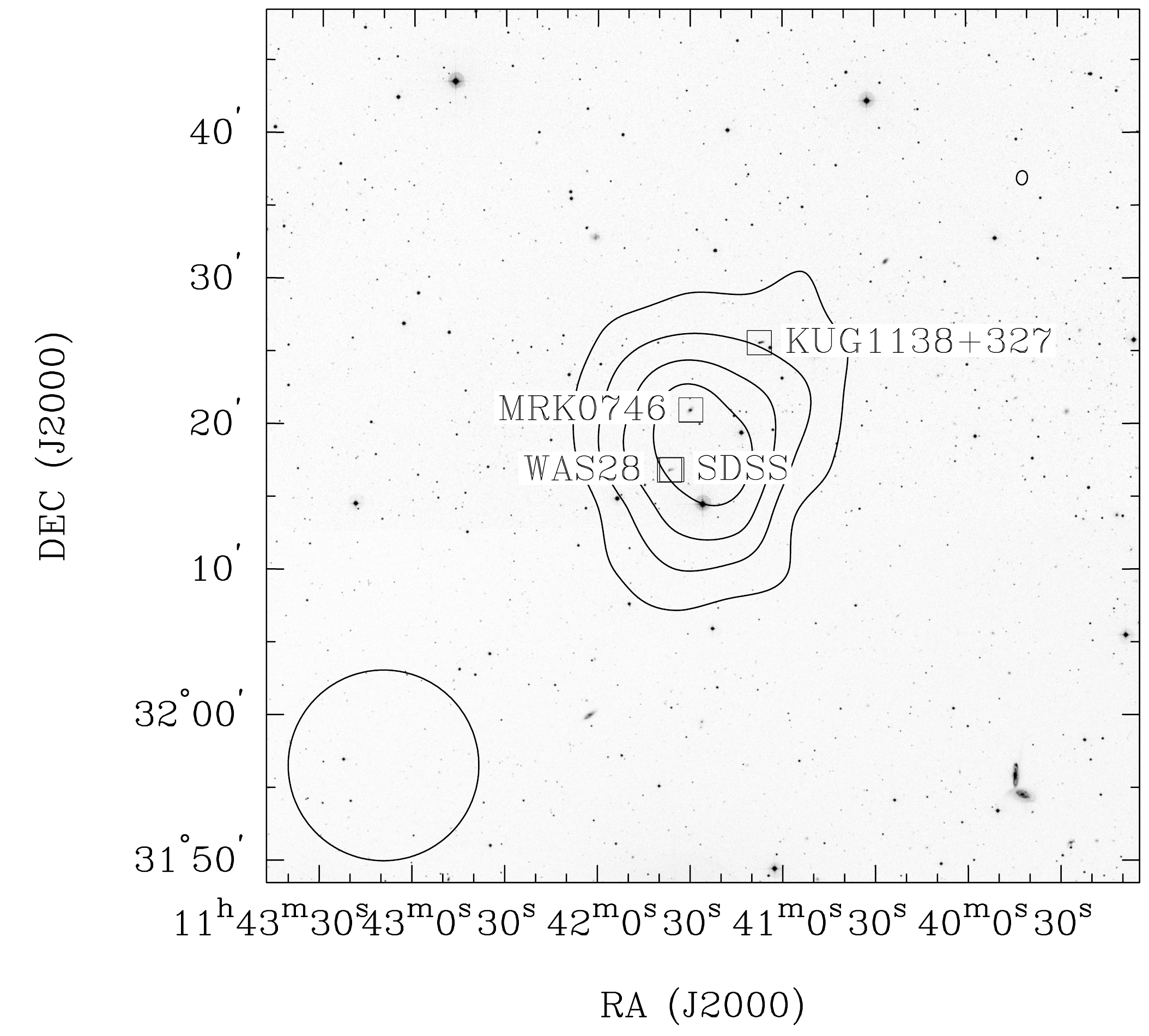} & \includegraphics[width=8cm]{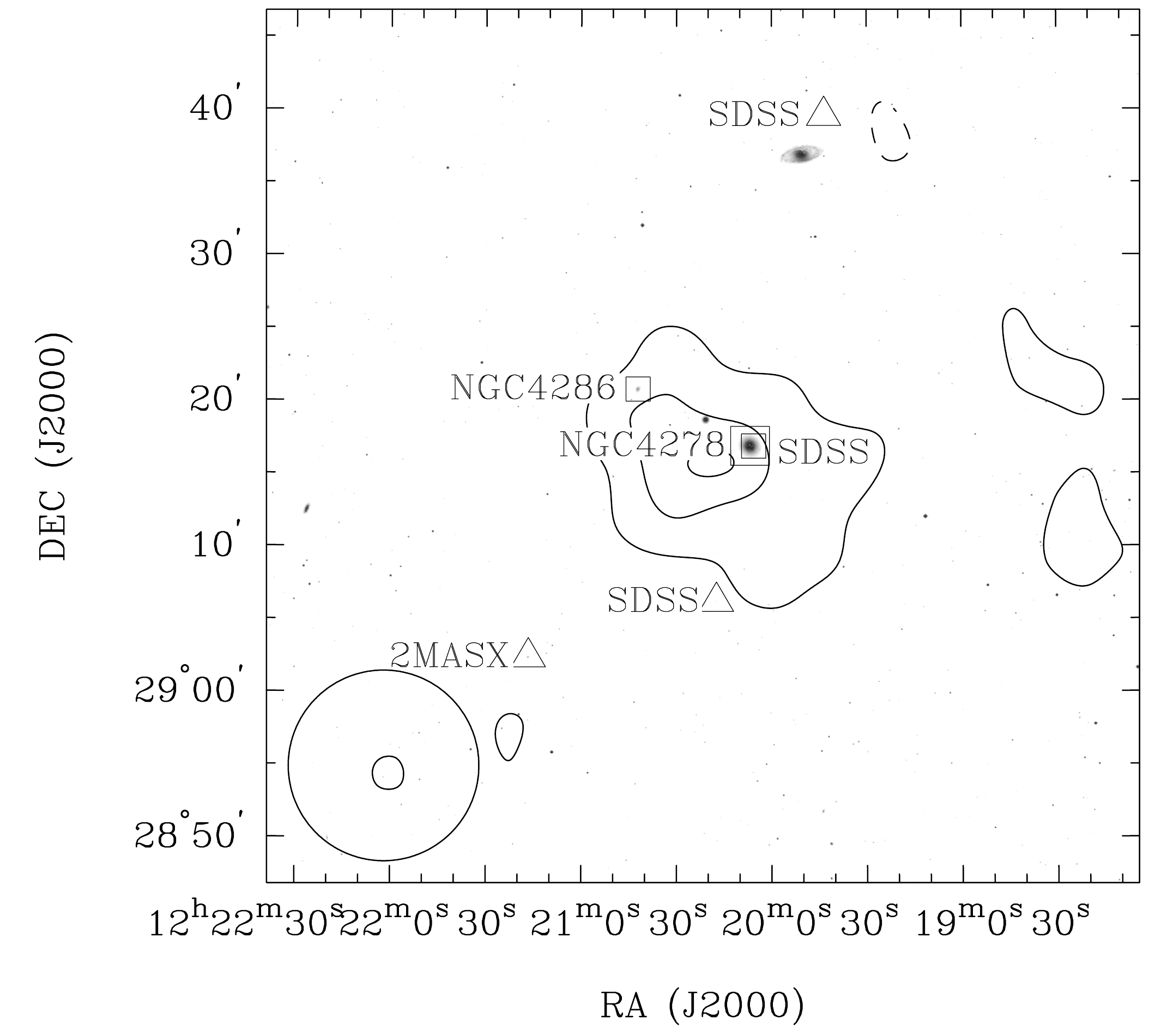}
\end{tabular}
\caption{Two candidate regions for galaxy-galaxy interactions are shown with DSS2 B-band images ($1^{\circ} \times 1^{\circ}$) and overlaid integrated flux levels (contours) starting at $\pm$1.5~Jy~beam$\mathrm{^{-1}}$~km~s$^{-1}$. Negative contour levels are indicated with dashed lines. The gridded beam size is indicated in the bottom left corner of each image. (\textit{left}) HIJASS~J1141+32: There are four dwarf galaxies at approximately $1700-1900$~km~s$^{-1}$ as indicated by boxes and labelled in the Figure. The increment of the contour levels is 2~Jy~beam$\mathrm{^{-1}}$~km~s$^{-1}$. (\textit{right}) HIJASS~J1220+29: The NGC4278 group at approximately $400-850$~km~s$^{-1}$. The contour lines include three previously catalogued galaxies as listed in NED (marked with boxes and labelled in the Figure) and increase with 1~Jy~beam$\mathrm{^{-1}}$~km~s$^{-1}$. We also show three galaxies (at similar velocities than the NGC4278 group) that are not detected in HIJASS (triangles).}
\label{fig:inter_one}
\end{figure*}

For example, \textbf{HIJASS~J1141+32} encompasses four previously known dwarf galaxies within 1700 and 1900~km~s$^{-1}$ (see Figure~\ref{fig:inter_one}): MRK0746, KUG1138+327, WAS28 and SDSSJ114135.98+321654.0 (which has no cross-identification in NED and is therefore not listed in our catalogue) -- there are no previous detections in the 21-cm emission line for any of these galaxies. High resolution follow up observations are required to study the \HI distribution, kinematics and possible interaction of these galaxies.

\textbf{HIJASS~J1220+29} encompasses three previously catalogued galaxies (see Figure~\ref{fig:inter_one}) at $610-710$~km~s$^{-1}$, namely NGC4278 (elliptical), SDSSJ122005.73+291650.0 (spiral) and NGC4286 (spiral) (ordered with increasing angular separation from the central \HI position). \citet{morganti2006} have conducted a WSRT observation of NGC4278, NGC4286 and SDSSJ122005.73+291650.0 -- the observations reveal two tail-like structures west and south-west of NGC4278.

Out of the 54 confused HIJASS detections, 33 galaxies appear to have dwarf companions (one or two small objects as listed in SDSS, 2MASX, SBS, MCG, HS, CGCG). This leaves us with 21 high-probability candidates for galaxy-galaxy interactions of which 11 HIJASS detections (52~per~cent) are within the Ursa Major cluster as defined by \citet{tully1996}. It is remarkable that the Ursa Major cluster as defined by \citet{tully1996} only comprises approximately 12~per~cent of the surveyed volume discussed in this paper but contains about half of all high-probablility candidates for galaxy-galaxy interactions. This is also the case (48~per~cent) if we include two nearby {\small DUCHAMP} double detections, i.e. we do not only consider galaxies within the gridded beam (as the physical size of the beam is smaller for nearby regions) but also merged \HI detections as candidates for high-probability galaxy-galaxy interactions. \citet{verheijen2001} have observed 10 of the high-probability candidates for galaxy-galaxy interactions in the Ursa Major cluster. The sample includes the most massive galaxy (Messier109), the largest (NGC3718) and some of the brightest systems (NGC4026, NGC4111 and NGC4088).

According to \citet{verheijen2001} the interacting systems are: (\textit{i}) \textbf{NGC3769} is strongly interacting with its dwarf companion NGC3769A. (\textit{ii}) \textbf{NGC3893} is an interacting pair with NGC3896. (\textit{iii}) Some tidal debris can be seen in the interacting pair \textbf{IC0750} and IC0749. (\textit{iv}) The largest galaxy in the cluster -- \textbf{NGC3718} -- shows distorted morphology and may be interacting with NGC3729. (\textit{v}) \textbf{NGC4088} is one of the brightest systems in the Ursa Major cluster with ongoing vigorous star formation. The system is strongly disturbed and in close vicinity of NGC4085. (\textit{vi}) The \textbf{NGC4111} and (\textit{vii}) \textbf{NGC4026} groups are also known to be interacting and discussed in more detail in Section~\ref{sec:extendedHI}.

Three of the high-probability candidates -- namely \textbf{NGC3917}, \textbf{Messier109} and \textbf{NGC4157} -- do not show any signs of interactions with their dwarf companions \citep{verheijen2001}. 

\citet{verheijen2001} find further interacting galaxy pairs: NGC3949 may be interacting with a small companion (SDSSJ115340.69+475156.4) and UGC06818 may be tidally interacting with a faint companion -- these systems are not included in our sample of high-probability candidates for galaxy-galaxy interactions as they include interactions with dwarf companions. Note that we quote UGC06818 as a clear single detection in our catalogue -- there are no galaxies listed in close vicinity of UGC06818 in NED/SDSS with know redshifts. \citet{verheijen2001} also find NGC3998 with disturbed \HI content, which lies just outside the HIJASS data cubes at Dec. $\delta \simeq 55^{\circ} 27' 13''$.

\subsection{HIJASS detections with extended HI envelopes}
\label{sec:extendedHI}

We find two \HI detections with extended \HI envelopes that do not have previously catalogued galaxies within the \HI extension. 

\begin{figure*}
\begin{tabular}{p{8cm} p{8cm}}
\includegraphics[width=8cm]{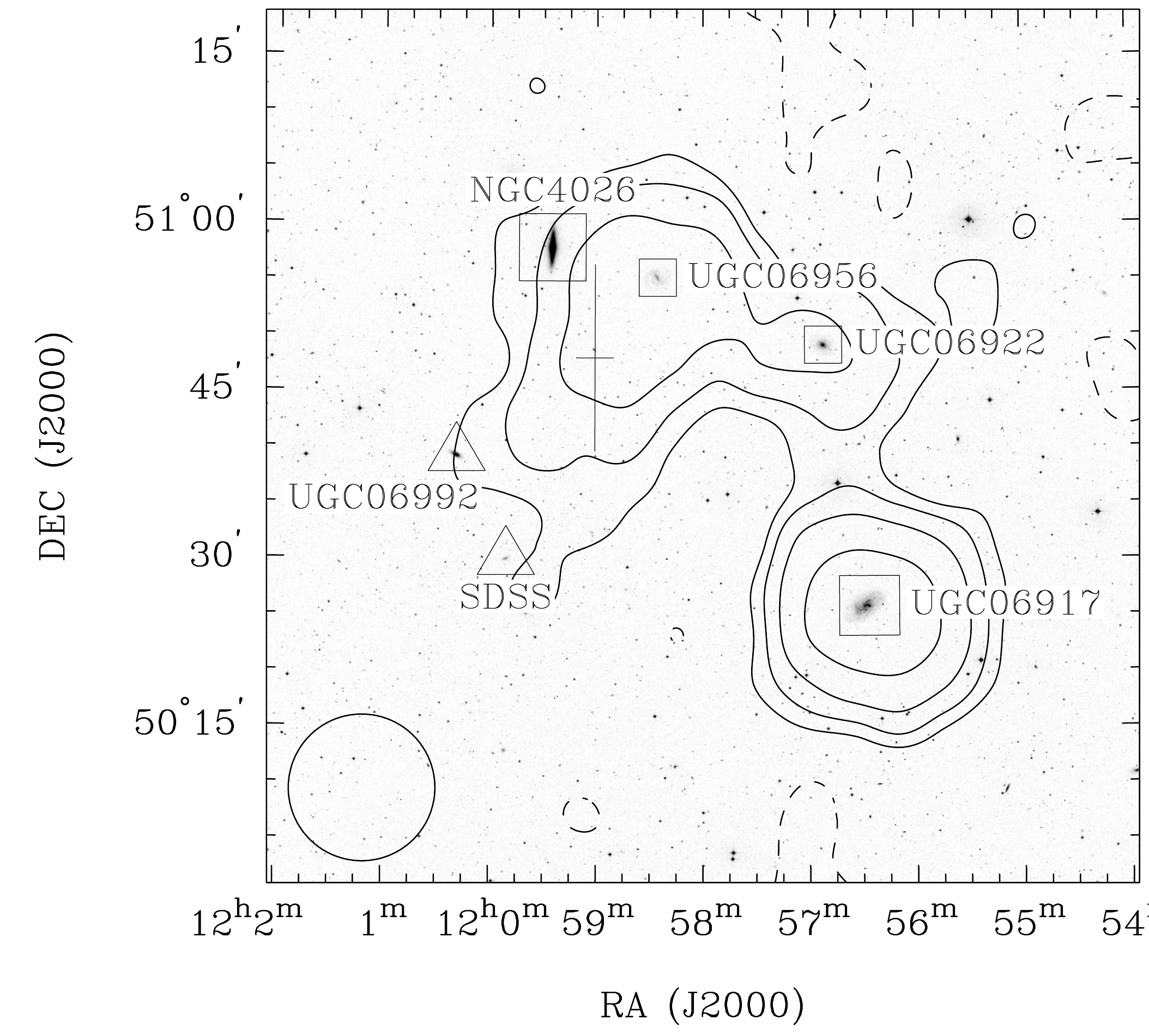} &\includegraphics[width=8cm]{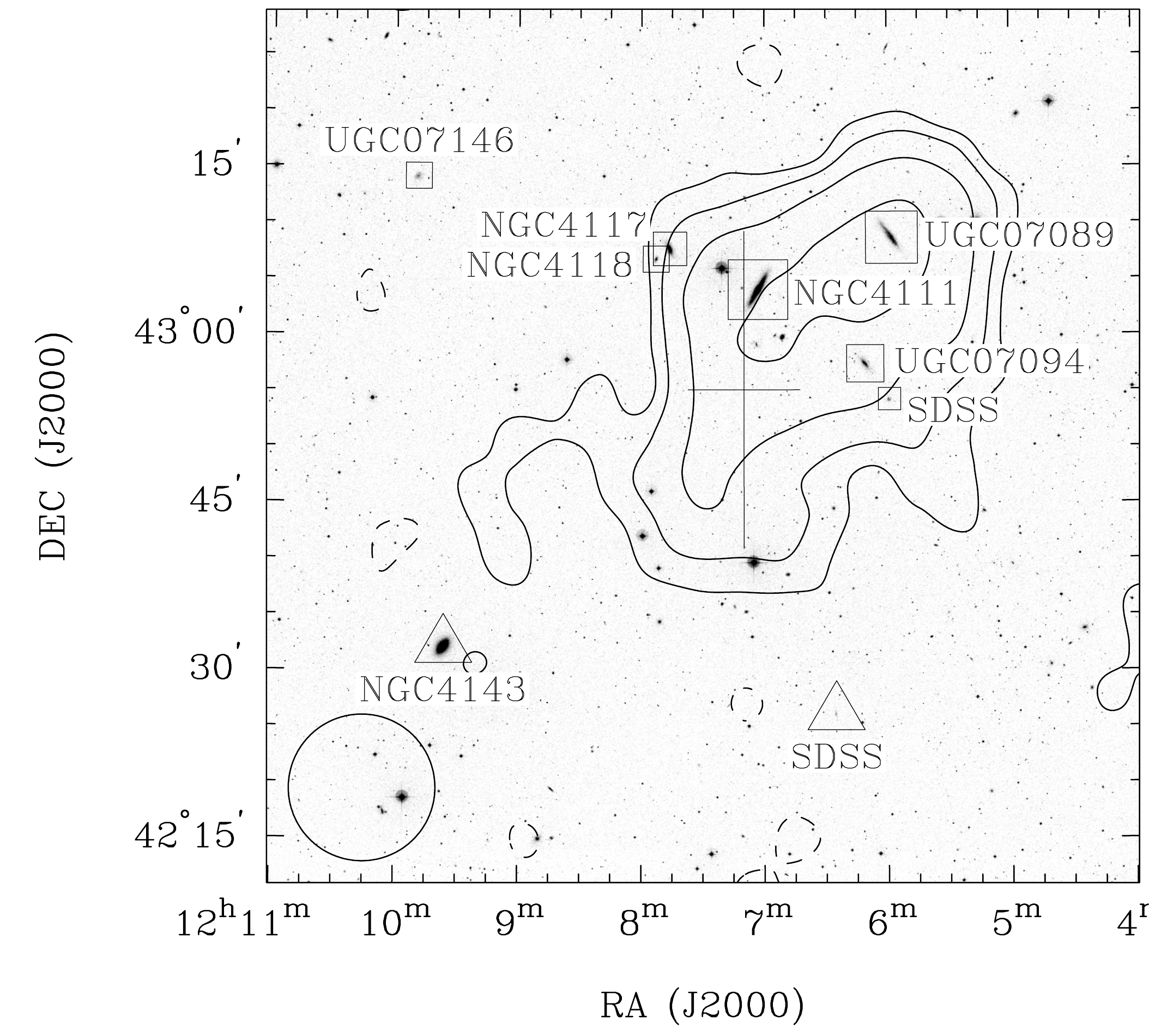}\\
\end{tabular}
\caption{\textbf{Two Ursa Major subregions with extended \HI content.} We show DSS2 B-band images ($1.3^{\circ} \times 1.3^{\circ}$) with overlaid \HI intensity contours at $\pm$1.5, 3, 6 and 12~Jy~beam$\mathrm{^{-1}}$~km~s$^{-1}$. Negative contour levels are indicated with dashed lines. The beam (13.1~arcmin) is indicated in the bottom left corner of each image. (\textit{left:}) The NGC4026 group at approximately $800-1050$~km~s$^{-1}$: The contour lines include 4 catalogued galaxies as indicated by boxes and labelled in the Figure. The galaxies marked with triangles, UGC06992 and SDSSJ115950.83+502955.0, are below the \HI sensitivity of this catalogue. The \HI filament is indicated by a cross - the extent of the \HI filament is similar to the cross size. (\textit{right:}) The NGC4111 group at approximately $650-950$~km~s$^{-1}$. The contour lines include six previously catalogued galaxies as listed in NED (marked with boxes and labelled in the Figure). The disturbed \HI content of NGC4111 and its extent is indicated by a cross. We also show two galaxies (at similar velocities than the NGC4111 group) that are not detected in HIJASS (triangles). We note that UGC07146 (in the top left corner) is detected for the first time in the 21-cm emission line (with a systemic velocity of 1057~km~s$^{-1}$ and therefore outside the displayed velocity range).}
\label{fig:inter_two}
\end{figure*}

Within the Ursa Major cluster lies the \HI extension/plume in the NGC4026 group (see Figure~\ref{fig:inter_two}), which has already been studied by Appleton (1983; Jodrell Bank), van Driel et al. (1988; WSRT) and Verheijen \& Zwaan (2001; VLA D-array). Six galaxies are listed in NED with heliocentric velocities $700 \leq v \leq 1100$~km~s$^{-1}$, four of which are included in the \HI envelope (boxes) and two of which are in vicinity of the \HI extensions (triangles; non-detection in HIJASS). The WSRT observations reveal an \HI filament at $\alpha \simeq 11^h 58^m 57^s$ and $\delta \simeq 50^{\circ} 46' 42''$ with a systemic velocity of $\sim 960$~km~s$^{-1}$ \citep{driel1988}. The \HI filament measures approximately 38~kpc in size (which corresponds to $\sim$7.5~arcmin at the distance of 17.1~Mpc) and contains $2.1 \times 10^{8}$~M$_{\odot}$. Furthermore, \citet{driel1988} determine the \HI mass of NGC4026$\leq 0.71 \times 10^{8}$~M$_{\odot}$, UGC06956 $=5.1 \times 10^{8}$~M$_{\odot}$, UGC06922$= 4.9 \times 10^{8}$~M$_{\odot}$ and UGC06917$=1.28 \times 10^{9}$~M$_{\odot}$. We detect NGC4026, UGC06956 and the \HI filament combined in HIJASS~J1158+50 and we measure an \HI mass of $7.94 \times 10^{8}$~M$_{\odot}$ -- this matches the total \HI mass measured by \citet{driel1988} within 1~per~cent. Regarding UGC06922 and UGC06917, we measure an \HI mass of $6.17 \times 10^{8}$~M$_{\odot}$ and $1.58 \times 10^{9}$~M$_{\odot}$, respectively. One may expect to measure a larger \HI flux using a single dish telescope compared to \citet{driel1988} using WSRT. However, the \HI content of UGC06922 is likely confused as UGC06956 lies in close vicinity. 

High resolution observations of this system conducted with the VLA D-array \citep{verheijenzwaan2001} reveal an \HI filament of $\sim$97~kpc in size (which corresponds to $\sim$18~arcmin and indicated by a cross in Figure~\ref{fig:inter_two}). The origin of the \HI filament is likely to be tidally stripped gas of NGC4026. \citet{verheijenzwaan2001} note that all three bright lenticular galaxies in the Ursa Major cluster including NGC4026 have disturbed H\,{\sevensize I}.

Another intriguing \HI extension in the Ursa Major cluster is seen in the NGC4111 system, which has already been studied by Verheijen \& Zwaan (2001; VLA D-array). There are six previously catalogued galaxies as listed in NED within 20~arcmin angular separation and within 150~km~s$^{-1}$ of the central \HI velocity (see Figure~\ref{fig:inter_two}). NGC4111 is also one of the three brightest lenticular galaxies in the Ursa Major cluster with disturbed \HI \citep{verheijenzwaan2001}. The \HI extends approximately 122~kpc (which corresponds to $\sim$28~arcmin at the distance of 15.0~Mpc; \citealt{tully2009} and indicated by a cross in Figure~\ref{fig:inter_two}) south of the projected central position of NGC4111 \citep{verheijenzwaan2001}. Furthermore, the disturbed \HI connects NGC4111 with its close neighbouring galaxies NGC4117 and NGC4118. 

The third of the brightest lenticular galaxy in the Ursa Major cluster with disturbed \HI content according to \citet{verheijenzwaan2001} -- NGC3998 -- lies just outside the HIJASS data cubes at Dec. $\delta \simeq 55^{\circ} 27' 13''$. 

We do not find any \HI detections with extended \HI envelopes outside the Ursa Major cluster as defined by \citet{tully1996}.

\subsection{Comparison to large area blind HI surveys}
\label{sec:blind}

Blind \HI surveys have the potential for detecting previously uncatalogued galaxies and \HI clouds. The two largest blind \HI surveys to date are HIPASS (\citealt{staveley-smith1996}, \citealt{barnes2001}, \citealt{koribalski2004}, \citealt{meyer2004}) and ALFALFA (\citealt{giovanelli2005}, \citealt{haynes2011}). While the HIPASS catalogues contain numerous sources without optical or infrared counterpart (e.g. in regions of high Galactic extinction), there is no evidence for any optically dark galaxies (\citealt{ryan-weber2002}, \citealt{koribalski2004}). One of the most massive star-less \HI clouds is HIPASS~J0731-69 ($M\mathrm{_{HI}} \sim 10^9$~M$_{\odot}$; \citealt{ryder2001}). It is located near the asymmetric spiral galaxy NGC~2442 in the NGC~2434 group. A possible sceanrio for its origin is discussed in \citet{bekki2005}. Australia Telescope Compact Array (ATCA) observations by \citet{ryder2004} reveal several \HI clumps that are probably embedded in a common envelope of diffuse H\,{\sevensize I}. A group of \HI clouds found around 400~km~s$^{-1}$, including HIPASS~J1712--64 \citep{kilborn2000}, HIPASS~J1718--59 and HIPASS~J1616--55 \citep{koribalski2004}, are likely tidal debris related to the Magellanic Clouds and the Leading Arm. In areas of high stellar density and dust obscuration, it can be difficult to detect optical or infrared counterparts to H\,{\sevensize I}-detected galaxies. Apart from the Galactic Plane, other sources --- like the Magellanic Clouds -- may hide the stellar component of galaxies. An example is HIPASS~J0546-68 ($v\mathrm{_{sys}} = 1306$~km~s$^{-1}$) behind the LMC \citep{ryan-weber2002}.

The ALFALFA 40~per~cent survey has detected nearly 16,000 \HI sources in the velocity range from $-470$ to $\sim 18,000$~km~s$^{-1}$ \citep{haynes2011}, among these are 1013~sources without assigned optical counterparts. Of the latter, 814 are at velocities from $-470$ to $\sim 340$~km~s$^{-1}$, i.e. likely to be HVCs. The remaining are \HI cloud candidates, with $\sim 75$~per~cent located near galaxies of similar redshifts (e.g. Leo Ring, \citealt{stierwalt2009}). About 50 candidates (0.3~per~cent) for isolated \HI clouds remain; these require follow-up optical and interferometric \HI observations to identify their nature.

In our survey of the Ursa Major region we were able to confirm just one \HI source without a definite optical counterpart (HIJASS~J1219+46; $v\mathrm{_{sys}} \sim 392$~km~s$^{-1}$). It is located near the peculiar galaxy NGC~4288 (HIJASS~J1220+46; $v\mathrm{_{sys}} \sim 520$~km~s$^{-1}$; also see \citealt{wilcots1996} and \citealt{kovac2009}) in the Canes Venatici region. This \HI source is discussed in more detail in Section~\ref{sec:new}. The nature of this \HI source could be a dwarf companion or tidally stipped gas. \HI imaging and deep optical follow-up will be required to determine the true nature.


\section{Summary}

The first blind \HI survey of the Ursa Major region was conducted in 2001--2002 as part of HIJASS. The nearby Ursa Major region is within a velocity range from 300 to 1900~km~s$^{-1}$ and covers an area of $\sim$480~deg$^2$. Using the automated source finder {\small DUCHAMP} we identify 166~\HI sources in the data based on their peak flux density. The measured \HI properties of the sources are presented in a comprehensive catalogue. We address the catalogue completeness and reliability -- we find that the catalogue is 95~per~cent complete above our peak flux density cut (33~mJy). The \HI mass limit of the catalogue is $2.5\times10^8$ M$_{\odot}$ (assuming a velocity width of 100~km~s$^{-1}$ and a distance of 17.1~Mpc). 

The catalogue includes 165~\HI sources for which optical counterparts are identified using NED and SDSS. We find multiple counterparts within the gridded beam for 33~per~cent of the \HI detections, which are candidates for galaxy-galaxy interactions. There are 21 high-probability interacting systems -- about half of these are within the Ursa Major cluster and have been studied by \citet{verheijen2001}. It is remarkable that the Ursa Major cluster as defined by \citet{tully1996} only comprises approximately 12~per~cent of the surveyed volume discussed in this paper but contains about half of all high-probablility candidates for galaxy-galaxy interactions. We discuss four examples of interacting systems, two of which show intriguing extended \HI content. The latter have been already been studied with higher resolution and revealed \HI filaments and \HI distortions of the bright lenticular galaxies.

The HIJASS data contain 10~\HI sources that are newly detected in the 21-cm emission line. One of these does not have a visible associate optical counterpart and is likely to be a candidate galaxy/\HI cloud. This previously uncatalogued \HI source requires follow-up observations to investigate its nature.

\section*{Acknowledgments}

We thank the HIJASS team for providing the data. HIJASS was funded by the UK PPARC (Grant No. PPA/G/S/1998/00620 to Cardiff). We also acknowledge the UK PPARC (Grant No. GR/K28237 to Jodrell Bank) that funded the multibeam receiving system at Jodrell Bank. We thank David Barnes and Mark Calabretta for their help with the fake source injection into the raw data. We thank Felix James `Jay' Lockman for his help with the Green Bank Telescope follow-up observations from setting up, execution to the data reduction. We thank Emma Ryan-Weber, Glenn Kacprzak and the anonymous referee for useful comments. This research has made use of the NASA/IPAC Extragalactic Database, which is operated by the Jet Propulsion Laboratory, California Institute of Technology, under contract with the National Aeronautics and Space Administration. Digitized Sky Survey material is acknowledged, which was produced at the Space Telescope Science Institute under US Government grant NAG W-2166. This research has also made use of the Sloan Digital Sky Survey (SDSS). Funding for the SDSS and SDSS-II was provided by the Alfred P. Sloan Foundation, the Participating Institutions, the National Science Foundation, the U.S. Department of Energy, the National Aeronautics and Space Administration, the Japanese Monbukagakusho, the Max Planck Society, and the Higher Education Funding Council for England. The SDSS was managed by the Astrophysical Research Consortium for the Participating Institutions.


\end{document}